\documentclass[prb,twocolumn,amsmath,amssymb,superscriptaddress]{revtex4-1}
\renewcommand{\figurename}{\textbf{Fig.~\!}}
\usepackage{graphicx}
\usepackage{todonotes} 
\usepackage{dcolumn}

\usepackage{hyperref}
\usepackage{nicefrac,xfrac}
\usepackage{dsfont}
\hypersetup{
    colorlinks= true,
    linkcolor= blue,
    citecolor = blue,
    filecolor= magenta,      
    urlcolor= blue,
    pdftitle={The superconducting grid-states qubit},
    pdfpagemode=FullScreen,
    }
\usepackage{xfrac}
\usepackage{float}
\newcommand{\altfrac}[2]{\ifmmode\def\tmp{$}\else\def\tmp{}\fi\mbox{%
    {\raisebox{.24\ht\strutbox}{\tmp#1\tmp}}%
    \kern-2.2pt\scalebox{1.6}[1.5]{/}\kern-1.8pt%
    {\tmp#2\tmp}%
    }}

\begin{document}
\title{The superconducting grid-states qubit}

\author{Long B. Nguyen}
\thanks{These authors contributed equally. Correspondence to \\ \url{longbnguyen@berkeley.edu} \& \url{hyunkim@berkeley.edu}}
\affiliation{Department of Physics, University of California, Berkeley, CA 94720, USA}
\affiliation{Computational Research Division, Lawrence Berkeley National Laboratory, Berkeley, CA 94720, USA}

\author{Hyunseong Kim}
\thanks{These authors contributed equally. Correspondence to \\ \url{longbnguyen@berkeley.edu} \& \url{hyunkim@berkeley.edu}}
\affiliation{Department of Physics, University of California, Berkeley, CA 94720, USA}
\affiliation{Computational Research Division, Lawrence Berkeley National Laboratory, Berkeley, CA 94720, USA}

\author{Dat T. Le}
\thanks{These authors contributed equally. Correspondence to \\ \url{longbnguyen@berkeley.edu} \& \url{hyunkim@berkeley.edu}}
\affiliation{School of Mathematics and Physics, University of Queensland, Brisbane, QLD 4072, Australia}

\author{Thomas Ersevim}
\affiliation{Department of Physics, University of California, Berkeley, CA 94720, USA}

\author{Sai P. Chitta}
\affiliation{Department of Physics and Astronomy, Northwestern University, Evanston, IL 60208, USA}

\author{Trevor Chistolini}
\affiliation{Department of Physics, University of California, Berkeley, CA 94720, USA}
\affiliation{Computational Research Division, Lawrence Berkeley National Laboratory, Berkeley, CA 94720, USA}

\author{\\ Christian Jünger}
\affiliation{Department of Physics, University of California, Berkeley, CA 94720, USA}
\affiliation{Computational Research Division, Lawrence Berkeley National Laboratory, Berkeley, CA 94720, USA}

\author{W. Clarke Smith}
\affiliation{Google Quantum AI, Santa Barbara, CA 93117, USA 
}

\author{T. M. Stace}
\affiliation{School of Mathematics and Physics, University of Queensland, Brisbane, QLD 4072, Australia}

\author{Jens Koch}
\affiliation{Department of Physics and Astronomy, Northwestern University, Evanston, IL 60208, USA}

\author{David I. Santiago}
\affiliation{Department of Physics, University of California, Berkeley, CA 94720, USA}
\affiliation{Computational Research Division, Lawrence Berkeley National Laboratory, Berkeley, CA 94720, USA}


\author{Irfan Siddiqi}
\affiliation{Department of Physics, University of California, Berkeley, CA 94720, USA}
\affiliation{Computational Research Division, Lawrence Berkeley National Laboratory, Berkeley, CA 94720, USA}


\begin{abstract}

Decoherence errors arising from noisy environments remain a central obstacle to progress in quantum computation and information processing. Quantum error correction (QEC) based on the Gottesman–Kitaev–Preskill (GKP) protocol~\cite{gottesman2001encoding} offers a powerful strategy to overcome this challenge, with successful demonstrations in trapped ions~\cite{fluhmann2019encoding,de2022error}, superconducting circuits~\cite{campagne2020quantum,sivak2023real}, and photonics~\cite{larsen2025integrated}. Beyond active QEC, a compelling alternative is to engineer Hamiltonians that intrinsically enforce stabilizers, offering passive protection akin to topological models~\cite{douccot2012physical,dodge2023hardware}. Inspired by the GKP encoding scheme, we implement a superconducting qubit whose eigenstates form protected grid states—long envisioned but not previously realized~\cite{le2019doubly,rymarz2021hardware}—by integrating an effective Cooper-quartet junction~\cite{bell2014protected,larsen2020parity,smith2022magnifying} with a quantum phase-slip element embedded in a high-impedance circuit~\cite{astafiev2012coherent,pechenezhskiy2020superconducting}. Spectroscopic measurements reveal pairs of degenerate states separated by large energy gaps, in excellent agreement with theoretical predictions. Remarkably, our observations indicate that the circuit tolerates small disorders and gains robustness against environmental noise as its parameters approach the ideal regime, establishing a new framework for exploring superconducting hardware. These findings also showcase the versatility of the superconducting circuit toolbox, setting the stage for future exploration of advanced solid-state devices with emergent properties.


\end{abstract}

\maketitle





\noindent The crux of quantum error correction, essential for achieving fault-tolerant quantum computation, involves encoding logical quantum information within the expansive Hilbert space of a physical system in such a way that any local errors that occur can be detected and corrected efficiently. Different encoding strategies can be implemented by utilizing a discrete-variable system comprising interconnected physical qubits~\cite{terhal2015quantum}, or through a continuous-variable platform such as a cavity coupled to a nonlinear ancilla~\cite{cai2021bosonic}. Executing quantum error correction protocols is, however, inherently complex, involving multiple error-prone steps. In addition, it is daunting to maintain high fidelity in both the preparation of the code space and the detection of the error syndromes as the system size increases.

On the other hand, one can construct a physical system governed by a Hamiltonian composed of syndrome operators that inhibit errors from happening, $\hat{\mathcal{H}}=-\sum_k E_k \hat{\mathcal{S}}_k$. Here, $\{\hat{\mathcal{S}}_k \}$ are the stabilizer generators, and $\{E_k\}$ are very large compared to the interaction rates between the system and the environmental noise sources~\cite{douccot2012physical,Vuillot2024homological}. This represents a hardware-efficient pathway toward fault tolerance, as the protection of the encoded quantum information is embedded directly within the system. Such an approach is often considered as passive quantum error correction~\cite{douccot2012physical,Vuillot2024homological}, with pioneering concepts encompassing topologically protected qubits~\cite{ioffe2002topologically,bacon2008stability,brooks2013protected} and topological quantum computing based on non-Abelian anyons~\cite{kitaev2003fault,nayak2008nonabelian}. However, these ideas are often considered impractical due to the requirement of system dimensions~\cite{bravyi2009no} or scaling complexity~\cite{gyenis2021experimental,dodge2023hardware}, akin to conventional quantum systems. Implementing hardware encoding of discrete-variable stabilizers has thus proven to be a significant challenge.

Bosonic continuous-variable QEC using the GKP code~\cite{gottesman2001encoding} has emerged as a hardware-efficient approach toward fault-tolerant quantum computation~\cite{fluhmann2019encoding,de2022error,campagne2020quantum,sivak2023real}. In this scheme, quantum information is encoded using the oscillator's states that form a grid of delta functions, such that small displacement errors can be corrected by reverting the lattice to the original positions. Notably, the computational GKP states are eigenstates of the stabilizers $\hat{\mathcal{S}}_Z=e^{i2\sqrt{\pi}\hat x }$ and $\hat{\mathcal{S}}_X=e^{-i2\sqrt{\pi}\hat{p}}$, where $[\hat{x},\hat{p}]=i$. This suggests that GKP states can be naturally encoded in a system described by the Hamiltonian
$
    \hat{\mathcal{H}}_\mathrm{GKP}=-E_x\cos(\sqrt{2\pi d}~\hat{x})-E_p\cos(\sqrt{2\pi d}~\hat{p}),
$
which has $d$-fold degenerate ground states forming grids in phase space, with the relative energy dispersion determined by the ratio $E_x/E_p$~\cite{gottesman2001encoding,rymarz2021hardware,brady2024advances}. 
Recent theoretical proposals have introduced the hardware encodings of $d$=1 and $d$=2 grid states in the superconducting dualmon~\cite{le2019doubly} and in fluxonium-gyrator-fluxonium circuits~\cite{rymarz2021hardware}, respectively. These concepts remain challenging to realize in practice.

In this work, we implement a superconducting architecture that hosts grid-like eigenstates by combining a Cooper-quartet-tunneling (CQT) circuit element~\cite{smith2022magnifying} with a quantum phase-slip (QPS) junction embedded within a high-impedance environment~\cite{pechenezhskiy2020superconducting}. The resulting circuit, \textit{gridium}, is fabricated, controlled, and readout using conventional circuit quantum electrodynamics (cQED) techniques, thereby promising scalability and compatibility. Spectroscopic characterization reveals excitation spectra that exhibit excellent agreement with theoretical predictions across multiple parameter regimes. Our comprehensive measurements, spanning diverse parameters and traversing across various flux points, confirm the circuit’s intrinsic resilience against the noisy environment.

\begin{figure}[t]
    \includegraphics[width=0.48\textwidth]{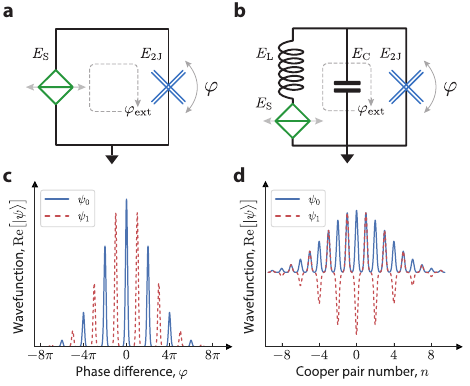}
    \caption{\label{fig1} \textbf{Gridium superconducting artificial atom}. \\ (\textbf{a}) A superconducting circuit model implementing the Hamiltonian given by Eq.~\ref{eqn:GKP_ideal}. The CQT junction (double \textcolor{blue}{blue} cross) allows charges to tunnel in units of $4e$. It is connected in parallel to the QPS junction (\textcolor{green}{green} diamond), which enables vortices to tunnel in and out of the loop. The circuit encloses an external magnetic flux $\varphi_\mathrm{ext}$. (\textbf{b}) Practical circuit model representing the gridium qubit, as described by Eq.~\ref{eqn:GKP_approx}. The circuit includes an inductive element with energy $E_\mathrm{L}$, and a capacitor with charging energy $E_\mathrm{C}$. (\textbf{c}) $\varphi$-basis wavefunctions of the two lowest eigenstates of the circuit shown in panel \textbf{b} for $E_\mathrm{2J},E_\mathrm{S}\gg E_\mathrm{C},E_\mathrm{L}$ and $\varphi_\mathrm{ext}=\large{\sfrac{\pi}{2}}$. (\textbf{d}) $n$-basis wavefunctions of the same two states.}
\end{figure}

\begin{figure*}[t]
    \includegraphics[width=0.95\textwidth]{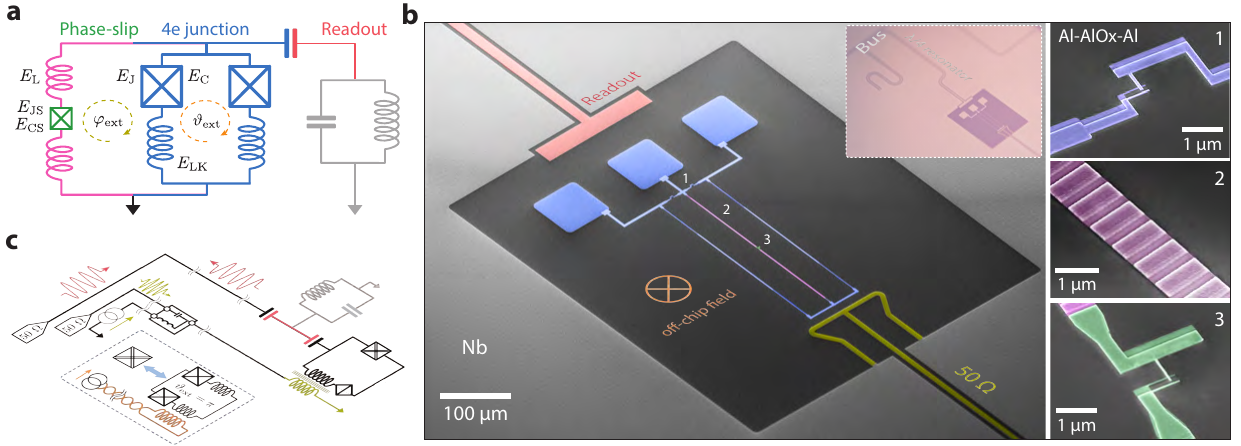}
    \caption{\label{fig2} \textbf{Device implementation}. (\textbf{a}) Circuit diagram of the gridium qubit. A small Josephson junction (\textcolor{green}{green}, Josephson energy $E_\mathrm{JS}$ and charging energy $E_\mathrm{CS}$)  embedded within the large inductor (\textcolor{pink}{pink}, inductive energy $E_\mathrm{L}$) facilitates coherent quantum phase-slips through the outer loop encircling external flux $\varphi_\mathrm{ext}$. The CQT junction is constructed by threading an external flux $\vartheta_\mathrm{ext}=\pi$ through a rhombus of pair-wise symmetric inductors (energy $E_\mathrm{LK}$) and capacitively shunted Josephson junctions with Josephson energy $E_\mathrm{J}$ and charging energy $E_\mathrm{C}$ (\textcolor{blue}{blue} loop).  The circuit is capacitively coupled to an ancillary resonator (\textcolor{gray}{gray}) for dispersive readout. (\textbf{b}) False-color scanning-electron micrograph of the device. The \textcolor{blue}{blue} electrodes and encircling inductors form the CQT junction. The \textcolor{pink}{pink} inductor, encompassing the QPS element,  splits the blue loop in equal halves. Inset: optical micrograph showing a gridium qubit coupled to external circuitry. Zoomed-in images of the circuit elements are shown in the rightmost numbered panels.  (\textbf{c}) Schematic depicting the experiment. An off-chip magnetic field (\textcolor{orange}{orange}) is applied to tune $\vartheta_\mathrm{ext}$. An integrated differential flux line (\textcolor{yellow}{yellow}) is used to tune $\varphi_\mathrm{ext}$ and irradiate the device at radio frequencies. To probe the state of the qubit, the capacitively coupled resonator (\textcolor{gray}{gray}) is measured using RF pulses (\textcolor{red}{red}).}
\end{figure*}

\begin{figure*}[t]
    \includegraphics[width=\textwidth]{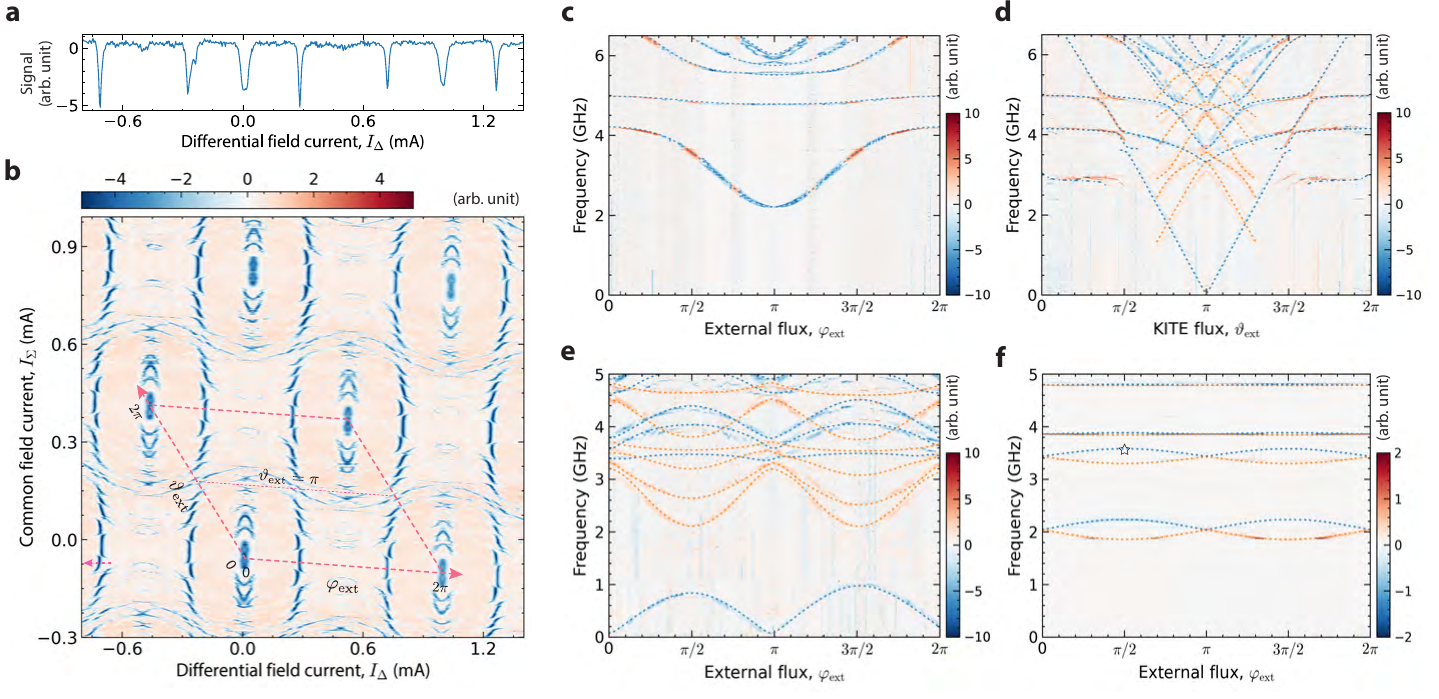}
    \caption{\label{fig3} \textbf{Radio-frequency spectroscopy}. (\textbf{a}) Variation of the normalized radio-frequency signal with respect to the differential control current, $I_\Delta$. The microwave drive is applied close to the resonator resonance frequency, $\omega_\mathrm{r}$. (\textbf{b}) Modulation of reflective resonator signal with respect to the common ($I_\Sigma$) and differential ($I_\Delta$) DC currents. The resonant coupling between the qubit and the resonator manifests as various divergent features. The dashed axes indicate the alignment
    for $\varphi_\mathrm{ext}$ and $\vartheta_\mathrm{ext}$. The symmetry reveals the anticipated regions of $\pi$-periodicity. 
    The dashed arrow indicates the $I_\Sigma$ value corresponding to the data in panel \textbf{a}. (\textbf{c}) Flux-dependent two-tone spectrum of a device with relatively small $E_\mathrm{J}$ at $\vartheta_\mathrm{ext}=0$. The dashed lines represent least-square fit using multi-mode model (SI Note~4). (\textbf{d}) Spectrum of the same device with respect to the KITE flux $\vartheta_\mathrm{ext}$ at fixed $\varphi_\mathrm{ext}=0$. The blue and red dashed lines show the numerically computed transitions from $|\psi_0\rangle$ and $|\psi_1\rangle$, respectively. The dash-dotted lines indicate transitions from higher levels such as $|\psi_2\rangle$. (\textbf{e}) Spectrum of the device at $\vartheta_\mathrm{ext}=\pi$. Numerical least-square fit using the extended GKP Hamiltonian given by Eq.~\ref{eqn:GKP_approx} yields effective circuit parameters $[E_\mathrm{2J},E_\mathrm{S},E_\mathrm{C},E_\mathrm{L}]/h\approx[3.97,2.95,0.42,0.41]~\mathrm{GHz}$. (\textbf{f}) Spectrum along $\vartheta_\mathrm{ext}=\pi$ of another device with larger energy ratios and increased protection. Least-square numerical parameter extraction using Eq.~\ref{eqn:GKP_approx} yields $[E_\mathrm{2J},E_\mathrm{S},E_\mathrm{C},E_\mathrm{L}]/h\approx[9.11,3.94,0.23,0.11]~\mathrm{GHz}$.}
\end{figure*}

\vspace{0.2em}

\noindent\textbf{Gridium concepts}

\noindent The GKP Hamiltonian, doubly periodic in both position and momentum space, describes the motion of a Bloch electron confined within a two-dimensional plane under the effect of a perpendicular magnetic field~\cite{gottesman2001encoding,rymarz2021hardware}. The eigenstates of the system can be characterized by Bloch quantum numbers in the Zak basis~\cite{zak1968dynamics,le2019doubly}, forming $d$-fold degenerate Landau levels, a well-known concept in quantum Hall physics~\cite{tong2016lecturesquantumhalleffect}. The motion of the electron via magnetic translation operators naturally implements the GKP code words~\cite{gottesman2001encoding,rymarz2021hardware}. In superconducting circuits, the equivalent Hamiltonian hosting grid states with two-fold degeneracy is given as
\begin{equation}
    \label{eqn:GKP_ideal}
    \hat{\mathcal{H}}_\mathrm{GKP}=-E_\mathrm{S}\cos(2\pi \hat{n}) + E_\mathrm{2J}\cos(2\hat{\varphi}),
\end{equation}
where $\hat{n}=\hat{q}/2e$ embodies the normalized charge on a superconducting electrode, $\hat{\varphi}$ is the conjugate phase operator satisfying the commutation relation $[\hat{\varphi},\hat{n}]=i$, and the signs preserve the potential maximum at $\varphi=\pi$~\cite{smith2022magnifying}.


The archetypal Hamiltonian in Eq.~\ref{eqn:GKP_ideal} emerges from a phenomenological electrical circuit formed by shunting a QPS junction with a CQT element, as shown in Fig.~\ref{fig1}\textbf{a}. The QPS junction facilitates the coherent tunneling of single-fluxon states (or circulating quantum vortices) with amplitude $E_\mathrm{S}$, leading to the nonlinear $\cos(2\pi \hat{n})$ term. Meanwhile, the CQT element allows charges to tunnel across the connected electrodes only in units of $4e$, or Cooper quartets, with effective energy $E_\mathrm{2J}$. The other charge transmission processes are forbidden. This circuit forms a loop that can enclose a finite external flux $\varphi_\mathrm{ext}$. Reminiscent of the quantum repetition error correction code, the eigenstates of the circuit are superpositions of infinitely squeezed charge and phase states, resembling combs of $\delta$-functions.

In practice, electrical circuits always contain capacitive and inductive elements. Therefore, a more practical superconducting artificial atom must inadvertently include elements corresponding to these effects, leading to the extended GKP Hamiltonian,
\begin{equation}
    \label{eqn:GKP_approx}
\hat{\mathcal{H}}=E_\mathrm{C}\hat{n}^2+\frac{1}{2}E_\mathrm{L}(\hat{\varphi}+\varphi_\mathrm{ext})^2-E_\mathrm{S}\cos(2\pi \hat{n}) + E_\mathrm{2J}\cos(2\hat{\varphi}),
\end{equation}
where $E_\mathrm{C}$ and $E_\mathrm{L}$ denote the charging and inductive energies, respectively. The additional quadratic terms reduce the squeezing of the wavefunctions in phase space, and at the same time form a parabolic confinement enveloping the grid states. We show in Fig.~\ref{fig1}\textbf{c,d} that when the quadratic corrections in  Hamiltonian~(\ref{eqn:GKP_approx}) are sufficiently small, the first two eigenstates still form large grids, albeit with a finite Gaussian envelope. Notably, as opposed to the ideal grid states spanning the entire phase space, the eigenstates of the circuit in Fig.~\ref{fig1}\textbf{b} form normalizable code words, analogous to the approximate GKP states prepared in cavities using active error correction~\cite{campagne2020quantum,sivak2023real}. We hereafter refer to circuits with this topology as gridium. 

Once encoded into the doubly degenerate eigenstates of the artificial atom, quantum information is inherently protected from local perturbations. The computational eigenstates, $|\psi_0\rangle$ and $|\psi_1\rangle$, are coherent superpositions of alternating phase and charge peaks with identical parities, as shown in Fig.~\ref{fig1}\textbf{c,d}. This configuration enforces the cancellation of linear matrix elements, thereby suppressing noise-induced transitions within the encoded subspace. Notably, the protection arises not simply from spatial disjointness of the wavefunctions, but rather from their selection rules as dictated by symmetry, which nullify dipole couplings even in the presence of wavefunction overlap. This principle extends to suppress phase and charge matrix elements throughout the entire flux period, exemplifying protection by parity (SI Note~1).

The superconducting circuit depicted in Fig.~\ref{fig1} remains vulnerable to common fluctuations inherent to solid-state environment. On one hand, the circuit loop encircles a finite magnetic flux, leading to possible flux-noise dephasing. On the other hand, the QPS junction represents a nonlinear capacitor interrupting the loop, introducing possible decoherence due to charge parity switching.  However, the artificial atom's eigenstates are superpositions of multiple charge and phase states, resembling the nonlocal encoding characteristic of quantum error-correcting codes. For small quadratic energy terms, this confers intrinsic protection against such fluctuations, manifested as suppressed frequency dispersions with respect to external flux and charge offset.

\vspace{0.2em}

\noindent\textbf{Circuit implementation}

\noindent The Josephson tunneling of Cooper pairs between two isolated superconducting grains forms the backbone of cQED~\cite{blais2021circuit}. The quantum duality between charge and flux dictates the formation of the exact counterpart of charge tunneling known as coherent QPS~\cite{mooij2006superconducting}, which has been experimentally observed in superconducting wires interrupted by small constrictions~\cite{pop2010measurement,astafiev2012coherent,manucharyan2012evidence}. Here, the state of the wires can be described by $|m\rangle$, where~$m$~is the number of flux quanta encircled by the loops, and the constrictions facilitate the tunneling of magnetic fluxes, shifting the enclosed flux in units of flux quantum, $|m\rangle \leftrightarrow |m\pm 1\rangle$\footnote{Or more generally, $|m\rangle \leftrightarrow |m\pm k\rangle$}, with an associated phase-slip energy $E_\mathrm{S}$. This paradigmatic framework can be applied to describe the dynamics of fluxonium qubit~\cite{manucharyan2012evidence}, the emergence of quasicharge effects~\cite{pechenezhskiy2020superconducting}, and the observation of quantized current Shapiro steps~\cite{shaikhaidarov2022quantized,crescini2023evidence}.

Notably, a hyperinductor embedding a miniaturized Josephson junction has shed light on the coherent insulating response of the junction~\cite{pechenezhskiy2020superconducting}. In this context, the Cooper-pair tunneling is analogous to Bragg reflection, the system exhibits a prominent $2e$-periodic charging energy, and the large inductance stabilizes the insulating behavior, satisfying the condition $E_\mathrm{S}\gg E_\mathrm{L}$ that corresponds to small quadratic corrections. Here, we employ a similar approach by adding a small Josephson junction with Josephson energy $E_\mathrm{JS}$ and charging energy $E_\mathrm{CS}$ to a nominal array of junctions with total inductive energy $E_\mathrm{L}$. The parameters are designed to reach a QPS amplitude $E_\mathrm{S}$ equivalent to a few $\mathrm{GHz}$ (SI Note~2). 

Meanwhile, the existence of higher-order Cooper-pair tunneling processes distorts the current-phase relation from the standard sinusoidal shape~\cite{willsch2024observation}. Remarkably, this paradigmatic behavior is also present in a series circuit consisting of a Josephson junction connected to an inductor~\cite{spanton2017current-phase}, where the effective amplitudes of higher harmonic terms can be enhanced by increasing the inductance. On the other hand, akin to photon interference in a Mach-Zehnder interferometer, the destructive interference via the Aharonov-Bohm effect dictates the cancellation of single-Cooper-pair transport across a loop consisting of two identical nonlinear elements when it is biased at optimal flux frustration~\cite{aharonov1959significance}. The combination of these properties leads to the observation and verification of $\cos(2\varphi)$-dominant phenomena in superconducting quantum interference devices with rhombus geometry~\cite{bell2014protected,smith2022magnifying}, which effectively implement CQT junctions (SI Note~3). 

Merging the QPS and the CQT elements results in a superconducting circuit as shown in Fig.~\ref{fig2}\textbf{a}. The interplay between the circuit parameters determines the properties of its eigenstates, which manifest as grid states with large Gaussian envelopes at $\vartheta_\mathrm{ext}=\pi$ for $E_\mathrm{L}\ll E_\mathrm{JS}<E_\mathrm{CS}$ and $E_\mathrm{J}\gg E_\mathrm{C},E_\mathrm{LK}$. This circuit model thus implements the extended GKP Hamiltonian given by Eq.~\ref{eqn:GKP_approx}, satisfying the condition $E_\mathrm{S},E_\mathrm{2J}\gg E_\mathrm{L},E_\mathrm{C}$. Notably, the inductors inside the rhombus on one hand tailor the effective CQT amplitude, and on the other hand contribute to the effective inductive energy of the circuit (SI Note~4). In addition, a resonator is connected to the qubit capacitively for dispersive readout (SI Note~5).

Figure~\ref{fig2}\textbf{b} illustrates the physical implementation of the device. Here, the blue superconducting electrodes serve as shunting capacitors with charging energy $E_\mathrm{C}$ (SI~Note~6). Each side electrode is connected to the central one via a Josephson junction with energy $E_\mathrm{J}$, and to each other via an array of Josephson junctions configured into a loop, through which an external magnetic field can be applied to tune $\vartheta_\mathrm{ext}$ and realize the CQT (blue rhombus in Fig.~\ref{fig2}\textbf{a}). Symmetrically splitting this loop is a superinductor that incorporates a quantum phase-slip (QPS) junction at its midpoint (pink and green elements in Fig.~\ref{fig2}\textbf{a}, respectively). This arrangement allows us to selectively tune $\varphi_\mathrm{ext}$ using a differential on-chip flux line. In addition, connecting this line to radio-frequency instruments enables coherent device control via microwave pulses. The circuit is capacitively coupled to a quarter-wave coplanar waveguide resonator for dispersive readout. The measurement and control circuitry is shown in Fig.~\ref{fig2}\textbf{c}. The inductive energies $E_\mathrm{L}$ and $E_\mathrm{LK}$ can be adjusted by varying the junction arrays (SI~Note~6). The device fabrication features a niobium ground plane layer integrated with sub-$\mathrm{\mu m}$ Dolan-style Al-AlO$_\mathrm{x}$-Al Josephson junctions (SI~Note~7).

\noindent\textbf{Spectral signatures}

\noindent To observe clear signatures of the gridium qubit and validate the theoretical framework, we begin the experiment with a device exhibiting strong responses to control and readout. This is achieved using a circuit configuration with relatively low $E_\mathrm{J}$ and high $E_\mathrm{L}$ (SI~Note~5). The properties of the device are probed using standard radio-frequency reflectometry. The coupling between qubit transitions with the resonator leads to a measurable shift in the resonator frequency, consistent with cQED principles. As the qubit is tuned with either flux bias, its transitions undergo periodic modulations, altering the microwave signal reflected from the resonator.  In the first measurement, a continuous-wave microwave tone is applied near the bare resonance frequency $\omega_\mathrm{r}$ of the readout resonator, and the reflected response is monitored as the differential flux current $I_\Delta$ is swept. A representative response is shown in Fig.~\ref{fig3}\textbf{a} to illustrate the qubit's flux-tunable behavior. By performing two-dimensional (2D) sweeps of both common and differential bias currents while monitoring the demodulated IQ trace near the dressed resonance, we can unveil the flux mapping parameters and reveal in detail the regions of interest.

Figure~\ref{fig3}\textbf{b} shows the microwave readout signal as both the common and differential control currents are varied, displaying a doubly periodic pattern. The resonant interaction between the qubit and the resonator gives rise to distinctive features, manifesting as avoided-crossing-like divergences. The differential flux generated by $I_\Delta$ predominantly tunes $\varphi_\mathrm{ext}$, a consequence of the symmetric splitting of the circuit loops. In contrast, the common current $I_\Sigma$ influences both $\vartheta_\mathrm{ext}$ and $\varphi_\mathrm{ext}$. Owing to the circuit’s geometry, $I_\Sigma$ adjusts $\vartheta_\mathrm{ext}$ at twice the rate of $\varphi_\mathrm{ext}$, resulting in a current-to-flux tuning relationship characterized by a parallelogram pattern. Quantifying this correlation enables precise tracking and compensation of the applied external fluxes (see SI Note~8). By appropriately combining $I_\Sigma$ and $I_\Delta$, we can navigate the 2D flux space to access regions of particular interest.

We next perform two-tone spectroscopy by sweeping the frequency of a microwave drive applied along the differential flux line. The resulting  $\varphi_\mathrm{ext}$-dependence spectrum of the device at fixed $\vartheta_\mathrm{ext}=0$, shown in  Fig.~\ref{fig3}\textbf{c}, embodies the experimental realization of the dualmon Hamiltonian~\cite{le2019doubly}, which essentially hosts $d$=1 grid states,
\begin{equation}\label{eqn:dualmon}
    \hat{\mathcal{H}}=4E_\mathrm{C}\hat{n}^2+\frac{1}{2}E_\mathrm{L}(\hat{\varphi}+\varphi_\mathrm{ext})^2-E_\mathrm{S}\cos(2\pi \hat{n}) - E_\mathrm{J}\cos\hat{\varphi}.
\end{equation}
Mapping the first transition to this model, we extract the effective circuit parameters $[E_\mathrm{J},E_\mathrm{S},E_\mathrm{C},E_\mathrm{L}] /h$ = $[5.4,2.26,0.47,0.341]~\mathrm{GHz}$. Notably, our implementation focuses on the regime corresponding to small quadratic contributions from the capacitive and inductive energies, as opposed to the parameter regime considered in the previous theoretical analysis~\cite{le2019doubly}.

Likewise, we sweep the KITE flux $\vartheta_\mathrm{ext}$ while holding the external flux fixed at $\varphi_\mathrm{ext} = 0$, yielding the spectrum shown in Fig.~\ref{fig3}\textbf{d}. The resulting $\vartheta_\mathrm{ext}$-dependent features closely resemble those of the heavy fluxonium qubit\cite{lin2018demonstration,earnest2018realization}. Remarkably, our circuit model provides an excellent fit to the observed transitions up to the readout resonator frequency, underscoring the strong agreement between theoretical predictions and experimental observations.

Building on this flux control, we access the spectral domain governed by the gridium Hamiltonian supporting $d$=2 grid states (Eq.~\ref{eqn:GKP_approx}) by biasing $\vartheta_\mathrm{ext}$ close to $\pi$. As shown in Fig.~\ref{fig3}\textbf{e}, the spectrum reveals doublets that become quasi-degenerate at symmetric flux points, with degeneracy consistently lifted away from these frustrations in a characteristic pattern. These doublets are separated from each other by large energy gaps, a hallmark of the dual evolution of the coherent quantum phase slip and Cooper-quartet tunneling dynamics. Mapping the visible transitions to Hamiltonian (\ref{eqn:GKP_approx}) allows us to extract the effective parameters which exhibit excellent agreement with our analysis (SI~Notes~4\&5). Interestingly, the spectral degeneracy near $\varphi_\mathrm{ext} = \pi$ is preserved even when $\vartheta_\mathrm{ext}$ deviates slightly from its ideal value of $\pi$. Rather than being lifted, the degeneracy point shifts modestly along the flux axis, indicating an inherent robustness against small disorder (Extended Data Fig.~\ref{efig1}).

We observe fluctuations in the visible qubit transitions, manifested as linewidth broadening and occasional discontinuities in the spectra. In this parameter regime, the eigenstates do not form extended grids in charge space, resulting in intrinsic charge sensitivity. We therefore attribute the observed fluctuations to charge noise. Using dispersive readout at fixed flux bias, we monitored the resonator response and recorded sporadic jumps. Analysis of these jumps yields a $1/f^\alpha$ noise power spectral density, consistent with the charge-noise model, as expected (Extended Data Fig.~\ref{efig2}).

We proceed to explore devices operating deeper within the GKP regime, characterized by increased ratios of $E_\mathrm{2J}/E_\mathrm{C}$ and $E_\mathrm{S}/E_\mathrm{L}$. This change essentially enhances protection from decoherence while simultaneously diminishing the visibility of the qubit transitions. Panel (\textbf{f}) in Fig.~\ref{fig3} showcases the spectrum of such a device. In this domain, the dipole moments of low-order transitions become negligible, rendering them undetectable in spectroscopy and preventing direct probing of the computational states (SI Note~5). Instead, transitions involving higher excited levels and multi-photon processes remain accessible, and can be leveraged to examine and characterize the circuit’s properties. 

Biasing the KITE flux $\vartheta_\mathrm{ext}$ at $\pi$ enforces the ideal $\pi$-periodicity along $\varphi_\mathrm{ext}$, realizing the hallmark behavior of a true gridium qubit. In this regime, the expanded grid support simultaneously suppresses flux and charge sensitivity: the quadratic inductive term is strongly quenched, producing broad phase superpositions and flat flux dispersion, while enhanced charge-space delocalization eliminates charge dispersion and yields exceptionally stable, narrow spectral lines (Extended Data Fig.~\ref{efig2}, SI Notes 1 and 5). In addition, we detect no random fluctuations of the readout resonator. These observations provide compelling experimental evidence that, as the system approaches the canonical GKP Hamiltonian~(\ref{eqn:GKP_ideal}), its computational subspace acquires intrinsic protection against decoherence.\\

\noindent\textbf{Temporal dynamics}

\begin{figure}[t]
    \includegraphics[width=0.43\textwidth]{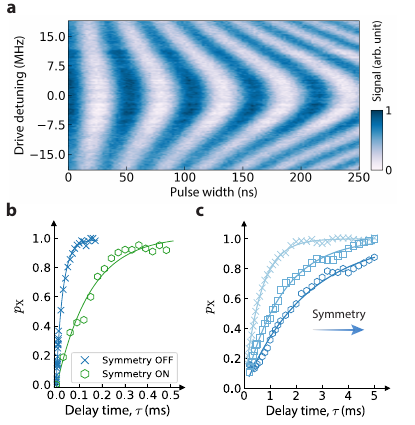}
    \caption{\label{fig4} \textbf{Dynamical response}. (\textbf{a}) Rabi chevron corresponding to a transition allowed by selection rules, marked by the star in Fig.~\ref{fig3}\textbf{f}. (\textbf{b}) Bit-flip probabilities of the less-protected device, whose spectra are shown in Fig.~\ref{fig3}\textbf{c-e}; the symmetry-off trace corresponds to a visible transition used for state preparation and readout, while the symmetry-on signal is obtained from fast-flux-assisted $T_X$ measurements. (\textbf{c}) Bit-flip probabilities of a more protected device (Extended Data Fig.~\ref{efig3}); the left-most trace is taken at the operation flux point, and predistorted baseband pulses are used to probe dynamics near the doubly symmetric flux bias.}
\end{figure}

\noindent The circuit’s dynamical behavior is accessed through tailored sequences of baseband and radio-frequency pulses. By connecting the chip to an additional integrated flux line (SI Note~8), we achieve fast control over both $\vartheta_\mathrm{ext}$ and $\varphi_\mathrm{ext}$.  Time-resolved characterization is particularly direct for transitions that are spectroscopically visible and symmetry-allowed, even for devices approaching the limit of Hamiltonian (\ref{eqn:GKP_ideal}). As a representative example, we perform a standard Rabi experiment on the transition that is visible in Fig.~\ref{fig3}\textbf{f}, observing rapid oscillations that form a quintessential chevron pattern as a function of pulse duration and drive detuning (Fig.~\ref{fig4}\textbf{a}). Notably, these allowed transitions facilitate efficient manipulation of the involved states, with a $\pi$-pulse duration of $\sim 25$~ns. Together, these results demonstrate that the circuit supports coherent quantum dynamics, matching the forecast behaviors (SI Note~5).

To complement the spectroscopic signatures of reduced charge and flux sensitivity, we investigate the relaxation dynamics of the circuit. Although the degenerate computational subspace cannot be accessed directly, the rich spectra enable the use of spectroscopically visible transitions in combination with fast-flux pulses for state preparation and readout (SI Note 10). For the less protected device, a standard relaxation measurement at a flux bias corresponding to a visible $0-1$ transition yields a bit-flip time of $29(1)~\mathrm{\mu s}$. With a predistorted flux pulse applied to idle the qubit at the doubly symmetric flux instead, the bit-flip time is enhanced fivefold to $0.15(2)~\mathrm{ms}$ (Fig.~\ref{fig4}\textbf{b}).

In a more protected device with relatively smaller quadratic confinements (Extended Data Fig.~\ref{efig3}), bit-flip resilience is already evident at the asymmetric flux point where the qubit can be probed directly, with $T_X=0.63(1)~\mathrm{ms}$. Applying predistorted flux pulses and idling the qubit closer to the symmetry point, we observe a notable enhancement of the bit-flip time, with $T_X$ increasing to $1.46(6)~\mathrm{ms}$ and reaching $2.4(2)~\mathrm{ms}$ as the qubit approaches the doubly symmetric flux bias (Fig.~\ref{fig4}\textbf{c}). This progression reveals robust protection from bit-flip processes as the device approaches its ideal operating regime, providing evidence that the gridium architecture achieves the improved noise resilience as theoretically envisioned.\\


\noindent\textbf{Discussion and Outlook}

\noindent Building on the principles of circuit quantum electrodynamics, we realize superconducting devices governed by the doubly nonlinear extended GKP Hamiltonian, achieving hardware-encoded grid states that confirm a long-standing theoretical prediction. By in-situ flux tuning, we access both the dualmon ($d$=1) and gridium ($d$=2) regimes, and characterize multiple devices spanning parameter space from less-protected, easily measurable configurations to more-protected, harder-to-control ones. The excellent agreement between experiment and theory underscores the feasibility of realizing protected qubits within this framework.

While the present discussion focuses on the doubly degenerate grid space ($d$=2), the dualmon Hamiltonian ($d$=1) has been proposed for noise-resilient encodings, motivating further study~\cite{le2019doubly}. A simpler realization of such a qubit would remove the need for the KITE component, thereby reducing hardware complexity and simplifying operation. In this context, the gridium’s Cooper-pair tunneling element could alternatively be realized using novel superconducting structures, such as stacked $\mathrm{d}$–$\mathrm{s}$ or $\mathrm{d}$–$\mathrm{d}$ flakes~\cite{brosco2024flowermon,patel2024dmon}. This approach trades some operational versatility for architectural simplicity, opening promising directions at the interface of device engineering and quantum materials research.

Importantly, fully harnessing the device’s intrinsic noise protection will require gate protocols that preserve this protection, particularly in regimes approaching the ideal GKP Hamiltonian (Eq.~\ref{eqn:GKP_ideal}), where error suppression is strongest. This challenge echoes braiding in topological qubits~\cite{kitaev2003fault} and noise-protected gates in the 0–$\pi$ architecture~\cite{brooks2013protected}, where operations must remain confined to a protected manifold to preserve the intrinsic noise immunity. Notably, the richness of the visible spectrum points to logical operations achievable through projective measurement, offering a natural route to realizing protected gates~\cite{aasen2025roadmap}. Meanwhile, symmetry-allowed transitions to higher levels suggest pathways to controlled-$\mathrm{Z}$ entangling gates, though mitigating leakage will require advanced techniques such as erasure conversion.

Beyond quantum computing applications, the doubly-nonlinear circuit introduced here provides a versatile platform to investigate intriguing physical phenomena, such as the dynamics of Bloch electrons confined to a two-dimensional lattice under a perpendicular homogeneous magnetic field, characterized by the distinctive Hofstadter butterfly spectrum~\cite{rymarz2021hardware}. The high impedance and geometry of our circuit, in combination with the observations of degeneracy of higher energy levels, suggest the emergence of novel topological features that may lack direct analogs in condensed matter systems, opening exciting opportunities for applications in quantum sensing and metrology~\cite{peyruchat2024spectral}. The presented framework thus offers a potentially transformative avenue for uncovering previously unexplored quantum landscapes and advancing fundamental quantum science and technology. 
\clearpage


\noindent\textbf{Acknowledgments}\\
\noindent The authors thank Bingcheng Qing, Ke Wang, Larry Chen, Gerwin Koolstra, Alessandro Ciani, and Philippe Campagne-Ibarcq for fruitful discussions. This material is based upon work supported by the U.S. Department of Energy, Office of Science, National Quantum Information Science Research Centers, Quantum Systems Accelerator under contract DE-AC02-05CH11231.\\

\noindent\textbf{Competing interest}\\
\noindent The authors declare no competing interests.\\

\noindent\textbf{Data availability}\\
\noindent The supporting data and numerical code used in this study are available from the corresponding authors upon reasonable request.




\bibliography{apssamp}

\begin{thebibliography}{57}%
\makeatletter
\providecommand \@ifxundefined [1]{%
 \@ifx{#1\undefined}
}%
\providecommand \@ifnum [1]{%
 \ifnum #1\expandafter \@firstoftwo
 \else \expandafter \@secondoftwo
 \fi
}%
\providecommand \@ifx [1]{%
 \ifx #1\expandafter \@firstoftwo
 \else \expandafter \@secondoftwo
 \fi
}%
\providecommand \natexlab [1]{#1}%
\providecommand \enquote  [1]{``#1''}%
\providecommand \bibnamefont  [1]{#1}%
\providecommand \bibfnamefont [1]{#1}%
\providecommand \citenamefont [1]{#1}%
\providecommand \href@noop [0]{\@secondoftwo}%
\providecommand \href [0]{\begingroup \@sanitize@url \@href}%
\providecommand \@href[1]{\@@startlink{#1}\@@href}%
\providecommand \@@href[1]{\endgroup#1\@@endlink}%
\providecommand \@sanitize@url [0]{\catcode `\\12\catcode `\$12\catcode `\&12\catcode `\#12\catcode `\^12\catcode `\_12\catcode `\%12\relax}%
\providecommand \@@startlink[1]{}%
\providecommand \@@endlink[0]{}%
\providecommand \url  [0]{\begingroup\@sanitize@url \@url }%
\providecommand \@url [1]{\endgroup\@href {#1}{\urlprefix }}%
\providecommand \urlprefix  [0]{URL }%
\providecommand \Eprint [0]{\href }%
\providecommand \doibase [0]{http://dx.doi.org/}%
\providecommand \selectlanguage [0]{\@gobble}%
\providecommand \bibinfo  [0]{\@secondoftwo}%
\providecommand \bibfield  [0]{\@secondoftwo}%
\providecommand \translation [1]{[#1]}%
\providecommand \BibitemOpen [0]{}%
\providecommand \bibitemStop [0]{}%
\providecommand \bibitemNoStop [0]{.\EOS\space}%
\providecommand \EOS [0]{\spacefactor3000\relax}%
\providecommand \BibitemShut  [1]{\csname bibitem#1\endcsname}%
\let\auto@bib@innerbib\@empty
\bibitem [{\citenamefont {Gottesman}\ \emph {et~al.}(2001)\citenamefont {Gottesman}, \citenamefont {Kitaev},\ and\ \citenamefont {Preskill}}]{gottesman2001encoding}%
  \BibitemOpen
  \bibfield  {author} {\bibinfo {author} {\bibfnamefont {D.}~\bibnamefont {Gottesman}}, \bibinfo {author} {\bibfnamefont {A.}~\bibnamefont {Kitaev}}, \ and\ \bibinfo {author} {\bibfnamefont {J.}~\bibnamefont {Preskill}},\ }\href {\doibase 10.1103/PhysRevA.64.012310} {\bibfield  {journal} {\bibinfo  {journal} {Phys. Rev. A}\ }\textbf {\bibinfo {volume} {64}},\ \bibinfo {pages} {012310} (\bibinfo {year} {2001})}\BibitemShut {NoStop}%
\bibitem [{\citenamefont {Fl{\"u}hmann}\ \emph {et~al.}(2019)\citenamefont {Fl{\"u}hmann}, \citenamefont {Nguyen}, \citenamefont {Marinelli}, \citenamefont {Negnevitsky}, \citenamefont {Mehta},\ and\ \citenamefont {Home}}]{fluhmann2019encoding}%
  \BibitemOpen
  \bibfield  {author} {\bibinfo {author} {\bibfnamefont {C.}~\bibnamefont {Fl{\"u}hmann}}, \bibinfo {author} {\bibfnamefont {T.~L.}\ \bibnamefont {Nguyen}}, \bibinfo {author} {\bibfnamefont {M.}~\bibnamefont {Marinelli}}, \bibinfo {author} {\bibfnamefont {V.}~\bibnamefont {Negnevitsky}}, \bibinfo {author} {\bibfnamefont {K.}~\bibnamefont {Mehta}}, \ and\ \bibinfo {author} {\bibfnamefont {J.~P.}\ \bibnamefont {Home}},\ }\href {\doibase 10.1038/s41586-019-0960-6} {\bibfield  {journal} {\bibinfo  {journal} {Nature}\ }\textbf {\bibinfo {volume} {566}},\ \bibinfo {pages} {513} (\bibinfo {year} {2019})}\BibitemShut {NoStop}%
\bibitem [{\citenamefont {De~Neeve}\ \emph {et~al.}(2022)\citenamefont {De~Neeve}, \citenamefont {Nguyen}, \citenamefont {Behrle},\ and\ \citenamefont {Home}}]{de2022error}%
  \BibitemOpen
  \bibfield  {author} {\bibinfo {author} {\bibfnamefont {B.}~\bibnamefont {De~Neeve}}, \bibinfo {author} {\bibfnamefont {T.-L.}\ \bibnamefont {Nguyen}}, \bibinfo {author} {\bibfnamefont {T.}~\bibnamefont {Behrle}}, \ and\ \bibinfo {author} {\bibfnamefont {J.~P.}\ \bibnamefont {Home}},\ }\href {\doibase 10.1038/s41567-021-01487-7} {\bibfield  {journal} {\bibinfo  {journal} {Nat. Phys.}\ }\textbf {\bibinfo {volume} {18}},\ \bibinfo {pages} {296} (\bibinfo {year} {2022})}\BibitemShut {NoStop}%
\bibitem [{\citenamefont {Campagne-Ibarcq}\ \emph {et~al.}(2020)\citenamefont {Campagne-Ibarcq}, \citenamefont {Eickbusch}, \citenamefont {Touzard}, \citenamefont {Zalys-Geller}, \citenamefont {Frattini}, \citenamefont {Sivak}, \citenamefont {Reinhold}, \citenamefont {Puri}, \citenamefont {Shankar}, \citenamefont {Schoelkopf} \emph {et~al.}}]{campagne2020quantum}%
  \BibitemOpen
  \bibfield  {author} {\bibinfo {author} {\bibfnamefont {P.}~\bibnamefont {Campagne-Ibarcq}}, \bibinfo {author} {\bibfnamefont {A.}~\bibnamefont {Eickbusch}}, \bibinfo {author} {\bibfnamefont {S.}~\bibnamefont {Touzard}}, \bibinfo {author} {\bibfnamefont {E.}~\bibnamefont {Zalys-Geller}}, \bibinfo {author} {\bibfnamefont {N.~E.}\ \bibnamefont {Frattini}}, \bibinfo {author} {\bibfnamefont {V.~V.}\ \bibnamefont {Sivak}}, \bibinfo {author} {\bibfnamefont {P.}~\bibnamefont {Reinhold}}, \bibinfo {author} {\bibfnamefont {S.}~\bibnamefont {Puri}}, \bibinfo {author} {\bibfnamefont {S.}~\bibnamefont {Shankar}}, \bibinfo {author} {\bibfnamefont {R.~J.}\ \bibnamefont {Schoelkopf}},  \emph {et~al.},\ }\href {\doibase 10.1038/s41586-020-2603-3} {\bibfield  {journal} {\bibinfo  {journal} {Nature}\ }\textbf {\bibinfo {volume} {584}},\ \bibinfo {pages} {368} (\bibinfo {year} {2020})}\BibitemShut {NoStop}%
\bibitem [{\citenamefont {Sivak}\ \emph {et~al.}(2023)\citenamefont {Sivak}, \citenamefont {Eickbusch}, \citenamefont {Royer}, \citenamefont {Singh}, \citenamefont {Tsioutsios}, \citenamefont {Ganjam}, \citenamefont {Miano}, \citenamefont {Brock}, \citenamefont {Ding}, \citenamefont {Frunzio} \emph {et~al.}}]{sivak2023real}%
  \BibitemOpen
  \bibfield  {author} {\bibinfo {author} {\bibfnamefont {V.~V.}\ \bibnamefont {Sivak}}, \bibinfo {author} {\bibfnamefont {A.}~\bibnamefont {Eickbusch}}, \bibinfo {author} {\bibfnamefont {B.}~\bibnamefont {Royer}}, \bibinfo {author} {\bibfnamefont {S.}~\bibnamefont {Singh}}, \bibinfo {author} {\bibfnamefont {I.}~\bibnamefont {Tsioutsios}}, \bibinfo {author} {\bibfnamefont {S.}~\bibnamefont {Ganjam}}, \bibinfo {author} {\bibfnamefont {A.}~\bibnamefont {Miano}}, \bibinfo {author} {\bibfnamefont {B.~L.}\ \bibnamefont {Brock}}, \bibinfo {author} {\bibfnamefont {A.~Z.}\ \bibnamefont {Ding}}, \bibinfo {author} {\bibfnamefont {L.}~\bibnamefont {Frunzio}},  \emph {et~al.},\ }\href {\doibase 10.1038/s41586-023-05782-6} {\bibfield  {journal} {\bibinfo  {journal} {Nature}\ }\textbf {\bibinfo {volume} {616}},\ \bibinfo {pages} {50} (\bibinfo {year} {2023})}\BibitemShut {NoStop}%
\bibitem [{\citenamefont {Larsen}\ \emph {et~al.}(2025)\citenamefont {Larsen}, \citenamefont {Bourassa}, \citenamefont {Kocsis}, \citenamefont {Tasker}, \citenamefont {Chadwick}, \citenamefont {Gonz{\'a}lez-Arciniegas}, \citenamefont {Hastrup}, \citenamefont {Lopetegui-Gonz{\'a}lez}, \citenamefont {Miatto}, \citenamefont {Motamedi} \emph {et~al.}}]{larsen2025integrated}%
  \BibitemOpen
  \bibfield  {author} {\bibinfo {author} {\bibfnamefont {M.~V.}\ \bibnamefont {Larsen}}, \bibinfo {author} {\bibfnamefont {J.~E.}\ \bibnamefont {Bourassa}}, \bibinfo {author} {\bibfnamefont {S.}~\bibnamefont {Kocsis}}, \bibinfo {author} {\bibfnamefont {J.~F.}\ \bibnamefont {Tasker}}, \bibinfo {author} {\bibfnamefont {R.~S.}\ \bibnamefont {Chadwick}}, \bibinfo {author} {\bibfnamefont {C.}~\bibnamefont {Gonz{\'a}lez-Arciniegas}}, \bibinfo {author} {\bibfnamefont {J.}~\bibnamefont {Hastrup}}, \bibinfo {author} {\bibfnamefont {C.~E.}\ \bibnamefont {Lopetegui-Gonz{\'a}lez}}, \bibinfo {author} {\bibfnamefont {F.~M.}\ \bibnamefont {Miatto}}, \bibinfo {author} {\bibfnamefont {A.}~\bibnamefont {Motamedi}},  \emph {et~al.},\ }\href {\doibase 10.1038/s41586-025-09044-5} {\bibfield  {journal} {\bibinfo  {journal} {Nature}\ }\textbf {\bibinfo {volume} {642}},\ \bibinfo {pages} {587} (\bibinfo {year} {2025})}\BibitemShut {NoStop}%
\bibitem [{\citenamefont {Dou{\c{c}}ot}\ and\ \citenamefont {Ioffe}(2012)}]{douccot2012physical}%
  \BibitemOpen
  \bibfield  {author} {\bibinfo {author} {\bibfnamefont {B.}~\bibnamefont {Dou{\c{c}}ot}}\ and\ \bibinfo {author} {\bibfnamefont {L.~B.}\ \bibnamefont {Ioffe}},\ }\href {\doibase 10.1088/0034-4885/75/7/072001} {\bibfield  {journal} {\bibinfo  {journal} {Reports on Progress in Physics}\ }\textbf {\bibinfo {volume} {75}},\ \bibinfo {pages} {072001} (\bibinfo {year} {2012})}\BibitemShut {NoStop}%
\bibitem [{\citenamefont {Dodge}\ \emph {et~al.}(2023)\citenamefont {Dodge}, \citenamefont {Liu}, \citenamefont {Klots}, \citenamefont {Cole}, \citenamefont {Shearrow}, \citenamefont {Senatore}, \citenamefont {Zhu}, \citenamefont {Ioffe}, \citenamefont {McDermott},\ and\ \citenamefont {Plourde}}]{dodge2023hardware}%
  \BibitemOpen
  \bibfield  {author} {\bibinfo {author} {\bibfnamefont {K.}~\bibnamefont {Dodge}}, \bibinfo {author} {\bibfnamefont {Y.}~\bibnamefont {Liu}}, \bibinfo {author} {\bibfnamefont {A.~R.}\ \bibnamefont {Klots}}, \bibinfo {author} {\bibfnamefont {B.}~\bibnamefont {Cole}}, \bibinfo {author} {\bibfnamefont {A.}~\bibnamefont {Shearrow}}, \bibinfo {author} {\bibfnamefont {M.}~\bibnamefont {Senatore}}, \bibinfo {author} {\bibfnamefont {S.}~\bibnamefont {Zhu}}, \bibinfo {author} {\bibfnamefont {L.~B.}\ \bibnamefont {Ioffe}}, \bibinfo {author} {\bibfnamefont {R.}~\bibnamefont {McDermott}}, \ and\ \bibinfo {author} {\bibfnamefont {B.~L.~T.}\ \bibnamefont {Plourde}},\ }\href {\doibase 10.1103/PhysRevLett.131.150602} {\bibfield  {journal} {\bibinfo  {journal} {Phys. Rev. Lett.}\ }\textbf {\bibinfo {volume} {131}},\ \bibinfo {pages} {150602} (\bibinfo {year} {2023})}\BibitemShut {NoStop}%
\bibitem [{\citenamefont {Le}\ \emph {et~al.}(2019)\citenamefont {Le}, \citenamefont {Grimsmo}, \citenamefont {M.{\"u}ller},\ and\ \citenamefont {Stace}}]{le2019doubly}%
  \BibitemOpen
  \bibfield  {author} {\bibinfo {author} {\bibfnamefont {D.~T.}\ \bibnamefont {Le}}, \bibinfo {author} {\bibfnamefont {A.}~\bibnamefont {Grimsmo}}, \bibinfo {author} {\bibfnamefont {C.}~\bibnamefont {M.{\"u}ller}}, \ and\ \bibinfo {author} {\bibfnamefont {T.~M.}\ \bibnamefont {Stace}},\ }\href {\doibase 10.1103/PhysRevA.100.062321} {\bibfield  {journal} {\bibinfo  {journal} {Phys. Rev. A}\ }\textbf {\bibinfo {volume} {100}},\ \bibinfo {pages} {062321} (\bibinfo {year} {2019})}\BibitemShut {NoStop}%
\bibitem [{\citenamefont {Rymarz}\ \emph {et~al.}(2021)\citenamefont {Rymarz}, \citenamefont {Bosco}, \citenamefont {Ciani},\ and\ \citenamefont {DiVincenzo}}]{rymarz2021hardware}%
  \BibitemOpen
  \bibfield  {author} {\bibinfo {author} {\bibfnamefont {M.}~\bibnamefont {Rymarz}}, \bibinfo {author} {\bibfnamefont {S.}~\bibnamefont {Bosco}}, \bibinfo {author} {\bibfnamefont {A.}~\bibnamefont {Ciani}}, \ and\ \bibinfo {author} {\bibfnamefont {D.~P.}\ \bibnamefont {DiVincenzo}},\ }\href {\doibase 10.1103/PhysRevX.11.011032} {\bibfield  {journal} {\bibinfo  {journal} {Phys. Rev. X}\ }\textbf {\bibinfo {volume} {11}},\ \bibinfo {pages} {011032} (\bibinfo {year} {2021})}\BibitemShut {NoStop}%
\bibitem [{\citenamefont {Bell}\ \emph {et~al.}(2014)\citenamefont {Bell}, \citenamefont {Paramanandam}, \citenamefont {Ioffe},\ and\ \citenamefont {Gershenson}}]{bell2014protected}%
  \BibitemOpen
  \bibfield  {author} {\bibinfo {author} {\bibfnamefont {M.~T.}\ \bibnamefont {Bell}}, \bibinfo {author} {\bibfnamefont {J.}~\bibnamefont {Paramanandam}}, \bibinfo {author} {\bibfnamefont {L.~B.}\ \bibnamefont {Ioffe}}, \ and\ \bibinfo {author} {\bibfnamefont {M.~E.}\ \bibnamefont {Gershenson}},\ }\href {\doibase 10.1103/PhysRevLett.112.167001} {\bibfield  {journal} {\bibinfo  {journal} {Phys. Rev. Lett.}\ }\textbf {\bibinfo {volume} {112}},\ \bibinfo {pages} {167001} (\bibinfo {year} {2014})}\BibitemShut {NoStop}%
\bibitem [{\citenamefont {Larsen}\ \emph {et~al.}(2020)\citenamefont {Larsen}, \citenamefont {Gershenson}, \citenamefont {Casparis}, \citenamefont {Kringh{\o}j}, \citenamefont {Pearson}, \citenamefont {McNeil}, \citenamefont {Kuemmeth}, \citenamefont {Krogstrup}, \citenamefont {Petersson},\ and\ \citenamefont {Marcus}}]{larsen2020parity}%
  \BibitemOpen
  \bibfield  {author} {\bibinfo {author} {\bibfnamefont {T.~W.}\ \bibnamefont {Larsen}}, \bibinfo {author} {\bibfnamefont {M.~E.}\ \bibnamefont {Gershenson}}, \bibinfo {author} {\bibfnamefont {L.}~\bibnamefont {Casparis}}, \bibinfo {author} {\bibfnamefont {A.}~\bibnamefont {Kringh{\o}j}}, \bibinfo {author} {\bibfnamefont {N.~J.}\ \bibnamefont {Pearson}}, \bibinfo {author} {\bibfnamefont {R.~P.~G.}\ \bibnamefont {McNeil}}, \bibinfo {author} {\bibfnamefont {F.}~\bibnamefont {Kuemmeth}}, \bibinfo {author} {\bibfnamefont {P.}~\bibnamefont {Krogstrup}}, \bibinfo {author} {\bibfnamefont {K.~D.}\ \bibnamefont {Petersson}}, \ and\ \bibinfo {author} {\bibfnamefont {C.~M.}\ \bibnamefont {Marcus}},\ }\href {\doibase 10.1103/PhysRevLett.125.056801} {\bibfield  {journal} {\bibinfo  {journal} {Phys. Rev. Lett.}\ }\textbf {\bibinfo {volume} {125}},\ \bibinfo {pages} {056801} (\bibinfo {year} {2020})}\BibitemShut {NoStop}%
\bibitem [{\citenamefont {Smith}\ \emph {et~al.}(2022)\citenamefont {Smith}, \citenamefont {Villiers}, \citenamefont {Marquet}, \citenamefont {Palomo}, \citenamefont {Delbecq}, \citenamefont {Kontos}, \citenamefont {Campagne-Ibarcq}, \citenamefont {Dou{\c{c}}ot},\ and\ \citenamefont {Leghtas}}]{smith2022magnifying}%
  \BibitemOpen
  \bibfield  {author} {\bibinfo {author} {\bibfnamefont {W.~C.}\ \bibnamefont {Smith}}, \bibinfo {author} {\bibfnamefont {M.}~\bibnamefont {Villiers}}, \bibinfo {author} {\bibfnamefont {A.}~\bibnamefont {Marquet}}, \bibinfo {author} {\bibfnamefont {J.}~\bibnamefont {Palomo}}, \bibinfo {author} {\bibfnamefont {M.~R.}\ \bibnamefont {Delbecq}}, \bibinfo {author} {\bibfnamefont {T.}~\bibnamefont {Kontos}}, \bibinfo {author} {\bibfnamefont {P.}~\bibnamefont {Campagne-Ibarcq}}, \bibinfo {author} {\bibfnamefont {B.}~\bibnamefont {Dou{\c{c}}ot}}, \ and\ \bibinfo {author} {\bibfnamefont {Z.}~\bibnamefont {Leghtas}},\ }\href {\doibase 10.1103/PhysRevX.12.021002} {\bibfield  {journal} {\bibinfo  {journal} {Phys. Rev. X}\ }\textbf {\bibinfo {volume} {12}},\ \bibinfo {pages} {021002} (\bibinfo {year} {2022})}\BibitemShut {NoStop}%
\bibitem [{\citenamefont {Astafiev}\ \emph {et~al.}(2012)\citenamefont {Astafiev}, \citenamefont {Ioffe}, \citenamefont {Kafanov}, \citenamefont {Pashkin}, \citenamefont {Arutyunov}, \citenamefont {Shahar}, \citenamefont {Cohen},\ and\ \citenamefont {Tsai}}]{astafiev2012coherent}%
  \BibitemOpen
  \bibfield  {author} {\bibinfo {author} {\bibfnamefont {O.~V.}\ \bibnamefont {Astafiev}}, \bibinfo {author} {\bibfnamefont {L.~B.}\ \bibnamefont {Ioffe}}, \bibinfo {author} {\bibfnamefont {S.}~\bibnamefont {Kafanov}}, \bibinfo {author} {\bibfnamefont {Y.~A.}\ \bibnamefont {Pashkin}}, \bibinfo {author} {\bibfnamefont {K.~Y.}\ \bibnamefont {Arutyunov}}, \bibinfo {author} {\bibfnamefont {D.}~\bibnamefont {Shahar}}, \bibinfo {author} {\bibfnamefont {O.}~\bibnamefont {Cohen}}, \ and\ \bibinfo {author} {\bibfnamefont {J.~S.}\ \bibnamefont {Tsai}},\ }\href {\doibase 10.1038/nature10930} {\bibfield  {journal} {\bibinfo  {journal} {Nature}\ }\textbf {\bibinfo {volume} {484}},\ \bibinfo {pages} {355} (\bibinfo {year} {2012})}\BibitemShut {NoStop}%
\bibitem [{\citenamefont {Pechenezhskiy}\ \emph {et~al.}(2020)\citenamefont {Pechenezhskiy}, \citenamefont {Mencia}, \citenamefont {Nguyen}, \citenamefont {Lin},\ and\ \citenamefont {Manucharyan}}]{pechenezhskiy2020superconducting}%
  \BibitemOpen
  \bibfield  {author} {\bibinfo {author} {\bibfnamefont {I.~V.}\ \bibnamefont {Pechenezhskiy}}, \bibinfo {author} {\bibfnamefont {R.~A.}\ \bibnamefont {Mencia}}, \bibinfo {author} {\bibfnamefont {L.~B.}\ \bibnamefont {Nguyen}}, \bibinfo {author} {\bibfnamefont {Y.-H.}\ \bibnamefont {Lin}}, \ and\ \bibinfo {author} {\bibfnamefont {V.~E.}\ \bibnamefont {Manucharyan}},\ }\href {\doibase 10.1038/s41586-020-2687-9} {\bibfield  {journal} {\bibinfo  {journal} {Nature}\ }\textbf {\bibinfo {volume} {585}},\ \bibinfo {pages} {368} (\bibinfo {year} {2020})}\BibitemShut {NoStop}%
\bibitem [{\citenamefont {Terhal}(2015)}]{terhal2015quantum}%
  \BibitemOpen
  \bibfield  {author} {\bibinfo {author} {\bibfnamefont {B.~M.}\ \bibnamefont {Terhal}},\ }\href {\doibase 10.1103/RevModPhys.87.307} {\bibfield  {journal} {\bibinfo  {journal} {Rev. Mod. Phys.}\ }\textbf {\bibinfo {volume} {87}},\ \bibinfo {pages} {307} (\bibinfo {year} {2015})}\BibitemShut {NoStop}%
\bibitem [{\citenamefont {Cai}\ \emph {et~al.}(2021)\citenamefont {Cai}, \citenamefont {Ma}, \citenamefont {Wang}, \citenamefont {Zou},\ and\ \citenamefont {Sun}}]{cai2021bosonic}%
  \BibitemOpen
  \bibfield  {author} {\bibinfo {author} {\bibfnamefont {W.}~\bibnamefont {Cai}}, \bibinfo {author} {\bibfnamefont {Y.}~\bibnamefont {Ma}}, \bibinfo {author} {\bibfnamefont {W.}~\bibnamefont {Wang}}, \bibinfo {author} {\bibfnamefont {C.-L.}\ \bibnamefont {Zou}}, \ and\ \bibinfo {author} {\bibfnamefont {L.}~\bibnamefont {Sun}},\ }\href {\doibase 10.1016/j.fmre.2020.12.006} {\bibfield  {journal} {\bibinfo  {journal} {Fundam. Res.}\ }\textbf {\bibinfo {volume} {1}},\ \bibinfo {pages} {50} (\bibinfo {year} {2021})}\BibitemShut {NoStop}%
\bibitem [{\citenamefont {Vuillot}\ \emph {et~al.}(2024)\citenamefont {Vuillot}, \citenamefont {Ciani},\ and\ \citenamefont {Terhal}}]{Vuillot2024homological}%
  \BibitemOpen
  \bibfield  {author} {\bibinfo {author} {\bibfnamefont {C.}~\bibnamefont {Vuillot}}, \bibinfo {author} {\bibfnamefont {A.}~\bibnamefont {Ciani}}, \ and\ \bibinfo {author} {\bibfnamefont {B.~M.}\ \bibnamefont {Terhal}},\ }\href {\doibase 10.1007/s00220-023-04905-4} {\bibfield  {journal} {\bibinfo  {journal} {Commun. Math. Phys.}\ }\textbf {\bibinfo {volume} {405}},\ \bibinfo {pages} {787} (\bibinfo {year} {2024})}\BibitemShut {NoStop}%
\bibitem [{\citenamefont {Ioffe}\ \emph {et~al.}(2002)\citenamefont {Ioffe}, \citenamefont {Feigel'man}, \citenamefont {Ioselevich}, \citenamefont {Ivanov}, \citenamefont {Troyer},\ and\ \citenamefont {Blatter}}]{ioffe2002topologically}%
  \BibitemOpen
  \bibfield  {author} {\bibinfo {author} {\bibfnamefont {L.~B.}\ \bibnamefont {Ioffe}}, \bibinfo {author} {\bibfnamefont {M.~V.}\ \bibnamefont {Feigel'man}}, \bibinfo {author} {\bibfnamefont {A.}~\bibnamefont {Ioselevich}}, \bibinfo {author} {\bibfnamefont {D.}~\bibnamefont {Ivanov}}, \bibinfo {author} {\bibfnamefont {M.}~\bibnamefont {Troyer}}, \ and\ \bibinfo {author} {\bibfnamefont {G.}~\bibnamefont {Blatter}},\ }\href {\doibase 10.1038/415503a} {\bibfield  {journal} {\bibinfo  {journal} {Nature}\ }\textbf {\bibinfo {volume} {415}},\ \bibinfo {pages} {503} (\bibinfo {year} {2002})}\BibitemShut {NoStop}%
\bibitem [{\citenamefont {Bacon}(2008)}]{bacon2008stability}%
  \BibitemOpen
  \bibfield  {author} {\bibinfo {author} {\bibfnamefont {D.}~\bibnamefont {Bacon}},\ }\href {\doibase 10.1103/PhysRevA.78.042324} {\bibfield  {journal} {\bibinfo  {journal} {Phys. Rev. A}\ }\textbf {\bibinfo {volume} {78}},\ \bibinfo {pages} {042324} (\bibinfo {year} {2008})}\BibitemShut {NoStop}%
\bibitem [{\citenamefont {Brooks}\ \emph {et~al.}(2013)\citenamefont {Brooks}, \citenamefont {Kitaev},\ and\ \citenamefont {Preskill}}]{brooks2013protected}%
  \BibitemOpen
  \bibfield  {author} {\bibinfo {author} {\bibfnamefont {P.}~\bibnamefont {Brooks}}, \bibinfo {author} {\bibfnamefont {A.}~\bibnamefont {Kitaev}}, \ and\ \bibinfo {author} {\bibfnamefont {J.}~\bibnamefont {Preskill}},\ }\href {\doibase 10.1103/PhysRevA.87.052306} {\bibfield  {journal} {\bibinfo  {journal} {Phys. Rev. A}\ }\textbf {\bibinfo {volume} {87}},\ \bibinfo {pages} {052306} (\bibinfo {year} {2013})}\BibitemShut {NoStop}%
\bibitem [{\citenamefont {Kitaev}(2003)}]{kitaev2003fault}%
  \BibitemOpen
  \bibfield  {author} {\bibinfo {author} {\bibfnamefont {A.~Y.}\ \bibnamefont {Kitaev}},\ }\href {\doibase 10.1016/S0003-4916(02)00018-0} {\bibfield  {journal} {\bibinfo  {journal} {Ann. Phys.}\ }\textbf {\bibinfo {volume} {303}},\ \bibinfo {pages} {2} (\bibinfo {year} {2003})}\BibitemShut {NoStop}%
\bibitem [{\citenamefont {Nayak}\ \emph {et~al.}(2008)\citenamefont {Nayak}, \citenamefont {Simon}, \citenamefont {Stern}, \citenamefont {Freedman},\ and\ \citenamefont {Das~Sarma}}]{nayak2008nonabelian}%
  \BibitemOpen
  \bibfield  {author} {\bibinfo {author} {\bibfnamefont {C.}~\bibnamefont {Nayak}}, \bibinfo {author} {\bibfnamefont {S.~H.}\ \bibnamefont {Simon}}, \bibinfo {author} {\bibfnamefont {A.}~\bibnamefont {Stern}}, \bibinfo {author} {\bibfnamefont {M.}~\bibnamefont {Freedman}}, \ and\ \bibinfo {author} {\bibfnamefont {S.}~\bibnamefont {Das~Sarma}},\ }\href {\doibase 10.1103/RevModPhys.80.1083} {\bibfield  {journal} {\bibinfo  {journal} {Rev. Mod. Phys.}\ }\textbf {\bibinfo {volume} {80}},\ \bibinfo {pages} {1083} (\bibinfo {year} {2008})}\BibitemShut {NoStop}%
\bibitem [{\citenamefont {Bravyi}\ and\ \citenamefont {Terhal}(2009)}]{bravyi2009no}%
  \BibitemOpen
  \bibfield  {author} {\bibinfo {author} {\bibfnamefont {S.}~\bibnamefont {Bravyi}}\ and\ \bibinfo {author} {\bibfnamefont {B.}~\bibnamefont {Terhal}},\ }\href {\doibase 10.1088/1367-2630/11/4/043029} {\bibfield  {journal} {\bibinfo  {journal} {New J. Phys.}\ }\textbf {\bibinfo {volume} {11}},\ \bibinfo {pages} {043029} (\bibinfo {year} {2009})}\BibitemShut {NoStop}%
\bibitem [{\citenamefont {Gyenis}\ \emph {et~al.}(2021)\citenamefont {Gyenis}, \citenamefont {Mundada}, \citenamefont {Di~Paolo}, \citenamefont {Hazard}, \citenamefont {You}, \citenamefont {Schuster}, \citenamefont {Koch}, \citenamefont {Blais},\ and\ \citenamefont {Houck}}]{gyenis2021experimental}%
  \BibitemOpen
  \bibfield  {author} {\bibinfo {author} {\bibfnamefont {A.}~\bibnamefont {Gyenis}}, \bibinfo {author} {\bibfnamefont {P.~S.}\ \bibnamefont {Mundada}}, \bibinfo {author} {\bibfnamefont {A.}~\bibnamefont {Di~Paolo}}, \bibinfo {author} {\bibfnamefont {T.~M.}\ \bibnamefont {Hazard}}, \bibinfo {author} {\bibfnamefont {X.}~\bibnamefont {You}}, \bibinfo {author} {\bibfnamefont {D.~I.}\ \bibnamefont {Schuster}}, \bibinfo {author} {\bibfnamefont {J.}~\bibnamefont {Koch}}, \bibinfo {author} {\bibfnamefont {A.}~\bibnamefont {Blais}}, \ and\ \bibinfo {author} {\bibfnamefont {A.~A.}\ \bibnamefont {Houck}},\ }\href {\doibase 10.1103/PRXQuantum.2.010339} {\bibfield  {journal} {\bibinfo  {journal} {PRX Quantum}\ }\textbf {\bibinfo {volume} {2}},\ \bibinfo {pages} {010339} (\bibinfo {year} {2021})}\BibitemShut {NoStop}%
\bibitem [{\citenamefont {Brady}\ \emph {et~al.}(2024)\citenamefont {Brady}, \citenamefont {Eickbusch}, \citenamefont {Singh}, \citenamefont {Wu},\ and\ \citenamefont {Zhuang}}]{brady2024advances}%
  \BibitemOpen
  \bibfield  {author} {\bibinfo {author} {\bibfnamefont {A.~J.}\ \bibnamefont {Brady}}, \bibinfo {author} {\bibfnamefont {A.}~\bibnamefont {Eickbusch}}, \bibinfo {author} {\bibfnamefont {S.}~\bibnamefont {Singh}}, \bibinfo {author} {\bibfnamefont {J.}~\bibnamefont {Wu}}, \ and\ \bibinfo {author} {\bibfnamefont {Q.}~\bibnamefont {Zhuang}},\ }\href {\doibase 10.1016/j.pquantelec.2023.100496} {\bibfield  {journal} {\bibinfo  {journal} {Prog. Quantum Electron.}\ }\textbf {\bibinfo {volume} {93}},\ \bibinfo {pages} {100496} (\bibinfo {year} {2024})}\BibitemShut {NoStop}%
\bibitem [{\citenamefont {Zak}(1968)}]{zak1968dynamics}%
  \BibitemOpen
  \bibfield  {author} {\bibinfo {author} {\bibfnamefont {J.}~\bibnamefont {Zak}},\ }\href {\doibase 10.1103/PhysRev.168.686} {\bibfield  {journal} {\bibinfo  {journal} {Phys. Rev.}\ }\textbf {\bibinfo {volume} {168}},\ \bibinfo {pages} {686} (\bibinfo {year} {1968})}\BibitemShut {NoStop}%
\bibitem [{\citenamefont {Tong}(2016)}]{tong2016lecturesquantumhalleffect}%
  \BibitemOpen
  \bibfield  {author} {\bibinfo {author} {\bibfnamefont {D.}~\bibnamefont {Tong}},\ }\href@noop {} {\enquote {\bibinfo {title} {Lectures on the {Quantum Hall Effect}},}\ } (\bibinfo {year} {2016}),\ \Eprint {http://arxiv.org/abs/1606.06687} {arXiv:1606.06687 [hep-th]} \BibitemShut {NoStop}%
\bibitem [{\citenamefont {Blais}\ \emph {et~al.}(2021)\citenamefont {Blais}, \citenamefont {Grimsmo}, \citenamefont {Girvin},\ and\ \citenamefont {Wallraff}}]{blais2021circuit}%
  \BibitemOpen
  \bibfield  {author} {\bibinfo {author} {\bibfnamefont {A.}~\bibnamefont {Blais}}, \bibinfo {author} {\bibfnamefont {A.~L.}\ \bibnamefont {Grimsmo}}, \bibinfo {author} {\bibfnamefont {S.~M.}\ \bibnamefont {Girvin}}, \ and\ \bibinfo {author} {\bibfnamefont {A.}~\bibnamefont {Wallraff}},\ }\href {\doibase 10.1103/RevModPhys.93.025005} {\bibfield  {journal} {\bibinfo  {journal} {Rev. Mod. Phys.}\ }\textbf {\bibinfo {volume} {93}},\ \bibinfo {pages} {025005} (\bibinfo {year} {2021})}\BibitemShut {NoStop}%
\bibitem [{\citenamefont {Mooij}\ and\ \citenamefont {Nazarov}(2006)}]{mooij2006superconducting}%
  \BibitemOpen
  \bibfield  {author} {\bibinfo {author} {\bibfnamefont {J.~E.}\ \bibnamefont {Mooij}}\ and\ \bibinfo {author} {\bibfnamefont {Y.~V.}\ \bibnamefont {Nazarov}},\ }\href {\doibase 10.1038/nphys234} {\bibfield  {journal} {\bibinfo  {journal} {Nat. Phys.}\ }\textbf {\bibinfo {volume} {2}},\ \bibinfo {pages} {169} (\bibinfo {year} {2006})}\BibitemShut {NoStop}%
\bibitem [{\citenamefont {Pop}\ \emph {et~al.}(2010)\citenamefont {Pop}, \citenamefont {Protopopov}, \citenamefont {Lecocq}, \citenamefont {Peng}, \citenamefont {Pannetier}, \citenamefont {Buisson},\ and\ \citenamefont {Guichard}}]{pop2010measurement}%
  \BibitemOpen
  \bibfield  {author} {\bibinfo {author} {\bibfnamefont {I.~M.}\ \bibnamefont {Pop}}, \bibinfo {author} {\bibfnamefont {I.}~\bibnamefont {Protopopov}}, \bibinfo {author} {\bibfnamefont {F.}~\bibnamefont {Lecocq}}, \bibinfo {author} {\bibfnamefont {Z.}~\bibnamefont {Peng}}, \bibinfo {author} {\bibfnamefont {B.}~\bibnamefont {Pannetier}}, \bibinfo {author} {\bibfnamefont {O.}~\bibnamefont {Buisson}}, \ and\ \bibinfo {author} {\bibfnamefont {W.}~\bibnamefont {Guichard}},\ }\href {\doibase 10.1038/nphys1697} {\bibfield  {journal} {\bibinfo  {journal} {Nat. Phys.}\ }\textbf {\bibinfo {volume} {6}},\ \bibinfo {pages} {589} (\bibinfo {year} {2010})}\BibitemShut {NoStop}%
\bibitem [{\citenamefont {Manucharyan}\ \emph {et~al.}(2012)\citenamefont {Manucharyan}, \citenamefont {Masluk}, \citenamefont {Kamal}, \citenamefont {Koch}, \citenamefont {Glazman},\ and\ \citenamefont {Devoret}}]{manucharyan2012evidence}%
  \BibitemOpen
  \bibfield  {author} {\bibinfo {author} {\bibfnamefont {V.~E.}\ \bibnamefont {Manucharyan}}, \bibinfo {author} {\bibfnamefont {N.~A.}\ \bibnamefont {Masluk}}, \bibinfo {author} {\bibfnamefont {A.}~\bibnamefont {Kamal}}, \bibinfo {author} {\bibfnamefont {J.}~\bibnamefont {Koch}}, \bibinfo {author} {\bibfnamefont {L.~I.}\ \bibnamefont {Glazman}}, \ and\ \bibinfo {author} {\bibfnamefont {M.~H.}\ \bibnamefont {Devoret}},\ }\href {\doibase 10.1103/PhysRevB.85.024521} {\bibfield  {journal} {\bibinfo  {journal} {Phys. Rev. B}\ }\textbf {\bibinfo {volume} {85}},\ \bibinfo {pages} {024521} (\bibinfo {year} {2012})}\BibitemShut {NoStop}%
\bibitem [{Note1()}]{Note1}%
  \BibitemOpen
  \bibinfo {note} {Or more generally, $|m\rangle \leftrightarrow |m\pm k\rangle $}\BibitemShut {NoStop}%
\bibitem [{\citenamefont {Shaikhaidarov}\ \emph {et~al.}(2022)\citenamefont {Shaikhaidarov}, \citenamefont {Kim}, \citenamefont {Dunstan}, \citenamefont {Antonov}, \citenamefont {Linzen}, \citenamefont {Ziegler}, \citenamefont {Golubev}, \citenamefont {Antonov}, \citenamefont {Il’ichev},\ and\ \citenamefont {Astafiev}}]{shaikhaidarov2022quantized}%
  \BibitemOpen
  \bibfield  {author} {\bibinfo {author} {\bibfnamefont {R.~S.}\ \bibnamefont {Shaikhaidarov}}, \bibinfo {author} {\bibfnamefont {K.~H.}\ \bibnamefont {Kim}}, \bibinfo {author} {\bibfnamefont {J.~W.}\ \bibnamefont {Dunstan}}, \bibinfo {author} {\bibfnamefont {I.~V.}\ \bibnamefont {Antonov}}, \bibinfo {author} {\bibfnamefont {S.}~\bibnamefont {Linzen}}, \bibinfo {author} {\bibfnamefont {M.}~\bibnamefont {Ziegler}}, \bibinfo {author} {\bibfnamefont {D.~S.}\ \bibnamefont {Golubev}}, \bibinfo {author} {\bibfnamefont {V.~N.}\ \bibnamefont {Antonov}}, \bibinfo {author} {\bibfnamefont {E.~V.}\ \bibnamefont {Il’ichev}}, \ and\ \bibinfo {author} {\bibfnamefont {O.~V.}\ \bibnamefont {Astafiev}},\ }\href {\doibase 10.1038/s41586-022-04947-z} {\bibfield  {journal} {\bibinfo  {journal} {Nature}\ }\textbf {\bibinfo {volume} {608}},\ \bibinfo {pages} {45} (\bibinfo {year} {2022})}\BibitemShut {NoStop}%
\bibitem [{\citenamefont {Crescini}\ \emph {et~al.}(2023)\citenamefont {Crescini}, \citenamefont {Cailleaux}, \citenamefont {Guichard}, \citenamefont {Naud}, \citenamefont {Buisson}, \citenamefont {Murch},\ and\ \citenamefont {Roch}}]{crescini2023evidence}%
  \BibitemOpen
  \bibfield  {author} {\bibinfo {author} {\bibfnamefont {N.}~\bibnamefont {Crescini}}, \bibinfo {author} {\bibfnamefont {S.}~\bibnamefont {Cailleaux}}, \bibinfo {author} {\bibfnamefont {W.}~\bibnamefont {Guichard}}, \bibinfo {author} {\bibfnamefont {C.}~\bibnamefont {Naud}}, \bibinfo {author} {\bibfnamefont {O.}~\bibnamefont {Buisson}}, \bibinfo {author} {\bibfnamefont {K.~W.}\ \bibnamefont {Murch}}, \ and\ \bibinfo {author} {\bibfnamefont {N.}~\bibnamefont {Roch}},\ }\href {\doibase 10.1038/s41567-023-01961-4} {\bibfield  {journal} {\bibinfo  {journal} {Nat. Phys.}\ }\textbf {\bibinfo {volume} {19}},\ \bibinfo {pages} {851} (\bibinfo {year} {2023})}\BibitemShut {NoStop}%
\bibitem [{\citenamefont {Willsch}\ \emph {et~al.}(2024)\citenamefont {Willsch}, \citenamefont {Rieger}, \citenamefont {Winkel}, \citenamefont {Willsch}, \citenamefont {Dickel}, \citenamefont {Krause}, \citenamefont {Ando}, \citenamefont {Lescanne}, \citenamefont {Leghtas}, \citenamefont {Bronn} \emph {et~al.}}]{willsch2024observation}%
  \BibitemOpen
  \bibfield  {author} {\bibinfo {author} {\bibfnamefont {D.}~\bibnamefont {Willsch}}, \bibinfo {author} {\bibfnamefont {D.}~\bibnamefont {Rieger}}, \bibinfo {author} {\bibfnamefont {P.}~\bibnamefont {Winkel}}, \bibinfo {author} {\bibfnamefont {M.}~\bibnamefont {Willsch}}, \bibinfo {author} {\bibfnamefont {C.}~\bibnamefont {Dickel}}, \bibinfo {author} {\bibfnamefont {J.}~\bibnamefont {Krause}}, \bibinfo {author} {\bibfnamefont {Y.}~\bibnamefont {Ando}}, \bibinfo {author} {\bibfnamefont {R.}~\bibnamefont {Lescanne}}, \bibinfo {author} {\bibfnamefont {Z.}~\bibnamefont {Leghtas}}, \bibinfo {author} {\bibfnamefont {N.~T.}\ \bibnamefont {Bronn}},  \emph {et~al.},\ }\href {\doibase 10.1038/s41567-024-02400-8} {\bibfield  {journal} {\bibinfo  {journal} {Nat. Phys.}\ }\textbf {\bibinfo {volume} {20}},\ \bibinfo {pages} {815} (\bibinfo {year} {2024})}\BibitemShut {NoStop}%
\bibitem [{\citenamefont {Spanton}\ \emph {et~al.}(2017)\citenamefont {Spanton}, \citenamefont {Deng}, \citenamefont {Vaitiekėnas}, \citenamefont {Krogstrup}, \citenamefont {Nyg{\aa}rd}, \citenamefont {Marcus},\ and\ \citenamefont {Moler}}]{spanton2017current-phase}%
  \BibitemOpen
  \bibfield  {author} {\bibinfo {author} {\bibfnamefont {E.~M.}\ \bibnamefont {Spanton}}, \bibinfo {author} {\bibfnamefont {M.}~\bibnamefont {Deng}}, \bibinfo {author} {\bibfnamefont {S.}~\bibnamefont {Vaitiekėnas}}, \bibinfo {author} {\bibfnamefont {P.}~\bibnamefont {Krogstrup}}, \bibinfo {author} {\bibfnamefont {J.}~\bibnamefont {Nyg{\aa}rd}}, \bibinfo {author} {\bibfnamefont {C.~M.}\ \bibnamefont {Marcus}}, \ and\ \bibinfo {author} {\bibfnamefont {K.~A.}\ \bibnamefont {Moler}},\ }\href {\doibase 10.1038/nphys4224} {\bibfield  {journal} {\bibinfo  {journal} {Nat. Phys.}\ }\textbf {\bibinfo {volume} {13}},\ \bibinfo {pages} {1177} (\bibinfo {year} {2017})}\BibitemShut {NoStop}%
\bibitem [{\citenamefont {Aharonov}\ and\ \citenamefont {Bohm}(1959)}]{aharonov1959significance}%
  \BibitemOpen
  \bibfield  {author} {\bibinfo {author} {\bibfnamefont {Y.}~\bibnamefont {Aharonov}}\ and\ \bibinfo {author} {\bibfnamefont {D.}~\bibnamefont {Bohm}},\ }\href {\doibase 10.1103/PhysRev.115.485} {\bibfield  {journal} {\bibinfo  {journal} {Phys. Rev.}\ }\textbf {\bibinfo {volume} {115}},\ \bibinfo {pages} {485} (\bibinfo {year} {1959})}\BibitemShut {NoStop}%
\bibitem [{\citenamefont {Lin}\ \emph {et~al.}(2018)\citenamefont {Lin}, \citenamefont {Nguyen}, \citenamefont {Grabon}, \citenamefont {San~Miguel}, \citenamefont {Pankratova},\ and\ \citenamefont {Manucharyan}}]{lin2018demonstration}%
  \BibitemOpen
  \bibfield  {author} {\bibinfo {author} {\bibfnamefont {Y.-H.}\ \bibnamefont {Lin}}, \bibinfo {author} {\bibfnamefont {L.~B.}\ \bibnamefont {Nguyen}}, \bibinfo {author} {\bibfnamefont {N.}~\bibnamefont {Grabon}}, \bibinfo {author} {\bibfnamefont {J.}~\bibnamefont {San~Miguel}}, \bibinfo {author} {\bibfnamefont {N.}~\bibnamefont {Pankratova}}, \ and\ \bibinfo {author} {\bibfnamefont {V.~E.}\ \bibnamefont {Manucharyan}},\ }\href {\doibase 10.1103/PhysRevLett.120.150503} {\bibfield  {journal} {\bibinfo  {journal} {Phys. Rev. Lett.}\ }\textbf {\bibinfo {volume} {120}},\ \bibinfo {pages} {150503} (\bibinfo {year} {2018})}\BibitemShut {NoStop}%
\bibitem [{\citenamefont {Earnest}\ \emph {et~al.}(2018)\citenamefont {Earnest}, \citenamefont {Chakram}, \citenamefont {Lu}, \citenamefont {Irons}, \citenamefont {Naik}, \citenamefont {Leung}, \citenamefont {Ocola}, \citenamefont {Czaplewski}, \citenamefont {Baker}, \citenamefont {Lawrence}, \citenamefont {Koch},\ and\ \citenamefont {Schuster}}]{earnest2018realization}%
  \BibitemOpen
  \bibfield  {author} {\bibinfo {author} {\bibfnamefont {N.}~\bibnamefont {Earnest}}, \bibinfo {author} {\bibfnamefont {S.}~\bibnamefont {Chakram}}, \bibinfo {author} {\bibfnamefont {Y.}~\bibnamefont {Lu}}, \bibinfo {author} {\bibfnamefont {N.}~\bibnamefont {Irons}}, \bibinfo {author} {\bibfnamefont {R.~K.}\ \bibnamefont {Naik}}, \bibinfo {author} {\bibfnamefont {N.}~\bibnamefont {Leung}}, \bibinfo {author} {\bibfnamefont {L.}~\bibnamefont {Ocola}}, \bibinfo {author} {\bibfnamefont {D.~A.}\ \bibnamefont {Czaplewski}}, \bibinfo {author} {\bibfnamefont {B.}~\bibnamefont {Baker}}, \bibinfo {author} {\bibfnamefont {J.}~\bibnamefont {Lawrence}}, \bibinfo {author} {\bibfnamefont {J.}~\bibnamefont {Koch}}, \ and\ \bibinfo {author} {\bibfnamefont {D.~I.}\ \bibnamefont {Schuster}},\ }\href {\doibase 10.1103/PhysRevLett.120.150504} {\bibfield  {journal} {\bibinfo  {journal} {Phys. Rev. Lett.}\ }\textbf {\bibinfo {volume} {120}},\ \bibinfo {pages} {150504} (\bibinfo {year} {2018})}\BibitemShut {NoStop}%
\bibitem [{\citenamefont {Brosco}\ \emph {et~al.}(2024)\citenamefont {Brosco}, \citenamefont {Serpico}, \citenamefont {Vinokur}, \citenamefont {Poccia},\ and\ \citenamefont {Vool}}]{brosco2024flowermon}%
  \BibitemOpen
  \bibfield  {author} {\bibinfo {author} {\bibfnamefont {V.}~\bibnamefont {Brosco}}, \bibinfo {author} {\bibfnamefont {G.}~\bibnamefont {Serpico}}, \bibinfo {author} {\bibfnamefont {V.}~\bibnamefont {Vinokur}}, \bibinfo {author} {\bibfnamefont {N.}~\bibnamefont {Poccia}}, \ and\ \bibinfo {author} {\bibfnamefont {U.}~\bibnamefont {Vool}},\ }\href {\doibase 10.1103/PhysRevLett.132.017003} {\bibfield  {journal} {\bibinfo  {journal} {Phys. Rev. Lett.}\ }\textbf {\bibinfo {volume} {132}},\ \bibinfo {pages} {017003} (\bibinfo {year} {2024})}\BibitemShut {NoStop}%
\bibitem [{\citenamefont {Patel}\ \emph {et~al.}(2024)\citenamefont {Patel}, \citenamefont {Pathak}, \citenamefont {Can}, \citenamefont {Potter},\ and\ \citenamefont {Franz}}]{patel2024dmon}%
  \BibitemOpen
  \bibfield  {author} {\bibinfo {author} {\bibfnamefont {H.}~\bibnamefont {Patel}}, \bibinfo {author} {\bibfnamefont {V.}~\bibnamefont {Pathak}}, \bibinfo {author} {\bibfnamefont {O.}~\bibnamefont {Can}}, \bibinfo {author} {\bibfnamefont {A.~C.}\ \bibnamefont {Potter}}, \ and\ \bibinfo {author} {\bibfnamefont {M.}~\bibnamefont {Franz}},\ }\href {\doibase 10.1103/PhysRevLett.132.017002} {\bibfield  {journal} {\bibinfo  {journal} {Phys. Rev. Lett.}\ }\textbf {\bibinfo {volume} {132}},\ \bibinfo {pages} {017002} (\bibinfo {year} {2024})}\BibitemShut {NoStop}%
\bibitem [{\citenamefont {Aasen}\ \emph {et~al.}(2025)\citenamefont {Aasen}, \citenamefont {Aghaee}, \citenamefont {Alam}, \citenamefont {Andrzejczuk}, \citenamefont {Antipov}, \citenamefont {Astafev}, \citenamefont {Avilovas}, \citenamefont {Barzegar}, \citenamefont {Bauer}, \citenamefont {Becker} \emph {et~al.}}]{aasen2025roadmap}%
  \BibitemOpen
  \bibfield  {author} {\bibinfo {author} {\bibfnamefont {D.}~\bibnamefont {Aasen}}, \bibinfo {author} {\bibfnamefont {M.}~\bibnamefont {Aghaee}}, \bibinfo {author} {\bibfnamefont {Z.}~\bibnamefont {Alam}}, \bibinfo {author} {\bibfnamefont {M.}~\bibnamefont {Andrzejczuk}}, \bibinfo {author} {\bibfnamefont {A.}~\bibnamefont {Antipov}}, \bibinfo {author} {\bibfnamefont {M.}~\bibnamefont {Astafev}}, \bibinfo {author} {\bibfnamefont {L.}~\bibnamefont {Avilovas}}, \bibinfo {author} {\bibfnamefont {A.}~\bibnamefont {Barzegar}}, \bibinfo {author} {\bibfnamefont {B.}~\bibnamefont {Bauer}}, \bibinfo {author} {\bibfnamefont {J.}~\bibnamefont {Becker}},  \emph {et~al.},\ }\href@noop {} {\enquote {\bibinfo {title} {Roadmap to fault tolerant quantum computation using topological qubit arrays},}\ } (\bibinfo {year} {2025}),\ \Eprint {http://arxiv.org/abs/2502.12252} {arXiv:2502.12252 [quant-ph]} \BibitemShut {NoStop}%
\bibitem [{\citenamefont {Peyruchat}\ \emph {et~al.}(2024)\citenamefont {Peyruchat}, \citenamefont {Rodriguez}, \citenamefont {Smirr}, \citenamefont {Leone},\ and\ \citenamefont {Girit}}]{peyruchat2024spectral}%
  \BibitemOpen
  \bibfield  {author} {\bibinfo {author} {\bibfnamefont {L.}~\bibnamefont {Peyruchat}}, \bibinfo {author} {\bibfnamefont {R.~H.}\ \bibnamefont {Rodriguez}}, \bibinfo {author} {\bibfnamefont {J.-L.}\ \bibnamefont {Smirr}}, \bibinfo {author} {\bibfnamefont {R.}~\bibnamefont {Leone}}, \ and\ \bibinfo {author} {\bibfnamefont {C.~{\"O.}.}\ \bibnamefont {Girit}},\ }\href {\doibase 10.1103/PhysRevX.14.041041} {\bibfield  {journal} {\bibinfo  {journal} {Phys. Rev. X}\ }\textbf {\bibinfo {volume} {14}},\ \bibinfo {pages} {041041} (\bibinfo {year} {2024})}\BibitemShut {NoStop}%
\bibitem [{\citenamefont {Matveev}\ \emph {et~al.}(2002)\citenamefont {Matveev}, \citenamefont {Larkin},\ and\ \citenamefont {Glazman}}]{matveev2002persistent}%
  \BibitemOpen
  \bibfield  {author} {\bibinfo {author} {\bibfnamefont {K.~A.}\ \bibnamefont {Matveev}}, \bibinfo {author} {\bibfnamefont {A.~I.}\ \bibnamefont {Larkin}}, \ and\ \bibinfo {author} {\bibfnamefont {L.~I.}\ \bibnamefont {Glazman}},\ }\href {\doibase 10.1103/PhysRevLett.89.096802} {\bibfield  {journal} {\bibinfo  {journal} {Phys. Rev. Lett.}\ }\textbf {\bibinfo {volume} {89}},\ \bibinfo {pages} {096802} (\bibinfo {year} {2002})}\BibitemShut {NoStop}%
\bibitem [{\citenamefont {Averin}\ \emph {et~al.}(1985)\citenamefont {Averin}, \citenamefont {Zorin},\ and\ \citenamefont {Likharev}}]{averin1985bloch}%
  \BibitemOpen
  \bibfield  {author} {\bibinfo {author} {\bibfnamefont {D.~V.}\ \bibnamefont {Averin}}, \bibinfo {author} {\bibfnamefont {A.~B.}\ \bibnamefont {Zorin}}, \ and\ \bibinfo {author} {\bibfnamefont {K.~K.}\ \bibnamefont {Likharev}},\ }\href {http://jetp.ras.ru/cgi-bin/dn/e_061_02_0407.pdf} {\bibfield  {journal} {\bibinfo  {journal} {Sov. Phys. JETP}\ }\textbf {\bibinfo {volume} {61}},\ \bibinfo {pages} {407} (\bibinfo {year} {1985})}\BibitemShut {NoStop}%
\bibitem [{\citenamefont {Koch}\ \emph {et~al.}(2009)\citenamefont {Koch}, \citenamefont {Manucharyan}, \citenamefont {Devoret},\ and\ \citenamefont {Glazman}}]{koch2009charging}%
  \BibitemOpen
  \bibfield  {author} {\bibinfo {author} {\bibfnamefont {J.}~\bibnamefont {Koch}}, \bibinfo {author} {\bibfnamefont {V.}~\bibnamefont {Manucharyan}}, \bibinfo {author} {\bibfnamefont {M.~H.}\ \bibnamefont {Devoret}}, \ and\ \bibinfo {author} {\bibfnamefont {L.~I.}\ \bibnamefont {Glazman}},\ }\href {\doibase 10.1103/PhysRevLett.103.217004} {\bibfield  {journal} {\bibinfo  {journal} {Phys. Rev. Lett.}\ }\textbf {\bibinfo {volume} {103}},\ \bibinfo {pages} {217004} (\bibinfo {year} {2009})}\BibitemShut {NoStop}%
\bibitem [{\citenamefont {de~Lange}\ \emph {et~al.}(2015)\citenamefont {de~Lange}, \citenamefont {van Heck}, \citenamefont {Bruno}, \citenamefont {van Woerkom}, \citenamefont {Geresdi}, \citenamefont {Plissard}, \citenamefont {Bakkers}, \citenamefont {Akhmerov},\ and\ \citenamefont {DiCarlo}}]{delange2015realization}%
  \BibitemOpen
  \bibfield  {author} {\bibinfo {author} {\bibfnamefont {G.}~\bibnamefont {de~Lange}}, \bibinfo {author} {\bibfnamefont {B.}~\bibnamefont {van Heck}}, \bibinfo {author} {\bibfnamefont {A.}~\bibnamefont {Bruno}}, \bibinfo {author} {\bibfnamefont {D.~J.}\ \bibnamefont {van Woerkom}}, \bibinfo {author} {\bibfnamefont {A.}~\bibnamefont {Geresdi}}, \bibinfo {author} {\bibfnamefont {S.~R.}\ \bibnamefont {Plissard}}, \bibinfo {author} {\bibfnamefont {E.~P. A.~M.}\ \bibnamefont {Bakkers}}, \bibinfo {author} {\bibfnamefont {A.~R.}\ \bibnamefont {Akhmerov}}, \ and\ \bibinfo {author} {\bibfnamefont {L.}~\bibnamefont {DiCarlo}},\ }\href {\doibase 10.1103/PhysRevLett.115.127002} {\bibfield  {journal} {\bibinfo  {journal} {Phys. Rev. Lett.}\ }\textbf {\bibinfo {volume} {115}},\ \bibinfo {pages} {127002} (\bibinfo {year} {2015})}\BibitemShut {NoStop}%
\bibitem [{\citenamefont {Silver}\ and\ \citenamefont {Zimmerman}(1967)}]{Silver67NonsinusoidalCPR}%
  \BibitemOpen
  \bibfield  {author} {\bibinfo {author} {\bibfnamefont {A.~H.}\ \bibnamefont {Silver}}\ and\ \bibinfo {author} {\bibfnamefont {J.~E.}\ \bibnamefont {Zimmerman}},\ }\href {\doibase 10.1103/PhysRev.157.317} {\bibfield  {journal} {\bibinfo  {journal} {Phys. Rev.}\ }\textbf {\bibinfo {volume} {157}},\ \bibinfo {pages} {317} (\bibinfo {year} {1967})}\BibitemShut {NoStop}%
\bibitem [{\citenamefont {Smith}\ \emph {et~al.}(2020)\citenamefont {Smith}, \citenamefont {Kou}, \citenamefont {Xiao}, \citenamefont {Vool},\ and\ \citenamefont {Devoret}}]{smith2020npj}%
  \BibitemOpen
  \bibfield  {author} {\bibinfo {author} {\bibfnamefont {W.~C.}\ \bibnamefont {Smith}}, \bibinfo {author} {\bibfnamefont {A.}~\bibnamefont {Kou}}, \bibinfo {author} {\bibfnamefont {X.}~\bibnamefont {Xiao}}, \bibinfo {author} {\bibfnamefont {U.}~\bibnamefont {Vool}}, \ and\ \bibinfo {author} {\bibfnamefont {M.~H.}\ \bibnamefont {Devoret}},\ }\href {\doibase 10.1038/s41534-019-0231-2} {\bibfield  {journal} {\bibinfo  {journal} {npj Quantum Inf.}\ }\textbf {\bibinfo {volume} {6}},\ \bibinfo {pages} {6} (\bibinfo {year} {2020})}\BibitemShut {NoStop}%
\bibitem [{\citenamefont {Groszkowski}\ and\ \citenamefont {Koch}(2021)}]{groszkowski2021scqubits}%
  \BibitemOpen
  \bibfield  {author} {\bibinfo {author} {\bibfnamefont {P.}~\bibnamefont {Groszkowski}}\ and\ \bibinfo {author} {\bibfnamefont {J.}~\bibnamefont {Koch}},\ }\href {\doibase 10.22331/q-2021-11-17-583} {\bibfield  {journal} {\bibinfo  {journal} {Quantum}\ }\textbf {\bibinfo {volume} {5}},\ \bibinfo {pages} {583} (\bibinfo {year} {2021})}\BibitemShut {NoStop}%
\bibitem [{\citenamefont {Chitta}\ \emph {et~al.}(2022)\citenamefont {Chitta}, \citenamefont {Zhao}, \citenamefont {Huang}, \citenamefont {Mondragon-Shem},\ and\ \citenamefont {Koch}}]{chitta2022computer}%
  \BibitemOpen
  \bibfield  {author} {\bibinfo {author} {\bibfnamefont {S.~P.}\ \bibnamefont {Chitta}}, \bibinfo {author} {\bibfnamefont {T.}~\bibnamefont {Zhao}}, \bibinfo {author} {\bibfnamefont {Z.}~\bibnamefont {Huang}}, \bibinfo {author} {\bibfnamefont {I.}~\bibnamefont {Mondragon-Shem}}, \ and\ \bibinfo {author} {\bibfnamefont {J.}~\bibnamefont {Koch}},\ }\href {\doibase 10.1088/1367-2630/ac94f2} {\bibfield  {journal} {\bibinfo  {journal} {New J. Phys.}\ }\textbf {\bibinfo {volume} {24}},\ \bibinfo {pages} {103020} (\bibinfo {year} {2022})}\BibitemShut {NoStop}%
\bibitem [{\citenamefont {Zhu}\ \emph {et~al.}(2013)\citenamefont {Zhu}, \citenamefont {Ferguson}, \citenamefont {Manucharyan},\ and\ \citenamefont {Koch}}]{zhu2013dispersive}%
  \BibitemOpen
  \bibfield  {author} {\bibinfo {author} {\bibfnamefont {G.}~\bibnamefont {Zhu}}, \bibinfo {author} {\bibfnamefont {D.~G.}\ \bibnamefont {Ferguson}}, \bibinfo {author} {\bibfnamefont {V.~E.}\ \bibnamefont {Manucharyan}}, \ and\ \bibinfo {author} {\bibfnamefont {J.}~\bibnamefont {Koch}},\ }\href {\doibase 10.1103/PhysRevB.87.024510} {\bibfield  {journal} {\bibinfo  {journal} {Phys. Rev. B}\ }\textbf {\bibinfo {volume} {87}},\ \bibinfo {pages} {024510} (\bibinfo {year} {2013})}\BibitemShut {NoStop}%
\bibitem [{\citenamefont {J{\"u}nger}\ \emph {et~al.}(2025)\citenamefont {J{\"u}nger}, \citenamefont {Chistolini}, \citenamefont {Nguyen}, \citenamefont {Kim}, \citenamefont {Chen}, \citenamefont {Ersevim}, \citenamefont {Livingston}, \citenamefont {Koolstra}, \citenamefont {Santiago},\ and\ \citenamefont {Siddiqi}}]{junger2025implementation}%
  \BibitemOpen
  \bibfield  {author} {\bibinfo {author} {\bibfnamefont {C.}~\bibnamefont {J{\"u}nger}}, \bibinfo {author} {\bibfnamefont {T.}~\bibnamefont {Chistolini}}, \bibinfo {author} {\bibfnamefont {L.~B.}\ \bibnamefont {Nguyen}}, \bibinfo {author} {\bibfnamefont {H.}~\bibnamefont {Kim}}, \bibinfo {author} {\bibfnamefont {L.}~\bibnamefont {Chen}}, \bibinfo {author} {\bibfnamefont {T.}~\bibnamefont {Ersevim}}, \bibinfo {author} {\bibfnamefont {W.}~\bibnamefont {Livingston}}, \bibinfo {author} {\bibfnamefont {G.}~\bibnamefont {Koolstra}}, \bibinfo {author} {\bibfnamefont {D.~I.}\ \bibnamefont {Santiago}}, \ and\ \bibinfo {author} {\bibfnamefont {I.}~\bibnamefont {Siddiqi}},\ }\href {\doibase 10.1063/5.0250341} {\bibfield  {journal} {\bibinfo  {journal} {Appl. Phys. Lett.}\ }\textbf {\bibinfo {volume} {126}},\ \bibinfo {pages} {042601} (\bibinfo {year} {2025})}\BibitemShut {NoStop}%
\bibitem [{\citenamefont {Masluk}\ \emph {et~al.}(2012)\citenamefont {Masluk}, \citenamefont {Pop}, \citenamefont {Kamal}, \citenamefont {Minev},\ and\ \citenamefont {Devoret}}]{masluk2012microwave}%
  \BibitemOpen
  \bibfield  {author} {\bibinfo {author} {\bibfnamefont {N.~A.}\ \bibnamefont {Masluk}}, \bibinfo {author} {\bibfnamefont {I.~M.}\ \bibnamefont {Pop}}, \bibinfo {author} {\bibfnamefont {A.}~\bibnamefont {Kamal}}, \bibinfo {author} {\bibfnamefont {Z.~K.}\ \bibnamefont {Minev}}, \ and\ \bibinfo {author} {\bibfnamefont {M.~H.}\ \bibnamefont {Devoret}},\ }\href {\doibase 10.1103/PhysRevLett.109.137002} {\bibfield  {journal} {\bibinfo  {journal} {Phys. Rev. Lett.}\ }\textbf {\bibinfo {volume} {109}},\ \bibinfo {pages} {137002} (\bibinfo {year} {2012})}\BibitemShut {NoStop}%
\bibitem [{\citenamefont {Gr{\"u}nhaupt}\ \emph {et~al.}(2017)\citenamefont {Gr{\"u}nhaupt}, \citenamefont {von L{\"u}pke}, \citenamefont {Gusenkova}, \citenamefont {Skacel}, \citenamefont {Maleeva}, \citenamefont {Schl{\"o}r}, \citenamefont {Bilmes}, \citenamefont {Rotzinger}, \citenamefont {Ustinov}, \citenamefont {Weides} \emph {et~al.}}]{grunhaupt2017argon}%
  \BibitemOpen
  \bibfield  {author} {\bibinfo {author} {\bibfnamefont {L.}~\bibnamefont {Gr{\"u}nhaupt}}, \bibinfo {author} {\bibfnamefont {U.}~\bibnamefont {von L{\"u}pke}}, \bibinfo {author} {\bibfnamefont {D.}~\bibnamefont {Gusenkova}}, \bibinfo {author} {\bibfnamefont {S.~T.}\ \bibnamefont {Skacel}}, \bibinfo {author} {\bibfnamefont {N.}~\bibnamefont {Maleeva}}, \bibinfo {author} {\bibfnamefont {S.}~\bibnamefont {Schl{\"o}r}}, \bibinfo {author} {\bibfnamefont {A.}~\bibnamefont {Bilmes}}, \bibinfo {author} {\bibfnamefont {H.}~\bibnamefont {Rotzinger}}, \bibinfo {author} {\bibfnamefont {A.~V.}\ \bibnamefont {Ustinov}}, \bibinfo {author} {\bibfnamefont {M.}~\bibnamefont {Weides}},  \emph {et~al.},\ }\href {\doibase 10.1063/1.4990491} {\bibfield  {journal} {\bibinfo  {journal} {Appl. Phys. Lett.}\ }\textbf {\bibinfo {volume} {111}},\ \bibinfo {pages} {072601} (\bibinfo {year} {2017})}\BibitemShut {NoStop}%
\bibitem [{\citenamefont {Hellings}\ \emph {et~al.}(2025)\citenamefont {Hellings}, \citenamefont {Lacroix}, \citenamefont {Remm}, \citenamefont {Boell}, \citenamefont {Herrmann}, \citenamefont {Laz{\u{a}}r}, \citenamefont {Krinner}, \citenamefont {Swiadek}, \citenamefont {Andersen}, \citenamefont {Eichler} \emph {et~al.}}]{hellings2025calibrating}%
  \BibitemOpen
  \bibfield  {author} {\bibinfo {author} {\bibfnamefont {C.}~\bibnamefont {Hellings}}, \bibinfo {author} {\bibfnamefont {N.}~\bibnamefont {Lacroix}}, \bibinfo {author} {\bibfnamefont {A.}~\bibnamefont {Remm}}, \bibinfo {author} {\bibfnamefont {R.}~\bibnamefont {Boell}}, \bibinfo {author} {\bibfnamefont {J.}~\bibnamefont {Herrmann}}, \bibinfo {author} {\bibfnamefont {S.}~\bibnamefont {Laz{\u{a}}r}}, \bibinfo {author} {\bibfnamefont {S.}~\bibnamefont {Krinner}}, \bibinfo {author} {\bibfnamefont {F.}~\bibnamefont {Swiadek}}, \bibinfo {author} {\bibfnamefont {C.~K.}\ \bibnamefont {Andersen}}, \bibinfo {author} {\bibfnamefont {C.}~\bibnamefont {Eichler}},  \emph {et~al.},\ }\href@noop {} {\enquote {\bibinfo {title} {Calibrating magnetic flux control in superconducting circuits by compensating distortions on time scales from nanoseconds up to tens of microseconds},}\ } (\bibinfo {year} {2025}),\ \Eprint {http://arxiv.org/abs/2503.04610} {arXiv:2503.04610 [quant-ph]} \BibitemShut {NoStop}%
\end{thebibliography}%
\newpage
\setcounter{figure}{0}
\renewcommand{\figurename}{\textbf{Extended Data Fig.~\!}}
    \begin{figure*}[t]
    \includegraphics[width=\textwidth]{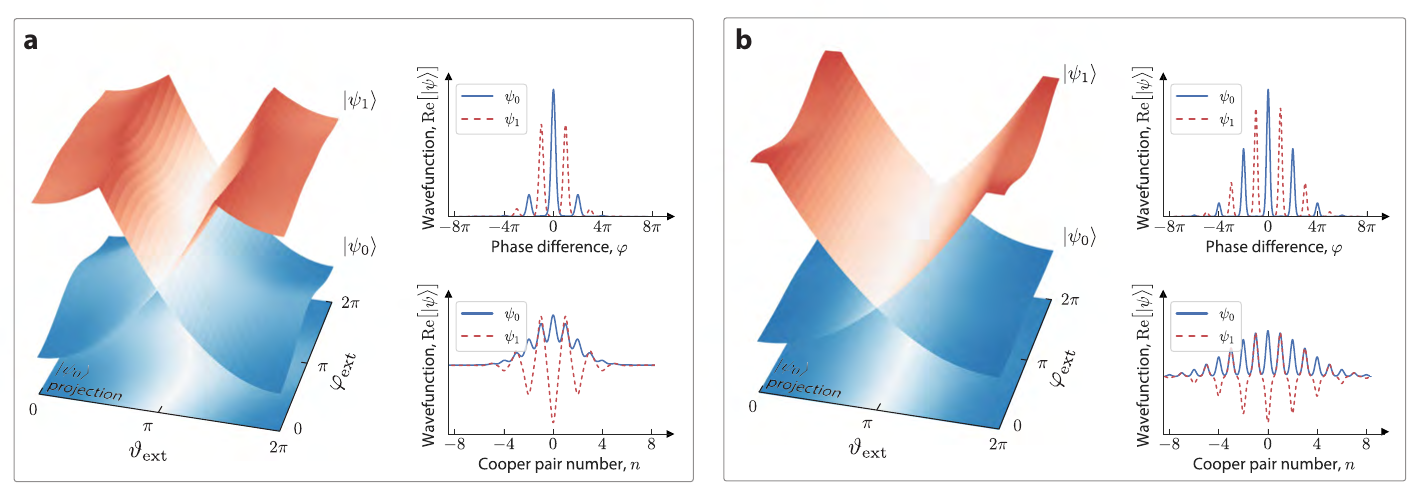}
    \caption{\label{efig1} \textbf{Interpretation of underlying device physics}.  (\textbf{a}) \textit{Left}: Energy bands as a function of control fluxes for the device studied in Fig.~\ref{fig3}\textbf{c-e}. The projection of the ground-state energy onto the XY plane reveals the underlying flux dependence responsible for the observed spectral features. \textit{Right}: Wavefunctions of the same qubit at $\vartheta_\mathrm{ext} = \pi$, $\varphi_\mathrm{ext} = \sfrac{\pi}{2}$, shown in both phase and charge bases according to Hamiltonian (\ref{eqn:GKP_approx}). This illustrates the emergence of effective grid-like states. (\textbf{b}) \textit{Left}: Energy bands for the more protected device corresponding to the spectrum in Fig.~\ref{fig3}\textbf{f}. The flux dispersion along $\varphi_\mathrm{ext}$ is markedly suppressed, reflecting enhanced robustness. \textit{Right}: Phase- and charge-basis wavefunctions for the same device, exhibiting broader grid supports, which underlies the increased resilience against environmental fluctuations.}
\end{figure*}

\begin{figure*}[t]
    \includegraphics[width=0.95\textwidth]{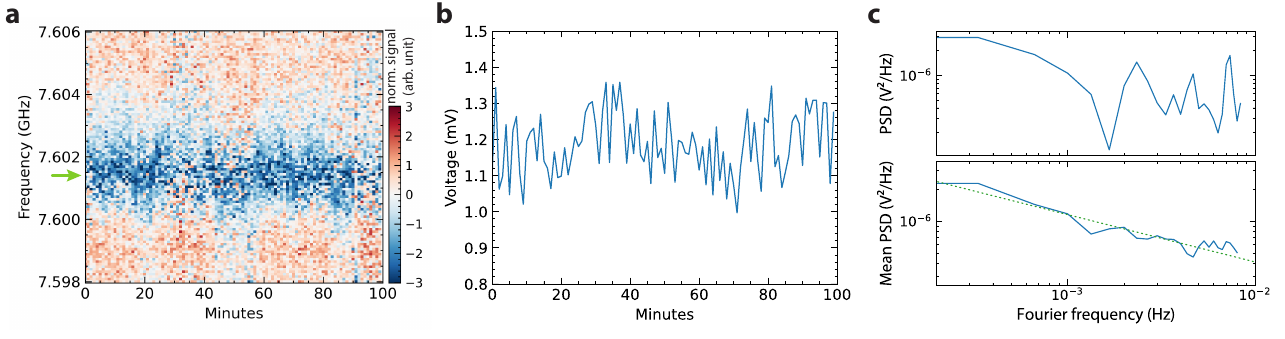}
    \caption{\label{efig2} \textbf{Fluctuation of the unprotected device}. (\textbf{a}) Time-series of resonator response measured from the unprotected device shown in Fig.~\ref{fig3}\textbf{c–e} at a fixed flux bias. Each vertical trace corresponds to a resonance sweep acquired in 1 minute. The measurement is repeated 100 times. (\textbf{b}) Integrated resonator signal at the frequency marked by the green arrow in panel \textbf{a}. (\textbf{c}) \textit{Top}: the Welch method is applied on the data presented in panel \textbf{b} to extract the effective noise power spectral density (PSD). \textit{Bottom}: Mean PSD obtained from analyzing the resonator fluctuations across the 7.6-7.604 GHz frequency range. The dashed line shows a fit to a simple noise model, $S(f)=K/f^\alpha$ noise spectrum, with $\alpha=0.45$ and $K=5.3\times 10^{-8}~\mathrm{V^2/ Hz^{\alpha}}$.}
\end{figure*}

\begin{figure*}[t]
    \includegraphics[width=\textwidth]{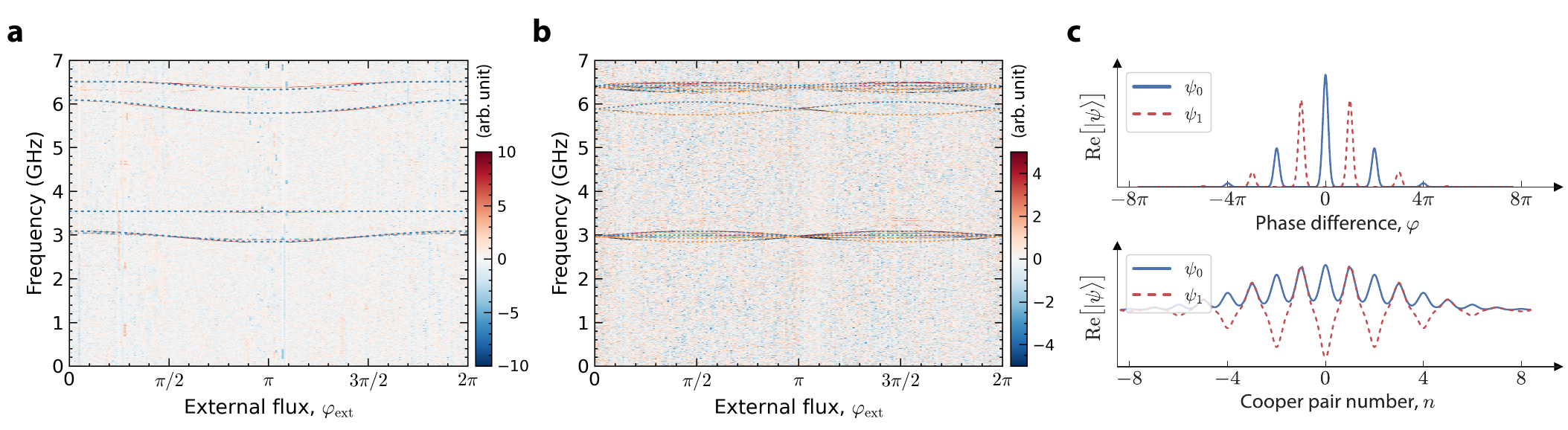}
    \caption{\label{efig3} \textbf{Device with protection and addressability}. In order to demonstrate the principle of protection, we characterize a device with increased $E_\mathrm{J}$ (and thus higher $E_\mathrm{2J}$) and smaller $E_\mathrm{JS}/E_\mathrm{CS}$ ratio (and thus higher $E_\mathrm{S}$).  (\textbf{a}) RF spectroscopy data along KITE flux $\vartheta_\mathrm{ext}=0$, corresponding to the dualmon regime where $d$=1 grids manifest. The dash-dotted line shows a second-order transition from the ground to the third excited state. The extracted dualmon Hamiltonian (Eq.~\ref{eqn:dualmon}) parameters are $[E_\mathrm{J},E_\mathrm{S},E_\mathrm{C},E_\mathrm{L}]/h\approx[7.97, 3.39, 0.21, 0.09]~\mathrm{GHz}$. (\textbf{b}) Device spectrum along KITE flux $\vartheta_\mathrm{ext}=\pi$, corresponding to the gridium regime where $d$=2 grids manifest. The blue and red colors represent transitions from the $|\psi_0\rangle$ and $|\psi_1\rangle$ states, respectively. Fitting to the gridium Hamiltonian (Eq.~\ref{eqn:GKP_approx}), we extract the effective parameters as $[E_\mathrm{2J},E_\mathrm{S},E_\mathrm{C},E_\mathrm{L}]/h\approx[7.11, 3.42, 0.18, 0.2]~\mathrm{GHz}$. Notably, the flux dispersion is heavily suppressed in both regimes, signifying increased protection against flux noise, or equivalently, broader grid support in the phase basis. Numerical simulation also shows exponentially suppressed charge sensitivity. (\textbf{c}) Wavefunctions of the ground and first excited state in the phase and charge bases.}
\end{figure*}

\clearpage
\section*{Supplementary Information Notes}

\renewcommand{\theequation}{S\arabic{equation}}
\setcounter{equation}{0}
\setcounter{figure}{0}
\setcounter{section}{1}

\makeatletter
\renewcommand \thetable{S\@arabic\c@table}
\renewcommand{\figurename}{\textbf{Fig.~\!}}
\renewcommand \thefigure{S\@arabic\c@figure}
\makeatother
\setcounter{page}{1}
    \begin{figure*}[t]
        \includegraphics[width=\textwidth]{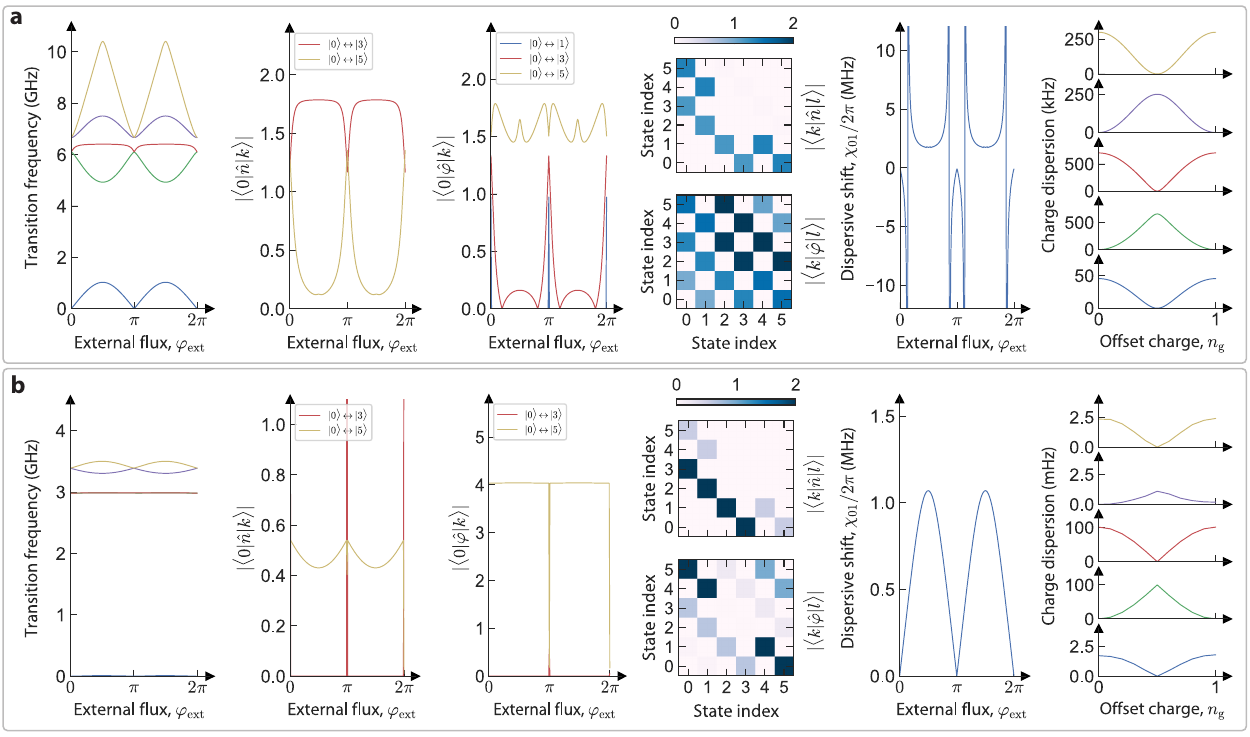}
        \caption{\label{figs1} \textbf{Quantum dynamics of the ideal gridium qubit}. (\textbf{a}) Dynamics of a gridium qubit described by Hamiltonian (\ref{eqn:GKP_approx}) with parameters $\{E_\mathrm{J}, E_\mathrm{S}, E_\mathrm{C}, E_\mathrm{L}\}/h = \{12, 4, 0.5, 0.5\}~\mathrm{GHz}$. From left to right are (i) the flux-dependent transition spectrum from the ground state $|0\rangle$ across a range of $2\pi$, (ii) charge and (iii) phase matrix elements from the ground state $|0\rangle$ across a flux interval of $2\pi$, (iv) selection rules at the symmetric flux bias $\varphi_\mathrm{ext}=0$, (v) dispersive shift of an ancilla resonator with capacitive coupling $g/2\pi=100~\mathrm{MHz}$ and frequency $\omega_\mathrm{r}/2\pi = 7.5~\mathrm{GHz}$, and (vi) charge dispersion of the transition spectrum from the ground state. (\textbf{b})  Gridium qubit dynamics for parameters $\{E_\mathrm{J}, E_\mathrm{S}, E_\mathrm{C}, E_\mathrm{L}\}/h = \{12, 4, 0.1, 0.1\}~\mathrm{GHz}$. For the dispersive shift simulation, we assume a capacitive coupling strength $g/2\pi=100~\mathrm{MHz}$ and resonator frequency $\omega_\mathrm{r}/2\pi = 3.8~\mathrm{GHz}$.}
    \end{figure*}
\subsection*{Supplementary Note 1 -- Ideal gridium qubit}

\noindent It follows conventional wisdom that analyzing a simplified physical model, free from the complexities of real-world systems, usually allows us to concentrate on the most significant features and uncover fundamental principles, paving the way for future creative developments. In addition, the idealized case often provides a more intuitive understanding that informs the study of more realistic scenarios. Furthermore, such an approach inspires the exploration of alternative methods to realize the desired model. With this perspective, we examine the key properties of an ideal single-mode gridium qubit with small quadratic corrections, as described by Hamiltonian~(\ref{eqn:GKP_approx}).

The inclusion of the quadratic terms allows us to diagonalize this Hamiltonian using the harmonic oscillator basis, with the zero-point-fluctuations of phase and charge given as
\begin{equation}
    \begin{split}
        \varphi_{_0} &= \frac{1}{\sqrt{2}}\left(\frac{2E_\mathrm{C}}{E_\mathrm{L}} \right)^{\sfrac{1}{4}}, \\
        n_{_0} &= \frac{1}{\sqrt{2}}\left(\frac{E_\mathrm{L}}{2E_\mathrm{C}} \right)^{\sfrac{1}{4}}.
    \end{split}
\end{equation}
The phase and charge operators are then written in terms of the ladder operators $\hat{a}$ and $\hat{a}^\dagger$ as $\hat{\varphi} = \varphi_{_0}(\hat{a}^\dagger+\hat{a})$ and $\hat{n} = i n_{_0}(\hat{a}^\dagger-\hat{a})$. We note that since the creation and annihilation operators are hermitian conjugates of each other, $[\hat{a},\hat{a}^\dagger]=1$, the circuit operators $\hat{\varphi}$ and $\hat{n}$ obey $[\hat{\varphi},\hat{n}]=i$. Although we observe that our numerical diagonalization generally converges for Hilbert space of dimension greater than 200, we truncate the dimension to at least 500 in all of the simulations to ensure accurate results as the values become exponentially small.

We compute the eigenspectrum and matrix elements of the circuit at various external flux values $\varphi_\mathrm{ext}$, and analyze the selection rules at the degenerate point, $\varphi_\mathrm{ext} = 0$. To explore the potential of dispersive readout, we evaluate the dispersive shift of a harmonic resonator with frequency $\omega_\mathrm{r}$ that is capacitively coupled to the qubit, with a coupling strength $g/2\pi = 100~\mathrm{MHz}$. Notably, the phase-slip junction functions as a nonlinear capacitor, creating a superconducting island within the circuit (Fig.~\ref{fig1}\textbf{b}). To estimate the charge dispersion of the qubit, we introduce an offset charge $n_g$, which modifies $\hat{n} \rightarrow \hat{n} + n_g$, and calculate the qubit spectrum's dependence on $n_g$.

Figure~\ref{figs1} shows the simulation results. We first consider 
 a circuit with capacitive and inductive energies that are easy to achieve, and analyze one with circuit parameters, specifically $[E_\mathrm{2J}, E_\mathrm{S}, E_\mathrm{C}, E_\mathrm{L}]/h = [12, 4, 0.5, 0.5]~\mathrm{GHz}$. The spectrum exhibits a $\pi$-periodic dependence on the external flux $\varphi_\mathrm{ext}$, with eigenstates becoming pairwise degenerate at $\varphi_\mathrm{ext} = 0$ and $\pi$. The matrix elements also reflect this $\pi$-periodicity, showing atypical values at symmetric flux points. For clarity, we focus on the appreciable matrix elements involving $|0\rangle$, as the others are exponentially small. Notably, $\varphi_{01} \equiv \langle 0 | \hat{\varphi} | 1 \rangle$ remains finite when the states $|0\rangle$ and $|1\rangle$ become degenerate. Inspecting the selection rules at $\varphi_\mathrm{ext} = 0$, we observe a checkerboard-like pattern in the phase matrix elements: $|0\rangle$ is phase-coupled to $|1\rangle$, $|3\rangle$, $|5\rangle$, and so on, while $|1\rangle$ is coupled to $|0\rangle$, $|2\rangle$, $|4\rangle$, and so forth. In the charge basis, the coupling dynamics are similar, with one notable exception: there is no charge coupling between degenerate states, such as $|0\rangle \leftrightarrow |1\rangle$, $|2\rangle \leftrightarrow |3\rangle$, $|4\rangle \leftrightarrow |5\rangle$, and so on. 

The charge matrix elements suggest that we can achieve substantial dispersive coupling to a capacitively coupled ancilla resonator via virtual transitions to higher eigenlevels, following
\begin{equation}
    \label{eqn:dispersive_shift}
    \chi_i = \sum_{j\neq i}g^2|{n_{ij}}|^2\frac{2\omega_{ij}}{\omega_{ij}^2-\omega_\mathrm{r}^2},
\end{equation}
where $\chi_i$ represents the shift of the resonator frequency when the qubit is in $|i\rangle$, $\omega_{ij}$ is the transition frequency between states $|i\rangle$ and $|j\rangle$, and $n_{ij} \equiv \langle i | \hat{n} | j \rangle$. Here, we specifically inspect the differential shift $\chi_{01} = \chi_1- \chi_0$ for resonator frequency $\omega_\mathrm{r}/2\pi = 7.5~\mathrm{GHz}$, observing substantial  shift except at the symmetric flux points. The divergent features arise from the resonant crossing between the $0 \leftrightarrow 5$ transition and the resonator.

Importantly, the qubit exhibits non-negligible charge dispersion due to the contribution from the quadratic energy terms, resulting in a frequency variation exceeding $50~\mathrm{kHz}$ in the $0 - 1$ subspace. Since the eigenstates can be understood as superpositions of multiple charge and phase states, the ratio $E_\mathrm{2J}/E_\mathrm{C}$ plays a crucial role in determining the number of contributing charge states, akin to the behavior observed in transmon. When the capacitive energy $E_\mathrm{C}$ is not sufficiently suppressed, the eigenstates are supported by only a few charge states, which makes the qubit more sensitive to charge fluctuations. Similarly, the relatively large flux dispersion is dictated by the inductive energy $E_\mathrm{L}$, which governs the qubit's sensitivity to flux variations.

We then consider a circuit with smaller capacitive and inductive energies, $[E_\mathrm{2J}, E_\mathrm{S}, E_\mathrm{C}, E_\mathrm{L}]/h = [12, 4, 0.1, 0.1]~\mathrm{GHz}$. As shown in Fig.~\ref{figs1}\textbf{b}, the energy dispersion with respect to both external flux $\varphi_\mathrm{ext}$ and offset charge $n_g$ is strongly suppressed, with $n_g$-variation of the $\omega_{01}$ frequency smaller than $2.5\times 10^{-3}~\mathrm{Hz}$. In addition, the phase matrix elements between degenerate states become exponentially small, while only the $0 \leftrightarrow 5$ transition has appreciable charge coupling across the entire flux range. These findings indicate that a gridium qubit with minimal quadratic corrections—closely approximating the ideal qubit described by Hamiltonian~(\ref{eqn:GKP_ideal}) in other words—is inherently robust against local charge and flux noise, making it a promising candidate for noise-resilient quantum computing. Interestingly, the higher levels exhibit negligible charge dispersions but finite variations with respect to the external flux, providing the qubit with valuable flux tunability.

For example, a capacitively coupled resonator with $g/2\pi = 100~\mathrm{MHz}$ and resonant frequency $\omega_\mathrm{r}/2\pi=3.8~\mathrm{GHz}$ displays a differential dispersive shift $\chi_{01}/2\pi\sim 1~\mathrm{MHz}$ away from the degenerate points, thanks to the finite coupling to the $|5\rangle$ state. This feature enables us to not only characterize the qubit's computational states across the entire flux range, but also to distinguish the states at the symmetric flux biases using fast flux pulses. The rich selection rules also allow the control of the qubit states in different fashions. For example, one may rapidly swap the computational states with surrogate states, $|0\rangle \leftrightarrow |3\rangle$ and $|1\rangle \leftrightarrow |4\rangle$, then perform a nominal operation between $|3\rangle$ and $|4\rangle$, then swap the states back. Fast flux pulses and Floquet driving shall open new avenues for quantum control in future works.

We note that the analytical assessment of the simple model described here is only representative. We choose to exclude other properties, such as flux-dependent transitions and matrix elements from the $|1\rangle$ state, since they share the general characteristics of those already presented. This simplified model already has multiple device parameters that can be used to tailor the dynamics, which deserves further exploration in the future.

\subsection*{Supplementary Note 2 -- Quantum phase slip}
    \begin{figure*}[t]
        \includegraphics[width=\textwidth]{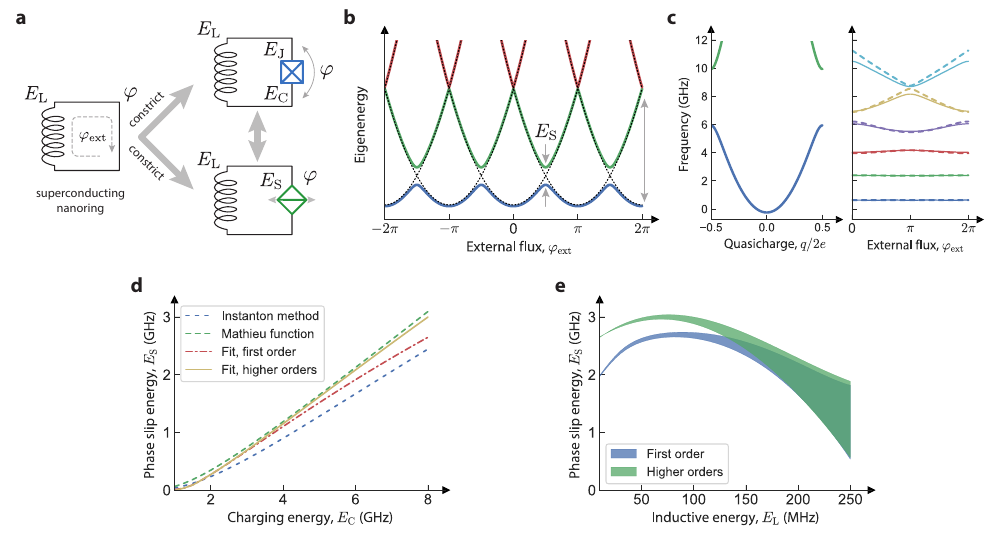}
        \caption{\label{figs2} \textbf{Quantum phase slip analysis}. (\textbf{a}) Circuit schematics of a superconducting nanoring featuring a constriction that facilitates coherent quantum phase slips. (\textbf{b}) Eigenenergy spectrum of a superconducting nanoring with inductive energy $E_\mathrm{L}$ interrupted by a Josephson junction. Under the condition $E_\mathrm{C}>E_\mathrm{J}\gg E_\mathrm{L}$, the interplay between quantum charge and phase fluctuations induces an effective quantum phase slip that opens an energy gap $E_\mathrm{S}$ at the first-order anticrossing. (\textbf{c}) Quasicharge Bloch bands (Eq.~\ref{eqn:bloch_bands}) and flux-dependent eigenfrequency spectrum of a blochnium qubit with $[E_\mathrm{J}, E_\mathrm{C}, E_\mathrm{L}]/h = [4, 8, 0.05]~\mathrm{GHz}$. The solid lines are obtained from Hamiltonian~(\ref{eqn:hamiltonian_flux}). The dashed lines correspond to numerical solutions using Hamiltonian~(\ref{eqn:hamiltonian_phaseslipcos}) with parameters $[E_\mathrm{S}^{(1)}, E_\mathrm{S}^{(2)}, E_\mathrm{S}^{(3)}, E_L]/h = [2.99, 0.68, 0.26, 0.05]~\mathrm{GHz}$. (\textbf{d}) Comparison of various methods to extract the lowest-order phase slip energy $E_\mathrm{S}^{(1)}$. The results correspond to the instanton method (Eq.~\ref{eqn:phaseslip_wkb}), Mathieu functions (Eq.~\ref{eqn:PhaseSlipRateEstimateMathieu}), and least-square fits using Eq.~\ref{eqn:hamiltonian_phaseslipcos}. The Josephson and inductive energies are fixed at $[E_\mathrm{J}, E_\mathrm{L}]/h = [4, 0.05]~\mathrm{GHz}$. (\textbf{e}) $E_\mathrm{L}$-dependence of the extracted phase slip energy $E_\mathrm{S}^{(1)}$. The shaded regions represent least-square fitting uncertainties. The Josephson and charging energies are fixed at $[E_\mathrm{J}, E_\mathrm{C}]/h = [4, 8]~\mathrm{GHz}$.}
    \end{figure*}
\noindent The quantum states of a sufficiently thick superconducting ring are persistent current states defined by the superconducting phase $\phi$ along the wire. As charges can move freely inside, the density of the supercurrent is proportional to this gauge invariant phase, which is linearly dependent on the position along the wire. The phase change accumulated over the length of the ring with circumference $l$ is $\varphi=\phi(l) - \phi(0)=2\pi m + \varphi_\mathrm{ext}$, where~$m\in \mathbb{Z}$,~and $\varphi_\mathrm{ext} = 2\pi \Phi_\mathrm{ext}/\Phi_o$ is the normalized external flux. The eigenenergies of the system are then quadratic parabolas with respect to $\varphi_\mathrm{ext}$~\cite{matveev2002persistent}.

In the presence of a constriction along the wire (Fig.~\ref{figs2}\textbf{a}), adjustment of the persistent current manifests through the change $m\rightarrow m\pm 1$~\cite{matveev2002persistent,astafiev2012coherent,manucharyan2012evidence}. This opens an energy gap equal to the single phase slip amplitude $E_\mathrm{S}$ at the crossings of the parabolic eigenenergies (Fig.~\ref{figs2}\textbf{b}). Modeling such a constriction as a Josephson junction with Josephson energy $E_\mathrm{J}$ and charging energy $E_\mathrm{C}$, we can estimate $E_\mathrm{S}$ in the regime $E_\mathrm{J}\gg E_\mathrm{C}$ by using a generalized  WKB technique known as instanton~\cite{matveev2002persistent,manucharyan2012evidence}, 

\begin{equation}\label{eqn:phaseslip_wkb}
    E_\mathrm{S} \approx \frac{4}{\sqrt{\pi}} (8 E_\mathrm{J}^3 E_\mathrm{C})^{1/4} \exp \left(- \sqrt{\frac{8 E_\mathrm{J}}{E_\mathrm{C}}}\right).
\end{equation}

In an alternative approach, we analyze the dynamics of an ultrasmall Josephson junction with $E_\mathrm{J} \lesssim E_\mathrm{C}$ embedded within a high impedance environment realized by a superinductance, $E_\mathrm{L} \ll E_\mathrm{J}, E_\mathrm{C}$. In this parameter regime, the charge fluctuation of the junction is suppressed, while the phase fluctuation is enhanced. Notably, due to the inductive shunt, the gauge invariant phase difference is a noncompact variable, $\varphi \in \mathbb{R}$, which dictates the translational property of the system Hamiltonian, $\hat{\mathcal{H}}(\varphi)\neq \hat{\mathcal{H}}(\varphi\pm 2\pi)$. The dynamics of the circuit thereby bears resemblance to that of a one-dimensional quantum particle with coordinate $\varphi$ moving in a periodic potential $U(\varphi)$. The Hamiltonian of such a system reads
\begin{equation}\label{eqn:hamiltonian_flux}
    \begin{split}
    \hat{\mathcal{H}} &= 4E_\mathrm{C}\hat{n}^2  - E_\mathrm{J}\cos\hat{\varphi} + \frac{1}{2}E_\mathrm{L}(\hat{\varphi}+\varphi_\mathrm{ext})^2 \\
    &= \hat{\mathcal{H}}_\mathrm{JJ} + \hat{\mathcal{H}}_\mathrm{L}.
    \end{split}
\end{equation}

To account for the noncompactness of $\varphi$ in the analysis, we can utilize the Bloch basis $\{|s,q\rangle \}$, where $s\in \mathbb{N}$ represents the Bloch band index, and $q\in [0,1)$ denotes the quasicharge that approximates the ``real'' charge~\cite{averin1985bloch,koch2009charging}. These Bloch states are the eigenstates of $\hat{\mathcal{H}}_\mathrm{JJ}$ in Eq.~\ref{eqn:hamiltonian_flux} with corresponding eigenenergies $E_\mathrm{S}(s,q)$. By separating the phase operator into $\hat{\varphi}=\hat{\varphi}_s + \hat{\varphi}_q$, where $[\hat{\varphi}_q,\hat{q}]=2ei$, $[\hat{\varphi}_s,\hat{q}]=0$, and confining the dynamics to the lowest band ($s=0$), Hamiltonian~(\ref{eqn:hamiltonian_flux}) can be reformulated as 
\begin{equation}\label{eqn:Hamiltonian_phaseslipm}
    \begin{split}
    \hat{\mathcal{H}} &= \frac{1}{2}E_\mathrm{L}(\hat{\varphi}_q+\varphi_\mathrm{ext})^2 + E_\mathrm{S}(\hat{q})\\
    &=\frac{1}{2}E_\mathrm{L}(2\pi |m\rangle \langle m| + \varphi_\mathrm{ext})^2\\
    &+ \frac{1}{2}\sum_{k=1}^\infty \sum_{m=-\infty}^\infty E_\mathrm{S}^{(k)}\Big[|m\rangle \langle m+k| + \mathrm{h.c.} \Big]. \\
    \end{split}
\end{equation}
Here, $\{|m\rangle\}$ are the local-minimum persistent current states, and $E_\mathrm{S}^{(k)}$ is the phase slip energy of order $k$~\cite{koch2009charging}. Therefore, this ultrasmall junction effectively functions as a phase slip element.

We emphasize that the solution above is valid only for an ultrasmall Josephson junction shunted by a large inductance with a characteristic impedance exceeding the resistance quantum. Under these conditions, charge fluctuations are strongly suppressed, interband dynamics can be neglected, and the operators simplify to $\hat{n}\sim\hat{q}/2e$, $\hat{\varphi}\sim\hat{\varphi}_q$~\cite{crescini2023evidence}. In addition, the dual nonlinearity of the phase-slip circuit becomes evident when Eq.~\ref{eqn:Hamiltonian_phaseslipm} is recast as
\begin{equation}\label{eqn:hamiltonian_phaseslipcos}
    \hat{\mathcal{H}} = \frac{1}{2}E_\mathrm{L}(\hat{\varphi}+\varphi_\mathrm{ext})^2 -\sum_{k=1}^{\infty} E_\mathrm{S}^{(k)}\cos(2\pi k \hat{n})\\
\end{equation}
In this regime, the Bloch theorem readily gives us the solution for the phase slip rates. The phase slip terms $E_\mathrm{S}(s,q)$ are given by the Mathieu functions,
\begin{equation}\label{eqn:bloch_bands}
    E_\mathrm{S} (s,q) = E_\mathrm{C}  \mathcal M_A \big( r(s,q),\lambda\big),
\end{equation}
where $\mathcal{M}_A (r,\lambda)$ is the Mathieu characteristic value for even Mathieu functions, $\lambda=-E_\mathrm{J}/2E_\mathrm{C}$, and $r$ is the characteristic exponent given by
\begin{equation}
    r(s,q) = s+1 - \text{mod}(s+1,2) + (-1)^s q/e.
\end{equation}
The lowest-order phase slip energy $E_\mathrm{S}^{(1)}$ is thereby given as the spectrum of the first Bloch band (see Fig.~\ref{figs2}\textbf{c}),
\begin{equation}\label{eqn:PhaseSlipRateEstimateMathieu}
\begin{split}
    E^{(1)}_\mathrm{S} &= \frac{1}{2} \left[ E_\mathrm{S}^{(1)}(q=e) -  E_\mathrm{S}^{(1)}(q=0)\right] \\
    &= \frac{1}{2} E_\mathrm{C}\Big[ \mathcal M_A \big(r(0,e), \lambda\big) -  \mathcal M_A \big(r(0,0), \lambda\big) \Big]. 
\end{split}
\end{equation}

We proceed to verify these approximations by performing a numerical least-squares fitting of the eigenspectrum of Hamiltonian~(\ref{eqn:hamiltonian_phaseslipcos}) to the exact circuit diagonalization using Hamiltonian~(\ref{eqn:hamiltonian_flux}). Overall, the two spectra show excellent agreement within the parameter regime of interest (Fig.~\ref{figs2}\textbf{c}).
For this analysis, we adopt a general model allowing for higher-order phase slips, with $k\leq 3$.
We then compare the solutions while varying the charging energy $E_\mathrm{C}$ of the Josephson junction. As illustrated in Fig.~\ref{figs2}\textbf{c}, the results derived using Mathieu functions exhibit close agreement with the numerical solution, whereas the instanton method becomes less accurate at high $E_\mathrm{C}$. Notably, when accounting for multiple phase-slip processes, $k>1$, the extracted single phase slip rate $E_\mathrm{S}^{(1)}$ is larger and aligns more closely with the analytical solution.

Finally, we examine the validity of our solutions when the large-impedance condition is not satisfied. Keeping the parameters of the Josephson junction fixed, we perform a numerical least-squares fit between Hamiltonians (\ref{eqn:hamiltonian_flux}) and (\ref{eqn:hamiltonian_phaseslipcos}) for varying inductive energy $E_\mathrm{L}$. As shown in Fig.~\ref{figs2}\textbf{e}, the fit becomes increasingly inaccurate when the large-inductance requirement is not met. Although the extracted phase-slip rate remains relatively high, increasing $E_\mathrm{L}$ disrupts the underlying assumptions of the model, as expected. Therefore, the inductance plays a critical role not only in shaping the quadratic correction to the GKP Hamiltonian but also in determining the behavior of the phase-slip element.

\subsection*{Supplementary Note 3 -- Cooper-quartet tunneling}

\noindent The electrodynamics of a superconducting weak link at DC can be described by the generalized Josephson current-phase relation (CPR), $I=\sum_{k}I_{c,k}\sin(k\varphi)$, where $k\in \mathbb{N}$ denotes the order of the Cooper-pair tunneling process, and $\varphi$ is the gauge-invariant phase difference across the junction~\cite{willsch2024observation}. It has been established that as the conduction channels within the weak link become more transparent, it is more likely for Cooper pairs to tunnel together in groups of $k$~\cite{larsen2020parity,delange2015realization}. In the tunneling limit of an opaque junction, the lowest order is dominant, resulting in the well-known sinusoidal equation $I=I_{c,1}\sin\varphi$, also known as the first Josephson relation or DC Josephson effect, where $I_{c,1}=(2e/\hbar)E_\mathrm{J}$ is the critical current of the junction. The inclusion of higher-order Josephson harmonics strongly alters this CPR, skewing and distorting it away from the nominal sinusoid. 

Alternative to modifying the conduction channels within the junction, this effect can be achieved by connecting an ordinary tunneling junction to a linear inductor \cite{Silver67NonsinusoidalCPR}. Kirchhoff's current law dictates that $E_\mathrm{J}\sin\varphi_\mathrm{J}=E_\mathrm{L}\varphi_\mathrm{L}$, where $\varphi_\mathrm{J}$ and $\varphi_\mathrm{L}$ represent the phase drops across the Josephson junction and the inductor, respectively. The phase difference across the series circuit is thus given by Kepler's transcendental equation,
\begin{equation}\label{eqn:kepler}
    \varphi = \varphi_\mathrm{J} + \frac{E_\mathrm{J}}{E_\mathrm{L}}\sin\varphi_\mathrm{J}\equiv F_1(\varphi_\mathrm{J}).
\end{equation}

\begin{figure*}[t]
        \includegraphics[width=\textwidth]{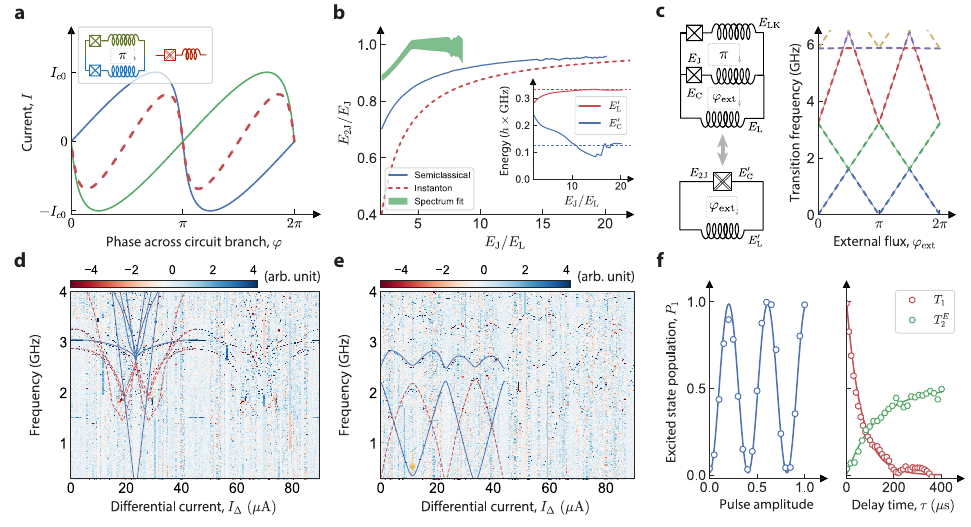}
        \caption{\label{figs3} \textbf{Tunneling of Cooper quartets}. (\textbf{a}) Relationship between the superconducting phase drop $\varphi$ across a series circuit (inset) and the supercurrent $I$ flowing through it. Destructive interference of currents in the two branches of a $\pi$-biased KITE results in a skewed current-phase relation (red dashed line), mimicking that of a CQT junction in series with an inductor. (\textbf{b}) Extracted $4e$-charge tunneling amplitude $E_\mathrm{2J}$ normalized by the Josephson energy $E_\mathrm{J}$ of the constituting junctions. The value is computed at various $E_\mathrm{J}/E_\mathrm{L}$ ratios using the semiclassical and the instanton approaches. The shaded region shows the  $E_\mathrm{2J}$ values extracted from numerical fitting of the spectra, where the uncertainty corresponds to the standard deviation errors on the circuit parameter. The other circuit parameters are kept at $[E_\mathrm{C},E_\mathrm{L}]/h= [0.5,0.5]~\mathrm{GHz}$ Inset: Inductive energy $E_\mathrm{L}'$ and charging energy $E_\mathrm{C}'$ extracted from numerical spectra fitting using Eq.~\ref{eqn:effective_cos2phi_1}. The dashed horizontal lines represent the expected convergence at large junction regime, where $E_\mathrm{J}/E_\mathrm{L}\gg1$. (\textbf{c}) Comparison between the inductively shunted KITE and the $\cos2\varphi$ fluxonium circuits. The numerically computed transition spectra, represented as dashed and solid lines respectively, exhibit excellent agreement at frequencies below 6~GHz. (\textbf{d}) $\varphi_\mathrm{ext}$-dependence transition spectrum of an inductively shunted KITE circuit at $\vartheta_\mathrm{ext}=0$. The solid and dashed lines represent the numerically computed transitions from $|0\rangle$ and $|1\rangle$, respectively. (\textbf{e}) $\varphi_\mathrm{ext}$-dependence transition spectrum of an inductively shunted KITE circuit at $\vartheta_\mathrm{ext}=\pi$. The numerically extracted circuit parameters corresponding to the lines are $[E_\mathrm{2J}, E_\mathrm{1J},E_\mathrm{C}',E_\mathrm{L}']/h = [1.87, 0.29, 0.23,0.58]~\mathrm{GHz}$, close to the predicted values. (\textbf{f}) (Left) Rabi oscillation  as a function of pulse amplitude for on-resonance driving of the qubit transition marked by the vertical arrow in panel \textbf{e}. (Right) Coherence measurement results including the energy relaxation $T_1$ and the Hahn echo decay $T_{2E}$ of the corresponding transition point.}
    \end{figure*}

Analytically inverting Eq.~\ref{eqn:kepler} to map $\varphi \rightarrow \varphi_\mathrm{J}$ is challenging, but it is possible to do so numerically. The effective CPR for the series circuit can thus be written as
\begin{equation}\label{eqn:jj+ind_CPR}
    I(\varphi) = \frac{2e}{\hbar}E_\mathrm{J}\sin\bigg(F_1^{-1}(\varphi)\bigg), 
\end{equation}
which corresponds to a skewed sinusoidal as shown by the solid lines in Fig.~\ref{figs3}\textbf{a}. This emulates a weak link with dominant first- and second-order harmonics.

By connecting two identical series circuits of this type in parallel to form a Kinetic Interference coTunneling Element (KITE), we can eliminate the lowest-order effects by applying an external flux bias $\vartheta_\mathrm{ext}=\pi$, making the second harmonic the dominant dynamics. The resulting CPR is represented by the dashed line in Fig.~\ref{figs3}\textbf{a}. The destructive interference cancels the $2\pi$-periodic behavior, while the skewness retains the $\pi$-periodic pattern. This modified CPR corresponds to an effective circuit comprising a second-order weak link connected in series with an inductor of half the original inductance,
\begin{equation}\label{eqn:2phijj+ind_CPR}
    I(\varphi) = -\frac{4e}{\hbar}E_\mathrm{2J} \sin \bigg(F_2^{-1}(\varphi) \bigg), 
\end{equation}
where $F_2(\varphi_\mathrm{J})\equiv \varphi=\varphi_\mathrm{J}-\tfrac{E_\mathrm{2J}}{E_\mathrm{L}}\sin(2\varphi_\mathrm{J})$ represents the total phase drop across the circuit, analogous to Eq.~\ref{eqn:kepler}. As before, the inverse function $F_2^{-1}(\varphi)$ is computed numerically. The effective second-order Josephson energy $E_\mathrm{2J}$ can be subsequently extracted by fitting the CPR in Eq.~(\ref{eqn:2phijj+ind_CPR}) to the CPR of the frustrated KITE loop.

We presently discuss the relation between the effective critical current $I_{c,k}$ and Josephson energy $E_{k\mathrm{J}}$. The phase-flux relation $\varphi=2e\phi/\hbar$ follows the AC Josephson effect,
$\dot{\varphi}=2eV/\hbar$, where $\phi$ and $V$ are the flux and voltage across the junction, respectively. For the first-order harmonic, the energy change across the junction is
    \begin{equation}
    \Delta E=\int_t IVdt=-\left(\hbar/2e\right)I_{c,1}\Delta\cos\varphi\equiv -E_\mathrm{J}\Delta\cos\varphi,
    \end{equation} 
which gives us the well-known Josephson energy $E_\mathrm{J}=(\hbar/2e)I_{c,1}$. By analogy, the Josephson energy corresponding to the $k$-order harmonic is given as 
\begin{equation}
    E_{k\mathrm{J}}=\frac{\hbar}{k\times 2e}I_{c,k}.
\end{equation}
Kirchhoff's current law for a $\cos2\varphi$ junction in series with an inductor $L'$ then dictates that $-2E_\mathrm{2J}\sin(2\varphi_\mathrm{J})=E_\mathrm{L}'\varphi_\mathrm{L}$. As the parallel inductors manifests as $L'=L/2$, we can define $F_2(\varphi_\mathrm{J})$ as shown above. 

For a circuit with $[E_\mathrm{J},E_\mathrm{L}]/h=[10,0.5]~\mathrm{GHz}$, the combined skewed CPR, illustrated by the red dashed line in Fig.~\ref{figs3}\textbf{a}, is numerically found to be equivalent to a series circuit characterized by $[E_\mathrm{2J},E_\mathrm{L}']/h=[9.26,1]~\mathrm{GHz}$. Extending this numerical analysis across various values of $E_\mathrm{J}$ allows us to map the relationship between the effective Cooper-quartet tunneling energy $E_\mathrm{2J}$ and the Josephson junction energy $E_\mathrm{J}$ of the junctions in the KITE circuit. The resulting trend is depicted as the solid line in Fig.~\ref{figs3}\textbf{b}. We observe that $E_\mathrm{2J}\rightarrow E_\mathrm{J}$ for large $E_\mathrm{J}/E_\mathrm{L}$ values, corresponding to large junctions and inductance in the KITE arms. 
    
The $\pi$-periodic behavior pertains to tunneling of pairs of Cooper pairs, or Cooper quartets, with charge unit of $4e$. To verify this, we recall the duality between phase and charge eigenbases associated with a Josephson junction,
\begin{equation}
    |\varphi\rangle = \sum_{n\in\mathbb{Z}} e^{in\varphi} |n\rangle, \hspace{1cm} |n\rangle = \int_0^{2\pi}\frac{d\varphi}{2\pi}e^{-in\varphi}|\varphi\rangle,
\end{equation}
 where $|\varphi\rangle$ and $|n\rangle$ obey the completeness relations, $\sum_{n\in\mathbb{Z}} |n\rangle \langle n|=\int _0^{2\pi}|\varphi\rangle \langle \varphi | \frac{d\varphi}{2\pi}=1$. The translational operation acts on the charge and phase bases as
 \begin{equation}
     e^{ik\hat{\varphi}}|n\rangle = |n+k\rangle, \hspace{1cm} e^{ik\hat{n}}|\varphi\rangle = |\varphi+k\rangle.
 \end{equation}
Hence, the operator $\cos(2\hat{\varphi})=\frac{1}{2}\sum_{n\in \mathbb{Z}}\big(|n\rangle \langle n+2| + |n+2\rangle \langle n|\big)$ indicates tunneling of two Cooper pairs, or Cooper quartets, across the junction. The modified rhombus, or KITE, thus effectively functions as a Cooper-Quartet-Tunneling (CQT) junction in series with an inductor.

Our discussion so far has focused on the dynamics in the DC limit, thereby neglecting charging effects from capacitors. In principle, this  approximation is valid at sufficiently low frequencies for small capacitive energies, such that quantum phase fluctuation is suppressed. The charging dynamics can be considered insignificant for small charging energy, which corresponds to diminished quantum phase fluctuations, and thus approaches the classical limit. Therefore, this technique is suitable for examining circuits with large capacitances.

In another approach, constraining the fluxon tunneling path through the effective potential at flux frustration $\vartheta_\mathrm{ext}=\pi$ gives $E_\mathrm{2J}\approx E_\mathrm{J}-5E_\mathrm{L}/4$~\cite{smith2020npj}. The result using this instanton trajectory method is presented as dashed line in Fig.~\ref{figs3}\textbf{b}. To further validate our approximation, we augment the KITE circuit biased at $\vartheta_\mathrm{ext}=\pi$ with an additional shunting superinductor with the same inductive energy $E_\mathrm{L}$, chosen for simplicity, to allows flux tuning (Fig.~\ref{figs3}\textbf{c}). The manifestation of the Cooper-quartet tunneling in the circuit can be modeled by the Hamiltonian
\begin{equation}\label{eqn:effective_cos2phi_1}
    \hat{\mathcal{H}}' = 4E_\mathrm{C}' \hat{n}^2 + \frac{1}{2}E_\mathrm{L}'(\hat{\varphi} + \varphi_\mathrm{ext})^2 + E_\mathrm{2J} \cos (2\hat{\varphi}),
\end{equation}
which corresponds to a $\cos2\varphi$ fluxonium qubit. Numerical fitting of the spectrum to Eq.~\ref{eqn:effective_cos2phi_1} thereby allows us to extract the effective circuit parameters. 

For the full circuit parameters $[E_\mathrm{J}, E_\mathrm{C}, E_\mathrm{L}]/h=[10,0.5,0.5]~\mathrm{GHz}$, the spectrum shows an excellent agreement below $6~\mathrm{GHz}$ with that of an effective circuit described by Eq.~\ref{eqn:effective_cos2phi_1} with  $[E_\mathrm{2J},E_\mathrm{L}']/h=[10,0.33]~\mathrm{GHz}$, as shown in Fig.~\ref{figs3}\textbf{c}. Our analysis thus consistently confirms that the effective Cooper-quartet tunneling amplitude $E_\mathrm{2J}$ approaches the Josephson energy $E_\mathrm{J}$ of the junction. We again sweep this parameter and show the extracted ratio $E_\mathrm{2J}/E_\mathrm{J}$ including fitting uncertainty in Fig.~\ref{figs3}\textbf{b}. This procedure also allows us to examine the effective charging energy $E_\mathrm{C}'$ and inductive energy $E_\mathrm{L}'$ of the toy circuit, which we show in the inset.

From the extracted parameters, we confirm that the effective inductive energy converges to the value predicted by simple inductive network analysis in the large junction regime, $E_\mathrm{J}\gg E_\mathrm{L}$, enabling efficient predictions of the inductance parameters. Meanwhile, the effective charging energy $E_\mathrm{C}'$ may vary from approximately $E_\mathrm{C}/2$ in the small junction limit to $E_\mathrm{C}/4$ in the large junction limit. Notably, in the large junction limit, the spectrum remains relatively insensitive to variations in $E_\mathrm{C}'$.

Following the standard convention in superconducting circuits, the charging energy associated with a Josephson junction is $\hat{\mathcal{H}}_C=\hat{Q}^2/2C$, where $Q$ is the accumulated charge on the electrodes, and $C$ is the effective capacitance. For the tunneling of Cooper pairs, the kinetic Hamiltonian is written as $\hat{\mathcal{H}}_C=4E_\mathrm{C}\hat{n}^2$, where $E_\mathrm{C}=e^2/(2C)$ is the conventional electrostatic charging energy to add/deplete single electrons to/from a superconducting electrode. The ratio between the Josephson energy and charging energy, $E_\mathrm{J}/E_\mathrm{C}$, largely determines the level of quantum phase fluctuations in the circuit. 

In the large fluctuation limit (relatively large $E_\mathrm{C}$), the capacitive network can be analyzed to simply combine the capacitors in the two branches, thereby cutting the resulting charging energy by half (see SI Note~4). In the semiclassical limit, the spectrum is relatively insensitive to absolute variations in the charging energy. Instead, we can reason that the spectra bear resemblance if the junction plasma frequency is kept the same by scaling the charging energy~\cite{smith2022magnifying}, which justifies our choice of the Hamiltonian in Eq.~\ref{eqn:GKP_approx}.

To validate this circuit analysis approach, we design, fabricate, and characterize a circuit with topology resembles that of gridium (Fig.~\ref{fig2}), excluding the phase slip. We  focus on the large phase fluctuation regime with target circuit parameters $[E_\mathrm{J},E_\mathrm{C},E_\mathrm{L}]/h = [2,0.5,1]~\mathrm{GHz}$. Figure~\ref{figs3}\textbf{d}(\textbf{e}) shows the spectrum of the circuit with respect to the external flux $\varphi_\mathrm{ext}$ when $\vartheta_\mathrm{ext}=0$($\pi$). Flux calibration is conducted using the same approach as described for the gridium circuit. Notably, going from the $\cos\varphi$ to the $\cos2\varphi$ domain, the flux periodicity is reduced from $44~\mathrm{\mu A}$ to $22~\mathrm{\mu A}$, characteristics of the CQT. The numerical least-square fit result shows excellent agreement with the spectral data. Notably, the finite circuit asymmetry leads to an observable residual $E_\mathrm{1J}$ term.

Finally, we characterize the coherence of the constructed quantum system at the symmetric flux bias point $[ \varphi_\mathrm{ext}, \vartheta_\mathrm{ext}]=[\pi,\pi ]$, which is first-order insensitive to flux noise. Figure~\ref{figs3}\textbf{f} shows the Rabi, $T_1$ time, and echo $T_{2E}$ experimental results, from which we extract coherence times $T_1=63(3)~\mathrm{\mu s}$ and $T_{2E}=128(9)~\mathrm{\mu s}$. Despite the presence of adjacent integrated flux lines and large device footprint, the system maintains robust phase coherence. This experimental validation also demonstrates that the inductively shunted CQT junction is a reliable prototype for the gridium architecture.

\subsection*{Supplementary Note 4 -- Circuit Hamiltonian}

\begin{figure}[h]
    \includegraphics[width=0.42\textwidth]{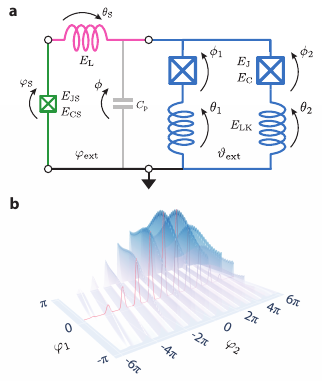}
    \caption{\label{figs4} \textbf{Circuit analysis}. (\textbf{a}) Electrical diagram of the gridium qubit using branch flux notations. The parasitic capacitance is denoted as $C_\mathrm{p}$. (\textbf{b}) Three-dimensional representation of the ground state wavefunction in the $\varphi_1$ and $\varphi_2$ bases. Here, as defined in Eq.~\ref{eqn:transformed_Hamiltonian}, $\varphi_{1}$ and $\varphi_2$ are compact and extended variables, respectively.}
\end{figure}

\noindent Although the circuit’s emergent behavior can be analyzed with a “divide-and-conquer” approach—treating the QPS element and the CQT junction separately—examining the full circuit is essential to capture spurious effects, such as parasitic modes and unintended contributions from the stray capacitance $C_\mathrm{p}$. We hereby proceed with analyzing the circuit using the canonical branch flux superconducting circuit theory with idealized circuit parameters and junction symmetry.

We denote the branch flux variables as shown in Fig.~\ref{figs4}\textbf{a}. The unconstrained circuit Lagrangian reads
\begin{equation}
    \begin{split}
        \mathcal{L} &= \frac{1}{2}\phi_0^2\Big[C_\mathrm{J}(\dot{\phi}_1^2+\dot{\phi}_2^2) + C_\mathrm{JS}\dot{\varphi}_\mathrm{S}^2+C_\mathrm{p}\dot{\phi}^2\Big] \\
         &+E_\mathrm{J}\big(\cos\phi_1 + \cos\phi_2\big) + E_\mathrm{JS}\cos\varphi_\mathrm{S}\\
         &-\frac{1}{2}E_\mathrm{LK}(\theta_1^2 + \theta_2^2)-\frac{1}{2}E_\mathrm{L}\theta_\mathrm{S}^2,
    \end{split}
\end{equation}
where $\phi_0=\hbar/(2e)$ is the reduced flux quantum and $E_{\mathrm{J}1} = E_{\mathrm{J}} (1-\epsilon)$ and $E_{\mathrm{J}2} = E_{\mathrm{J}} (1+\epsilon)$ are the two Josephson energies of the CQT junction with an asymmetry given by $\epsilon$. We can impose Kirchoff's voltage law onto the two closed loops to eliminate $\theta_1$, $\theta_2$, and $\theta_\mathrm{S}$. Next, we introduce the variable transformation $\varphi_\Sigma=(\phi_1+\phi_2)/2$, $\varphi_\Delta=(\phi_1-\phi_2)/2$ to find the constrained Lagrangian 
\begin{equation}\label{eqn:full_lagrangian}
    \begin{split}
        \mathcal{L} &= \frac{1}{2}\phi_0^2\Big[2C_\mathrm{J}(\dot{\varphi}_\Sigma^2+\dot{\varphi}_\Delta^2)  + C_\mathrm{JS}\dot{\varphi}_\mathrm{S}^2+C_\mathrm{p}\dot{\phi}^2\Big] \\
         &+2E_\mathrm{J}\cos\varphi_\Sigma \cos\varphi_\Delta+ E_\mathrm{JS}\cos\varphi_\mathrm{S}\\
         &-E_\mathrm{LK}\Big[ (\varphi_\Sigma-\phi)^2+(\varphi_\Delta-\frac{1}{2}\vartheta_\mathrm{ext})^2 \Big]\\&-E_\mathrm{L}(\varphi_\mathrm{S}-\phi+\varphi_\mathrm{ext})^2.
    \end{split}
\end{equation}
We proceed to compute the conjugate charge $n_j=\partial \mathcal{L}/\partial \dot{\varphi}_j$, where $j\in \{\Sigma,\Delta,\mathrm{S} \}$, use the Legendre transformation, and follow canonical quantization to derive the 4-mode gridium Hamiltonian,
\begin{equation}\label{eqn:gridium_4mode}
    \begin{split}  \hat{\mathcal{H}}&=2E_\mathrm{C}(\hat{n}^2_\Sigma+\hat{n}^2_\Delta)+4E_\mathrm{CS}\hat{n}_\mathrm{S}^2 + 4\varepsilon_\mathrm{p}\hat{n}^2 \\
        &- 2E_\mathrm{J}\cos\hat{\varphi}_\Sigma\cos\hat{\varphi}_\Delta - E_\mathrm{JS}\cos\hat{\varphi}_\mathrm{S}\\
        &+E_\mathrm{LK}\Big[(\hat{\varphi}_\Sigma-\hat{\phi})^2 + (\hat{\varphi}_\Delta-\frac{1}{2}\vartheta_\mathrm{ext})^2\Big]
        \\&+ E_\mathrm{L}(\hat{\varphi}_\mathrm{S}-\hat{\phi}+\varphi_\mathrm{ext})^2,
    \end{split}
\end{equation}
where $E_\mathrm{C}=e^2/2C_\mathrm{J}$, $\varepsilon_\mathrm{p}=e^2/2C_\mathrm{p}$, and $E_\mathrm{CS}=e^2/2C_\mathrm{JS}$. We note that the presence of the arranged Josephson junctions leads to an isolated superconducting island. The Hamiltonian can be rewritten via a simple coordinate transformation to include an offset charge $n_\mathrm{g}$, accounting for this island in the chosen circuit topology. 


When the stray capacitance $C_\mathrm{p}$ is vanishingly small, $\varepsilon_\mathrm{p}$ is the largest energy scale in the circuit. This allows us to eliminate the $\phi$-mode using Born-Oppenheimer approximation, which was originally employed to treat wavefunctions of molecules separately, given that a constituent  particle is much heavier than the other. Note that within our assumption, the corresponding high-energy Hamiltonian, given as
\begin{equation}\label{eqn:gridium_phimode}
    \hat{\mathcal{H}}_\phi= 4\varepsilon_\mathrm{p}\hat{n}^2 + E_\mathrm{L}(\hat{\phi}-\hat{\varphi}_\Sigma)^2+\frac{1}{2}E_\mathrm{LS}(\hat{\phi}-\hat{\varphi}_\Delta)^2,
\end{equation}
simply describes a harmonic oscillator with equilibrium position $\langle \phi\rangle$ defined as
\begin{equation}
    \langle \phi\rangle = \frac{2E_\mathrm{LK}\hat{\varphi}_\Sigma+E_\mathrm{L}\hat{\varphi}_\mathrm{S}}{2E_\mathrm{LK}+E_\mathrm{L}}.
\end{equation}
This is consistent with Kirchhoff's current law. This relation subsequently leads to the low-energy 3-mode Hamiltonian
\begin{equation}\label{eqn:gridium_3mode}
    \begin{split}
    \hat{\mathcal{H}}&=2E_\mathrm{C}(\hat{n}^2_\Sigma+\hat{n}^2_\Delta)+4E_\mathrm{CS}\hat{n}_\mathrm{S}^2 \\
        &- 2E_\mathrm{J}\cos\hat{\varphi}_\Sigma\cos\hat{\varphi}_\Delta - E_\mathrm{JS}\cos(\hat{\varphi}_\mathrm{S}-\varphi_\mathrm{ext})\\
        &+ E'_\mathrm{L}(\hat{\varphi}_\Sigma-\hat{\varphi}_\mathrm{S})^2+E_\mathrm{LK}(\hat{\varphi}_\Delta-\tfrac{1}{2}\vartheta_\mathrm{ext})^2,
    \end{split}
\end{equation}
where $E_\mathrm{L}'=2E_\mathrm{LK}E_\mathrm{L}/(2E_\mathrm{LK}+E_\mathrm{L})$ is the normalized inductive energy. To take into consideration the superconducting island in the numerical simulation of the full circuit, we perform a coordinate transformation and rewrite the phase variable as
\begin{equation}\label{eqn:3mode_phi_operators}
\varphi_1 = \varphi_\mathrm{S},~~\varphi_2= \varphi_\mathrm{S}- \varphi_\Sigma,~\mathrm{and} ~~\varphi_3=\varphi_\Delta.
\end{equation}
The corresponding charge variables are then written using the Lagrangian given by Eq.~\ref{eqn:full_lagrangian},
\begin{equation}\label{eqn:3mode_n_operators}
n_1 = n_\mathrm{S}+n_\Sigma,~~ n_2= - n_\Sigma, ~\mathrm{and}~~n_3=n_\Delta.
\end{equation}
The 3-mode Hamiltonian (Eq.~\ref{eqn:gridium_3mode}) is then written using the new operators as 
\begin{equation}\label{eqn:transformed_Hamiltonian}
    \begin{split}
    \hat{\mathcal{H}}&=4E_\mathrm{CS}(\hat n_1+ \hat n_2 + n_\mathrm{g})^2+2E_\mathrm{C}(\hat n_2^2+\hat n_3^2)\\
    &- 2E_\mathrm{J}\cos(\hat{\varphi}_1 - \hat \varphi_2)\cos\hat{\varphi}_3 - E_\mathrm{JS}\cos(\hat{\varphi}_\mathrm{1}-\varphi_\mathrm{ext})\\
        &+ \frac{E_\mathrm{LK}E_\mathrm{L}}{2E_\mathrm{LK}+E_\mathrm{L}}\hat\varphi_2^2+E_\mathrm{LK}(\hat{\varphi}_3-\tfrac{1}{2}\vartheta_\mathrm{ext})^2,
    \end{split}
\end{equation}
 which renders $\varphi_1$ a compact variable, invariant under integer shifts of $2\pi$. The resulting variable compactness despite the presence of the inductive term arises from the doubly nonlinear nature of the Hamiltonian, which prevents the elimination of these offset charges via unitary transformation. Figure~\ref{figs4}\textbf{b} shows the ground-state wavefunction of Hamiltonian (\ref{eqn:transformed_Hamiltonian}) in the $\varphi_1$–$\varphi_2$ plane, forming a two-dimensional grid structure for suitable circuit parameters. A slice along the $\varphi_2$ direction reveals a one-dimensional grid state, analogous to that of the extended GKP Hamiltonian given by Eq.~\ref{eqn:GKP_approx} (see Fig.~\ref{fig1}).

Examining Hamiltonian~(\ref{eqn:gridium_3mode}) within the Born-Oppenheimer framework, we can separate the circuit dynamics into two separate but weakly coupled subsystems, $\hat{\mathcal{H}}=\hat{\mathcal{H}}_\mathrm{S}+\hat{\mathcal{H}}_{\Sigma\Delta}+\hat{\mathcal{H}}_\mathrm{S-\Sigma}$, where the coupling term is written as $\hat{\mathcal{H}}_\mathrm{S-\Sigma} = E'_\mathrm{L}(\hat \varphi_\Sigma-\hat \varphi_\mathrm{S})^2$. Analyzing each system in the context of subsystem hierarchy, we note that the phase-slip part is described in the Bloch basis $|m\rangle$ as $\hat{\mathcal{H}}_\mathrm{S} = E_\mathrm{L}'\hat \varphi_\mathrm{S}^2-E_\mathrm{S}\cos(2\pi \hat{n})$ (see SI Note 2). Meanwhile, the dynamics of the KITE circuit encompassing the common $\Sigma$ and differential $\Delta$ modes are dictated by the formation of a 2D potential that resembles an egg carton surrounded by a taco shell. Crucially, at $\vartheta_\mathrm{ext}=\pm\pi$, the adjacent minimal coordinates within the potential are different by $\pi$. Therefore, the instanton trajectory emerges as $\pi$-periodic, equivalent to $4e$-charge tunneling described by $\hat {\mathcal H}_{\Sigma\Delta} \approx  E_\mathrm{C} \hat n_{\Sigma}^2 + E_\mathrm{J} \cos (2 \hat \varphi_{\Sigma}) + E'_{\rm S} \hat \varphi_{\Sigma}^2$~\cite{smith2020npj,smith2022magnifying}. 

Performing the variable transformation given by Eqs.~\ref{eqn:3mode_phi_operators} \& \ref{eqn:3mode_n_operators} on the combined Hamiltonian, we can thus approximate Eq.~\ref{eqn:transformed_Hamiltonian} as
    $\hat{\mathcal H}_{\rm approx} = E_{\rm C} \hat n_2^2 + E'_{\rm L} \hat \varphi_2^2
    - E_{\rm S} \cos [2\pi (\hat n_1 + \hat n_2)] + E_{\rm{2J}} \cos [2 (\hat \varphi_2 - \hat \varphi_1)]$.
Notably, since $\cos(2\pi \hat n_1)$, $\sin(2\pi \hat n_1)$, $\cos(2 \hat \varphi_1)$, and $\sin(2 \hat \varphi_1)$ commute with $\hat {\mathcal H}_{\rm approx}$, we can treat mode-1 as a frozen mode, and $\hat n_1$ and $\hat \varphi_1$ as static parameters set by external DC controls. The emergent dynamics of the multimode circuit is then encapsulated by mode-2, which forms grid in phase space (Fig.~\ref{figs4}\textbf{b}). Rearranging the external terms, we arrive at the single-mode Hamiltonian equivalent to the extended GKP Hamiltonian given by Eq.~\ref{eqn:GKP_approx},
\begin{equation}\label{eqn:1mode_hamiltonian}
\begin{split}
    \hat{\mathcal{H}}&=E_\mathrm{C}(\hat{n}+n_{\rm g})^2+E'_\mathrm{L}(\hat \varphi+\varphi_\mathrm{ext})^2\\ 
    &-E_\mathrm{S}\cos(2\pi \hat{n}) + E_\mathrm{2J}\cos(2\hat\varphi).
    \end{split}
\end{equation}

\subsection*{Supplementary Note 5 -- Circuit properties}
\begin{figure*}[ph]
\includegraphics[width=\textwidth]{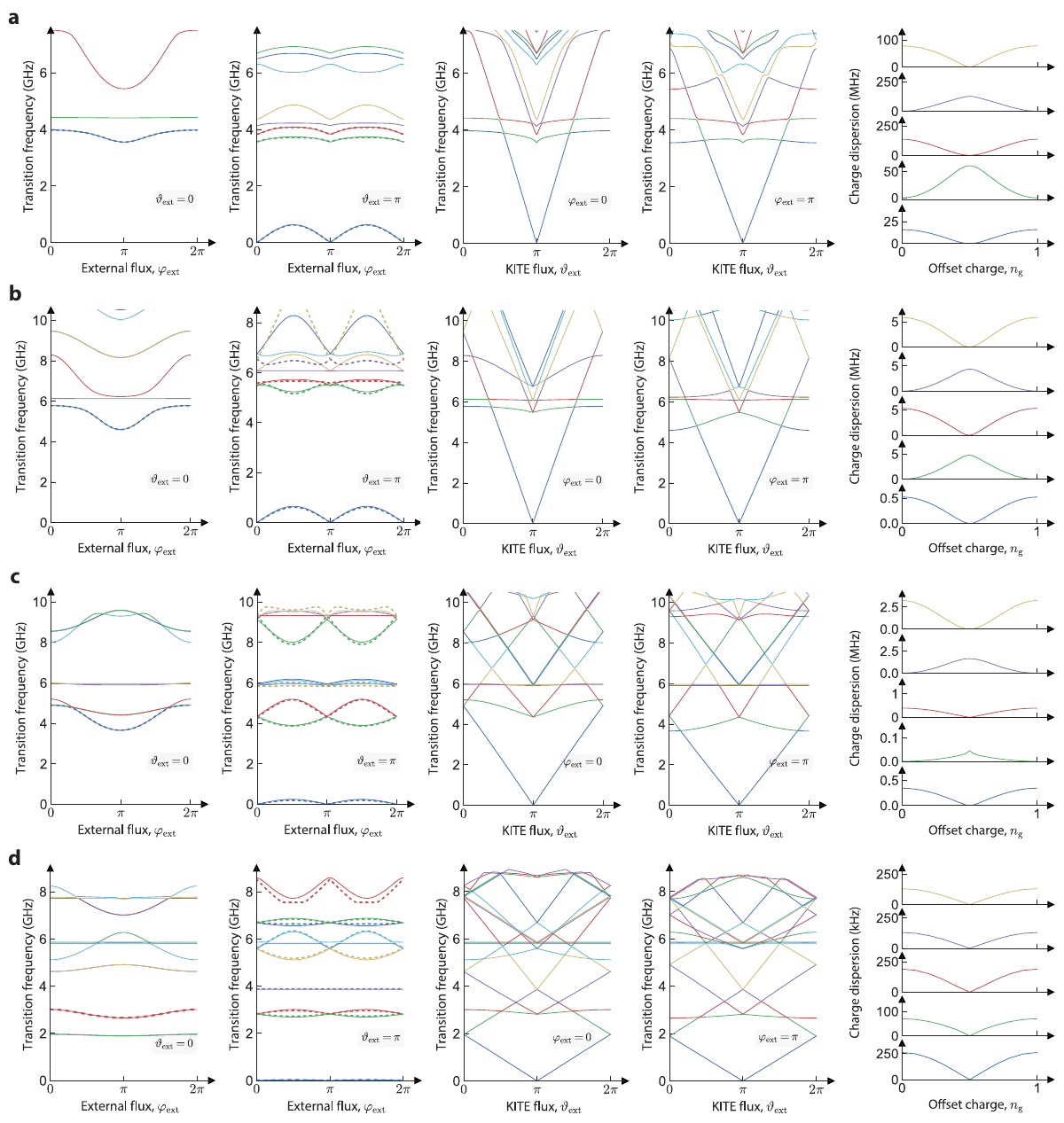}
    \caption{\label{figs5} \textbf{Multimode circuit spectra}. Transition frequency spectra from the ground state $|0\rangle$ of the gridium circuit described by Hamiltonian~(\ref{eqn:gridium_3mode}) are shown for various flux configurations. The corresponding circuit topology is depicted in Fig.~\ref{figs4}, and the circuit parameters associated with each row of panels, from (\textbf{a}) to (\textbf{d}), are listed in Table~\ref{tab:circuit_parameters}. The columns from left to right present the circuit behavior under the following conditions. (i) Transition spectra as a function of external flux $\varphi_\mathrm{ext}$ at fixed KITE flux $\vartheta_\mathrm{ext}=0$. Dashed lines indicate the transition frequencies predicted by the effective dualmon Hamiltonian~(Eq.~\ref{eqn:dualmon}). (ii) Transition spectra versus $\varphi_\mathrm{ext}$ at $\vartheta_\mathrm{ext}=\pi$. Dashed lines correspond to the effective gridium Hamiltonian~(Eq.~\ref{eqn:GKP_approx}). (iii) Transition spectra as a function of $\vartheta_\mathrm{ext}$ with $\varphi_\mathrm{ext}=0$. (iv) Transition spectra versus $\vartheta_\mathrm{ext}$ with $\varphi_\mathrm{ext}=\pi$. (v) Charge dispersion of the first five transitions at the flux bias point $\varphi_\mathrm{ext}=0$, $\vartheta_\mathrm{ext}=\pi$.}
\end{figure*}

\noindent The essence of our approach is to implement the extended GKP Hamiltonian given by Eq.~\ref{eqn:GKP_approx} through the three-mode circuit described by Hamiltonian~\ref{eqn:gridium_3mode}. Presently, we proceed to numerically simulate the properties of this circuit using the \texttt{scQubits} python package~\cite{groszkowski2021scqubits,chitta2022computer}, with a focus on spectral features and allowed transitions. We evaluate the additional complexities introduced by spurious capacitance and asymmetric parameters based on these simulations. To gain a comprehensive understanding of this circuit, we examine four distinct parameter regimes, as detailed on the left side of Table~\ref{tab:circuit_parameters}. The right side of the table shows the equivalent gridium circuit parameters obtained from least square fitting. The simulated spectral results and selection rules are presented in Fig.~\ref{figs5} and Fig.~\ref{figs6}, respectively.

We present the spectral data as transitions from the ground state $|0\rangle$, with the following variations (Fig.~\ref{figs5}): (i) the spectrum as a function of $\varphi_\mathrm{ext}$ when $\vartheta_\mathrm{ext}=0$, which in principle encodes the $d$=1 grid-states; (ii) $\varphi_\mathrm{ext}$-dependent spectrum when $\vartheta_\mathrm{ext}=\pi$, which shows manifestation from the gridium Hamiltonian, the focus of this work; (iii) $\vartheta_\mathrm{ext}$-dependent spectrum when $\varphi_\mathrm{ext}=0$, which highlight the important change of the energy landscape; (iv) $\vartheta_\mathrm{ext}$-dependent spectrum when $\varphi_\mathrm{ext}=\pi$; and (v) $n_\mathrm{g}$-dependent spectrum when $\varphi_\mathrm{ext}=\vartheta_\mathrm{ext}=0$, which represents the charge dispersion.

It is important to note that a unitary gauge transformation of the form $\hat{\mathcal{U}}=e^{-in_\mathrm{g}\hat{\varphi}}$, which eliminates the $n_\mathrm{g}$-dependence from the Hamiltonian, is only valid when the system’s wavefunctions are non-periodic. This interplay is elegantly captured by Bloch’s theorem, where the Bloch wavevector corresponds to the offset charge. Properly accounting for the compactness of the circuit variables is crucial. Notably, the extended GKP Hamiltonian presents a special case, where the presence of the inductive term is insufficient to eliminate charge sensitivity. This originates from the nonlinear capacitive, or so-called phase-slip, effect.

\begin{table}[b]
\centering
\caption{\label{tab:circuit_parameters} Circuit parameters corresponding to the simulation data shown in Fig.~\ref{figs5}. The unit is $h\cdot \mathrm{GHz}$.}
\renewcommand{\arraystretch}{1.3}
\begin{tabular}{c|cccccc|cccc}
\hline
\hline
\multicolumn{1}{c|}{} & \multicolumn{6}{c|}{Hamiltonian (\ref{eqn:gridium_3mode})} & \multicolumn{4}{c}{Hamiltonian (\ref{eqn:GKP_approx})} \\
\cline{2-11}
\textbf{Regime} & $E_\mathrm{J}$ & $E_\mathrm{C}$ & $E_\mathrm{L}$ & $E_\mathrm{LK}$ & $E_\mathrm{JS}$ & $E_\mathrm{CS}$ & $E_\mathrm{2J}$ & $E_\mathrm{S}$ & $E_\mathrm{C}$ & $E_\mathrm{L}$ \\
\hline\hline
\textbf{a} & 5 & 0.5 & 1 & 1 & 4 & 8 & 4.29 &4.21 & 0.56 & 0.41 \\
\textbf{b} & 10 & 0.5 & 1 & 1 & 4 & 8 & 9.81 &4.77 & 0.48 & 0.42 \\
\textbf{c} & 10 & 0.5 & 0.5 & 0.5 & 4 & 8 & 9.89 & 4.53 & 0.5 & 0.24 \\
\textbf{d} & 10 & 0.5 & 0.2 & 0.2 & 4 & 8 & 9.89 & 3.75 & 0.21 & 0.11 \\
\hline
\end{tabular}
\end{table}

Alongside spectral simulation, we evaluate the selection rules that dictate circuit control and readout. From the current-flow topology, and noting that the relevant phases are given by gradients of node fluxes, we find that the matrix elements associated with the differential flux induced by $I_\Delta$ couple predominantly to the operators $\hat{\varphi}_1$ and $\hat{\varphi}_2$. Given that $\hat \varphi_2$ inherently includes contributions from $\hat \varphi_1$, we focus our simulations solely on the matrix elements of $\hat \varphi_2$ (see Eq.~\ref{eqn:3mode_phi_operators}).

By analyzing node charges, we find that the resonator predominantly couples to the operator $\hat{n}_1$ (see Eq.~\ref{eqn:3mode_n_operators}). This coupling induces a dispersive interaction, causing the resonator’s frequency to shift depending on the qubit’s state. Notably, states with frequencies far away from the resonators would still induce a finite frequency shift through virtual transitions, analogous to the mechanism observed in fluxonium readout. This dispersive frequency shift can be estimated using second-order perturbation theory~\cite{zhu2013dispersive},
\begin{equation}\label{eqn:dispersive_shift}
        \chi_{i} = g^2\sum_{j\neq i} |\langle i |\hat n_1|j\rangle|^2 \frac{2\omega_{ij}}{\omega_{ij}^2 - \omega_\mathrm{r}^2}.
    \end{equation}
Here $g$ is the geometric coupling constant, defined via $\hat{\mathcal{H}}_\mathrm{q-r}=g\hat n_1 (\hat a_\mathrm{r} + \hat{a}_\mathrm{r}^\dagger)$, $\omega_\mathrm{r}$ is the resonator frequency, and $\omega_{ij}$ is the transition frequency between $|i\rangle$ and $|j\rangle$. For our simulations, we take the nominal parameter values $g/2\pi=100~\mathrm{MHz}$ and $\omega_\mathrm{r}/2\pi=7.5~\mathrm{GHz}$ unless specified otherwise. To improve clarity of the plots, we omit transitions with negligible matrix elements, and apply a gaussian filter to suppress the divergent points of the dispersive shifts in Fig.~\ref{figs6} arising from levels crossing between the qubit and the resonator. Quantum states with distinguishable dispersive shifts can be measured via the reflectometry microwave setup. 

The spectrum corresponding to the gridium regime (w.r.t. $\varphi_\mathrm{ext}$ when $\vartheta_\mathrm{ext}=\pi$) is fitted to the extended GKP Hamiltonian (given by Eq.~\ref{eqn:GKP_approx}) using least-square method to extract the corresponding parameters, summarized on the right-hand side of Table~\ref{tab:circuit_parameters}. Figure~\ref{figs5} displays the transitions corresponding to the extended GKP Hamiltonian in dashed lines. Our conversion starts with the first transition, and gradually include higher levels until the variances of the extracted parameters become exceedingly large or the levels no longer appear similar. The transitions corresponding to Hamiltonian~(\ref{eqn:GKP_approx}) that fit well are positioned close to their equivalent counterparts, and the extracted parameters are shown on the right-hand-side of Table.~\ref{tab:circuit_parameters}. Notably, we also observe good agreement between the first transitions in the $\vartheta_\mathrm{ext}=0$, $\varphi_\mathrm{ext}$-dependent spectra with the dualmon Hamiltonian (Eq.~\ref{eqn:dualmon}). However, since the KITE circuit biased at $\vartheta_\mathrm{ext}=0$ does not faithfully convey the quantum electrodynamics of a single Josephson junction in series with an inductor~\cite{smith2022magnifying}, we do not expect the transitions of the models to agree above 5 GHz.

Based on the simulation results presented in Fig.~\ref{figs5} and Fig.~\ref{figs6}, we discuss the circuit properties associated with each parameter regime detailed in Table~\ref{tab:circuit_parameters}. For regime (\textbf{a}), we expect to observe a few transitions below 5 GHz when $\vartheta_\mathrm{ext}=0$, enabling straightforward fitting procedures. Similarly, the spectrum at $\vartheta_\mathrm{ext}=\pi$ should exhibit multiple transitions with comparable dispersive shifts; transitions below 4 GHz align closely with predictions from the extended GKP model (Eq.~\ref{eqn:GKP_approx}). Notably, $\vartheta_\mathrm{ext}$-dependent spectra reveal distinct characteristics for $\varphi_\mathrm{ext}=0$ and $\pi$, and we include plots for both scenarios. Given the relatively small Josephson energy $E_\mathrm{J}$ in this regime, we expect pronounced charge dispersion, which can equivalently be attributed to limited grid support in the charge basis. In summary, this parameter regime yields clear experimental signatures that facilitate probing and theoretical comparison, albeit with the trade-off of reduced coherence times.

Turning now to regime (\textbf{b}), where $E_\mathrm{J}$ is doubled, we observe that the spectral features shift upward in frequency while largely retaining their original characteristics. Consequently, the extended GKP Hamiltonian remains accurate up to approximately 7 GHz. The most significant effect of this increased Josephson energy is a noticeable suppression of charge dispersion, which naturally results from enhanced grid support in the charge basis. Additionally, the larger $E_\mathrm{J}/E_\mathrm{C}$ ratio reduces the magnitude of the matrix elements, leading us to expect fewer observable transitions.

By doubling the inductance values and consequently lowering the inductive energies, we transition to regime (\textbf{c}). Relative to regime (\textbf{b}), the circuit transitions shift to lower frequencies, while the energy gaps between successive doublets become significantly larger than in regime (\textbf{a}). Charge dispersion further decreases to a level comparable to that of transmon qubits. Additionally, flux dispersion within the gridium regime is notably suppressed due to enhanced grid support in the phase basis. At this point, most matrix elements are negligible, leaving only a few observable transitions in the $\varphi_\mathrm{ext}$-dependent spectra. Despite this reduction, dispersive shifts remain distinguishable across various qubit states, facilitating experimental measurement and validation. Importantly, we observe strong agreement with the extended GKP Hamiltonian (Eq.~\ref{eqn:GKP_approx}) for frequencies up to 10 GHz in the gridium regime, further confirming the efficacy of our multimode circuit implementation. Thus, circuits with even larger $E_\mathrm{J}/E_\mathrm{C}$ ratios and smaller $E_\mathrm{L}$ than those in regime (\textbf{c}) are expected to faithfully realize the extended GKP Hamiltonian.

We now explore regime (\textbf{d}), which features even higher inductance and implements the GKP Hamiltonian better, offering larger grid support in the phase basis. The flux dispersion is further suppressed, and the $0$–$1$ transition becomes nearly degenerate across the entire flux period for $\vartheta_\mathrm{ext}=\pi$. Additionally, the transition frequencies are lowered throughout due to the reduced inductive energy. Notably, most matrix elements are now strongly suppressed, with only a few finite transitions remaining. The dispersive shifts are also largely uniform, which may hinder the ability to resolve and measure individual transitions. Compared to regime (\textbf{c}), the qubit in this regime offers enhanced protection, with energy ratios approaching those of the ideal GKP Hamiltonian (Eq.~\ref{eqn:GKP_ideal} and Table~\ref{tab:circuit_parameters}) at the cost of progressively restrictive selection rules. Characterizing such a circuit will likely require advanced techniques beyond the current repertoire.


\begin{figure*}[ph]
\includegraphics[width=\textwidth]{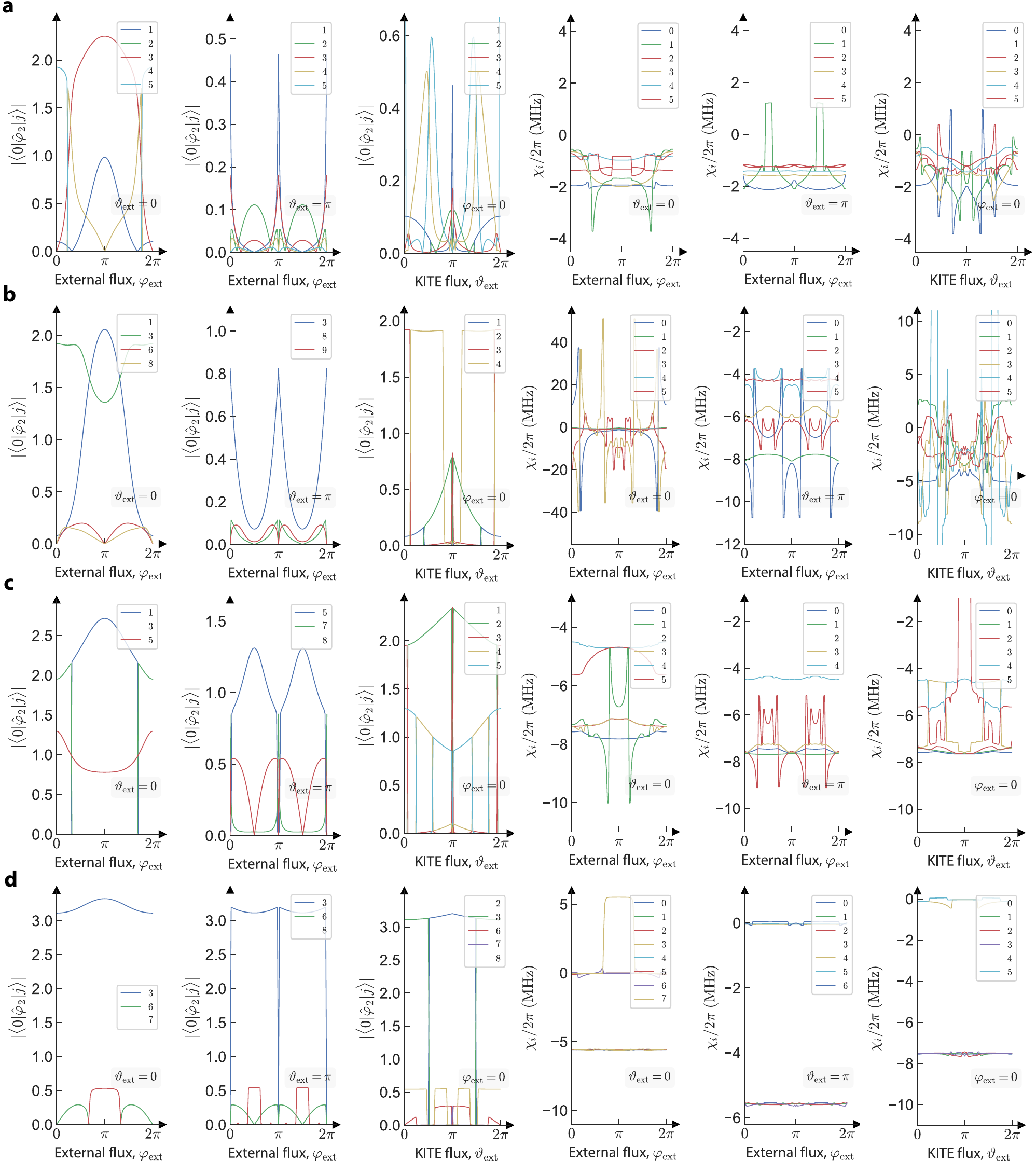}
    \caption{\label{figs6} \textbf{Selection rules of the multimode circuit}. The three left columns display the dominant phase matrix elements from the ground state $|0\rangle$ with respect to operator $\hat \varphi_2$, along with the corresponding spectra from Fig.~\ref{figs5}. The three right columns show the computed dispersive shift $\chi_i$ of a resonator at frequency $\omega_\mathrm{r}/2\pi=7.5~\mathrm{GHz}$,  capacitively connected to the qubit via operator $\hat n_1$ with coupling strength $g/2\pi=100~\mathrm{MHz}$, when the circuit is in eigenstate $|i\rangle$.  Quantum states with distinguishable dispersive shifts can be measured. Rows (\textbf{a})–(\textbf{d}) correspond to the circuit parameters listed in Table~\ref{tab:circuit_parameters}. The other external flux variable is fixed at the specified values in each plot.}
\end{figure*}
\begin{figure*}[ht]
 \includegraphics[width=0.9\textwidth]{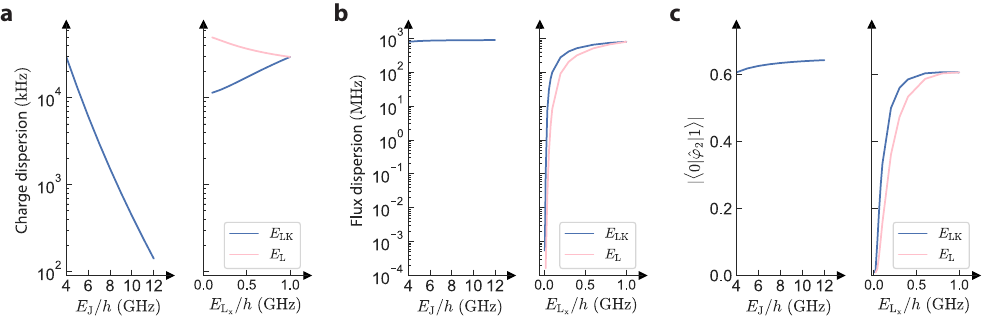}
    \caption{\label{figs7} \textbf{Engineering gridium qubit}. (\textbf{a}) Charge dispersion. (\textbf{b}) Flux dispersion. (\textbf{c}) Matrix element corresponding to $\hat \varphi_2$ at the symmetric flux bias, $\varphi_\mathrm{ext}=0,\vartheta_\mathrm{ext}=\pi$.}
\end{figure*}
We summarize the circuit’s grid support in both the charge and phase bases by evaluating the charge and flux dispersion for $\vartheta_\mathrm{ext} = \pi$ across varying circuit parameters. Additionally, the degree of protection against energy decay can be represented by the matrix element $\langle 0 | \hat \varphi_2 | 1 \rangle$. The results are presented in Fig.~\ref{figs7}. The charge dispersion is efficiently suppressed by increasing the Josephson energy $E_\mathrm{J}$, which enhances grid support in the charge basis. This effect corresponds to an increasing ratio of $E_\mathrm{2J}/E_\mathrm{C}$ in Hamiltonian~(\ref{eqn:GKP_approx}). Modifying the inductive terms also impacts the effective CQT amplitude $E_\mathrm{2J}$ (see SI Note 3), thereby influencing charge dispersion.
Similarly, the flux dispersion can be reduced by lowering the inductive energies. It rapidly approaches zero for small $E_\mathrm{L_x}$ as the circuit nears the non-loop topology limit. The phase matrix element $\langle 0 | \hat \varphi_2 | 1 \rangle$ exhibits the same qualitative trend. These results confirm the theoretical prediction that a three-mode circuit approaching the ideal GKP limit—characterized by enormous capacitance and inductance—offers strong protection against environmental noise.

\subsubsection*{Stray capacitance}  

\noindent Based on the analysis of the three-mode circuit governed by Hamiltonian~(\ref{eqn:gridium_3mode}), we now examine the impact of practical circuit imperfections, beginning with spurious capacitance. In any realistic implementation, stray electric fields between conducting components are inevitable, giving rise to unintended parasitic capacitances. The prominent case, visible in the device image (Fig.~\ref{fig2}\textbf{b}), is the parasitic capacitor $C_\mathrm{K}$ formed between the two outer conducting pads. This element introduces an additional charging energy term, $\varepsilon_\mathrm{K} = e^2 / 2C_\mathrm{K}$, which normalizes the kinetic energy of the differential mode.

Following the variable transformation introduced in Eq.~\ref{eqn:full_lagrangian}, we arrive at 

\begin{equation}
\begin{aligned}
    \mathcal{L} = & \frac{1}{2} \phi_0^2 \left[ 2C_\mathrm{J} \dot{\varphi}_\Sigma^2 + (2C_\mathrm{J} + C_\mathrm{K}) \dot{\varphi}_\Delta^2 + C_\mathrm{JS} \dot{\varphi}_\mathrm{S}^2 + C_\mathrm{p} \dot{\phi}^2 \right] \\
    & + 2E_\mathrm{J} \cos \varphi_\Sigma \cos \varphi_\Delta + E_\mathrm{JS} \cos \varphi_\mathrm{S} \\
    & - E_\mathrm{LK} \left[ (\varphi_\Sigma - \phi)^2 + \left(\varphi_\Delta - \frac{1}{2} \vartheta_{\text{ext}} \right)^2 \right] \\
    & - E_\mathrm{L} (\varphi_\mathrm{S} - \phi + \varphi_{\text{ext}})^2.
\end{aligned}
\end{equation}
The Legendre transformation then leads to 
\begin{equation}\label{eqn:gridium_4mode_eCK}
    \begin{split}  \hat{\mathcal{H}}&=2E_\mathrm{C}\hat{n}^2_\Sigma+2E'_\mathrm{C}\hat{n}^2_\Delta+4E_\mathrm{CS}\hat{n}_\mathrm{S}^2 + 4\varepsilon_\mathrm{p}\hat{n}^2 \\
        &- 2E_\mathrm{J}\cos\hat{\varphi}_\Sigma\cos\hat{\varphi}_\Delta - E_\mathrm{JS}\cos\hat{\varphi}_\mathrm{S}\\
        &+E_\mathrm{LK}\Big[(\hat{\varphi}_\Sigma-\hat{\phi})^2 + (\hat{\varphi}_\Delta-\frac{1}{2}\vartheta_\mathrm{ext})^2\Big]
        \\&+ E_\mathrm{L}(\hat{\varphi}_\mathrm{S}-\hat{\phi}+\varphi_\mathrm{ext})^2,
    \end{split}
\end{equation}
where $E'_\mathrm{C}=e^2/2(C_\mathrm{J}+0.5C_\mathrm{K})=2E_\mathrm{C}\varepsilon_\mathrm{K}/(E_\mathrm{C}+2\varepsilon_\mathrm{CK})$. Hence, we see that the cross-KITE capacitance reshapes the charging energy associated with mode $\Delta$, effectively lowering it. In the limit $C_\mathrm{K}\rightarrow 0$, $E'_\mathrm{C}\rightarrow E_\mathrm{C}$. The 4-mode circuit Hamiltonian in Eq.~\ref{eqn:gridium_4mode_eCK} embodies the general description used to explore spurious capacitive effects.

\begin{table}[b]
\centering
\caption{\label{tab:spurious_capacitance} Effects from cross-KITE capacitance $C_\mathrm{K}$ and stray capacitance $C_\mathrm{p}$. The parameter regimes are defined in Table~\ref{tab:circuit_parameters}. The unit of the extracted parameters is $h\cdot\mathrm{GHz}$.}
\renewcommand{\arraystretch}{1.3}
\begin{tabular}{c|cccc|cccc}
\hline
\hline
\multicolumn{1}{c|}{} & \multicolumn{4}{c|}{With $\varepsilon_\mathrm{K}/h=2.5~\mathrm{GHz}$} & \multicolumn{4}{c}{With $\varepsilon_\mathrm{p}/h=5.5~\mathrm{GHz}$} \\
\cline{2-9}
\textbf{Regime} & $E_\mathrm{2J}$ & $E_\mathrm{S}$ & $E_\mathrm{C}$ & $E_\mathrm{L}$ & $E_\mathrm{2J}$ & $E_\mathrm{S}$ & $E_\mathrm{C}$ & $E_\mathrm{L}$ \\
\hline\hline
\textbf{a} & 4.53 &3.54 &0.47 & 0.37 & 4.23 &3.75 & 0.53 & 0.4 \\
\textbf{b} & 9.81 & 4.95 & 0.41 &0.45 & 9.81 &4.23 & 0.4 & 0.44 \\
\textbf{c} & 9.89 & 4.71 & 0.39 & 0.23 & 9.89 & 4.2 & 0.51 & 0.22 \\
\textbf{d} & 9.89 & 3.35 & 0.21 & 0.11 & 9.89 & 3.39 & 0.21 & 0.11 \\
\hline
\end{tabular}
\end{table}

We begin by examining the gridium circuit spectra (with $\vartheta_\mathrm{ext}=\pi$) across the four parameter regimes in the presence of two types of capacitive imperfections: a cross-KITE charging energy term $\varepsilon_\mathrm{K}/h = 2.5~\mathrm{GHz}$ and a spurious capacitance across the circuit corresponding to $\varepsilon_\mathrm{p}/h = 5.5~\mathrm{GHz}$. The resulting spectra are fitted to the extended GKP Hamiltonian to extract the effective circuit parameters, as summarized in Table~\ref{tab:spurious_capacitance}. The fit reveals that the presence of $\varepsilon_\mathrm{K}$ reduces the effective charging energy, while $\varepsilon_\mathrm{p}$ predominantly contributes to the $E_\mathrm{CS}$ term, thereby lowering the effective phase-slip amplitude $E_\mathrm{S}$. In addition, at frequency above 6 GHz, we observe that a finite $\varepsilon_\mathrm{K}$ slightly lifts the degeneracy between the levels, and the value of $\varepsilon_\mathrm{p}$ introduces an additional mode.

The spectra across the four regimes remain largely unchanged, as expected, since adding capacitive elements mainly leads to a renormalization of the kinetic energy terms. The finite cross-circuit capacitance associated with $\varepsilon_\mathrm{p}$ introduces an additional mode; however, for sufficiently large $\varepsilon_\mathrm{p}$, this mode lies well above the frequency range of interest and does not affect the low-energy spectrum. Likewise, we perform numerical simulations of the matrix elements corresponding to the differential flux operator across all four regimes with varying external flux values in the presence of the spurious capacitive terms. Although the absolute magnitudes of the matrix elements are modified, their functional dependence on the external parameters remains invariant, consistent with our anticipation.

\subsubsection*{Asymmetry}

\noindent Our analysis thus far has neglected the impact of asymmetry in the KITE circuit. We now consider the effects arising from small fluctuations in the junction energies, which can be parameterized by the asymmetry ratio $\epsilon_\mathrm{J}$ such that $E_\mathrm{J_{1,2}}={E}^\mathrm{mean}_\mathrm{J}(1\pm \epsilon_\mathrm{J})$. This asymmetry modifies the Lagrangian and consequently the Hamiltonian of the system, which becomes $\hat{\mathcal{H}} = \hat{\mathcal{H}}_0 + \hat{\mathcal{H}}_{\epsilon_\mathrm{J}}$, where $\hat{\mathcal{H}}_0$ denotes the original three-mode Hamiltonian given in Eq.~\ref{eqn:gridium_3mode}, and the correction due to asymmetry is 
\begin{equation}\label{eqn:asymmetric_hamiltonian}
    \hat{\mathcal{H}}_{\epsilon_\mathrm{J}} = 2{\epsilon_\mathrm{J}} E_\mathrm{J}\sin \hat\varphi_\Sigma \sin\hat\varphi_\Delta.
\end{equation}
This term introduces additional coupling between the $\Sigma$ and $\Delta$ modes of the circuit. Given that these modes are already coupled with strength on the order of $E_\mathrm{J}$, we expect the additional interaction from small asymmetry to only weakly perturb the circuit dynamics.

Similarly, an asymmetry in the KITE inductors, parameterized as $E_\mathrm{LK_{1,2}} = E^\mathrm{mean}_\mathrm{LK}(1 \pm \epsilon_\mathrm{LK})$, perturbs the system Hamiltonian as $\hat{\mathcal{H}} = \hat{\mathcal{H}}_0 + \hat{\mathcal{H}}_{\epsilon_\mathrm{LK}}$, where the first-order correction is given by
\begin{equation}\label{eqn:asymmetric_hamiltonian_LK}
\hat{\mathcal{H}}_{\epsilon_\mathrm{LK}} = 2\epsilon_\mathrm{LK} \frac{E_\mathrm{L} E_\mathrm{LK}}{E_\mathrm{L} + E_\mathrm{LK}} \hat\varphi_\Delta (\hat\varphi_\mathrm{S} - \hat\varphi_\Sigma).
\end{equation}
This term enhances the coupling between the circuit modes, particularly between the slow $\Delta$-mode and the fast $\mathrm{S}$-mode. If the asymmetry is sufficiently large, the $\Delta$-mode may strongly hybridize with the $\mathrm{S}$-mode, thereby invalidating the Born–Oppenheimer approximation. A second-order correction to the effective inductive energy of the $\Delta$-mode also arises, though its contribution is comparatively minor.

We numerically simulate the impact of asymmetry on the circuit spectra and transition matrix elements. By introducing a controlled energy imbalance corresponding to increasing asymmetry $\epsilon$—up to 10\%—we compare the results across the four parameter regimes. While certain eigenfrequencies and transition amplitudes exhibit observable shifts, the overall spectral structure and selection rules remain largely consistent with those of the symmetric case. These findings indicate that the circuit is intrinsically robust against realistic levels of fabrication-induced asymmetry.

\begin{figure}[t]
    \includegraphics[width=0.45\textwidth]{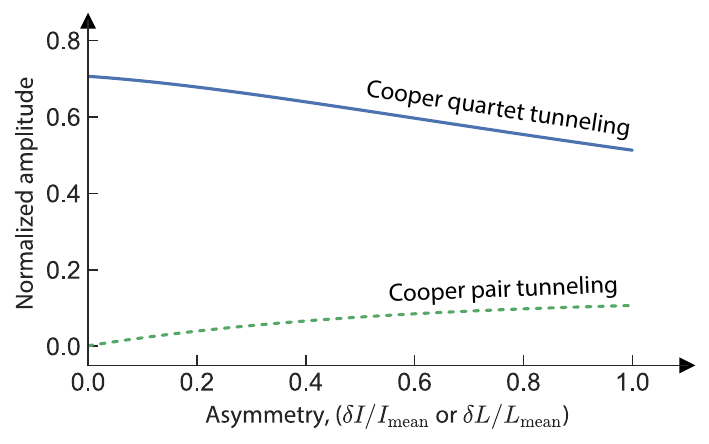}
    \caption{\label{figs_asymmetry} \textbf{Effects from asymmetry}. Changes of the effective tunneling harmonics resulted from asymmetry in either the KITE junctions or inductors. The result is limited to the non-hysteretic regime.}
\end{figure}

Increasing asymmetry inherently relaxes the suppression of single-Cooper-pair tunneling within the KITE, thereby perturbing the circuit’s selection rules. To quantitatively assess this effect, we revisit the semiclassical analysis outlined in SI Note 2 and examine the CPR of the KITE. Through Fourier decomposition, we extract the first and second harmonics of the CPR as a function of asymmetry level (Fig.~\ref{figs_asymmetry}). As expected, asymmetries in either the inductive or Josephson elements reintroduce single-Cooper-pair tunneling processes, concomitantly suppressing coherent Cooper-quartet tunneling. Incorporating up to 10\% asymmetry into the toy model simulations (see SI Note 1), we observe no appreciable deviation in the spectral features or matrix elements, thereby affirming the intrinsic resilience of the gridium qubit architecture against moderate fabrication-induced imperfections.

\begin{figure*}[ht]
    \includegraphics[width=\textwidth]{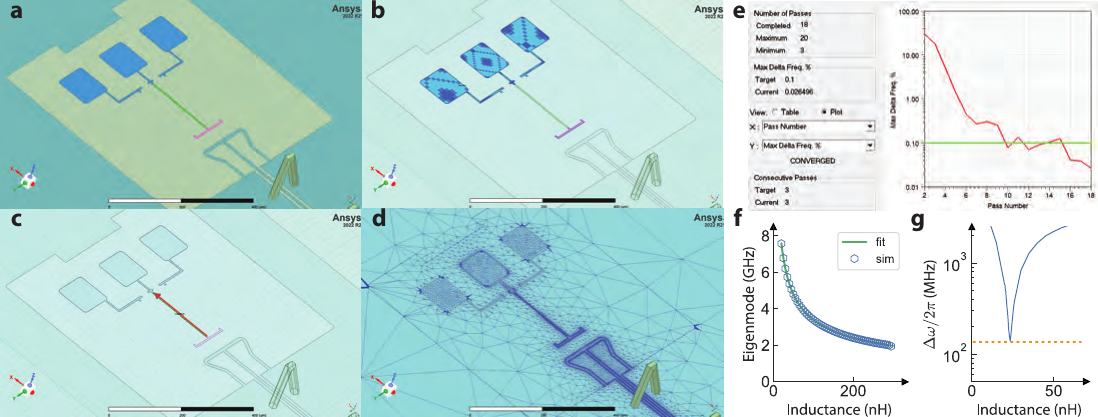}
    \caption{\label{figs_design} \textbf{Simulation of circuit parameters}. (\textbf{a}) Import of design and layer definition in Ansys HFSS. (\textbf{b}) Ground plane and metal electrodes are given the boundary condition as perfect electrical conductors. (\textbf{c}) An artificial lumped inductor is defined between the two electrodes of interest. The arrow indicates the flow of current. (\textbf{d}) A planar display of the final tetrahedral mesh upon completion of one simulation run. (\textbf{e}) Convergence of the adaptive solution setup. The left side presents the setting parameters, and the right side shows the simulation accuracy after each pass. (\textbf{f}) The effective shunting capacitance is extracted using least-square fit to the eigenmode frequencies. (\textbf{g}) Simulated avoided crossing between the resonator mode and linear qubit mode. The dash line indicates the amplitude of the coupling, which is equal to $g/\pi$.
    }
\end{figure*}

\subsection*{Supplementary Note 6 -- Circuit design}

\noindent Once the circuit parameters have been determined, we proceed to design the superconducting device as follows. First, the effective capacitive shunts between the metal electrodes are estimated using \texttt{Ansys HFSS} 3D high-frequency finite-element simulation software. The \texttt{gds} design file is imported into an HFSS platform via Ansys Electronics Desktop, and the simulation steps are visualized in Fig.~\ref{figs_design}\textbf{a-d}. We next draw the silicon substrate and define the package space within the software, define the ground plane and metal electrodes, and add the wirebonds across the coplanar-waveguide flux line manually to eliminate the associated slot-line modes. A shunt inductor is artificially defined across the circuit branch we wish to find the associated capacitance. Then, we specify a strict convergence criteria of 0.1\% and sweep the inductance value to avoid possible fluctuation effects. 

Afterwards, we inspect the converged results (Fig.~\ref{figs_design}\textbf{e}), and fit the eigenmodes using the simple resonance relation $\omega=(LC)^{-1/2}$ to extract the capacitance value (Fig.~\ref{figs_design}\textbf{f}).  The procedure is repeated for different pairs of electrodes. The coupling between the resonator and the qubit's $\Sigma$ mode is extracted by sweeping the inductor value between the outer pad and middle pad such that the lumped circuit's resonance crosses that of the readout resonator, manifesting into an avoided crossing with a frequency gap of $2g$ (Fig.~\ref{figs_design}\textbf{g}). For the presented design, we achieve $[E_\mathrm{C},\varepsilon_\mathrm{p},\varepsilon_\mathrm{K}]/h\sim [0.75, 5.43, 2.45]~\mathrm{GHz}$, and $g/2\pi\sim120~\mathrm{MHz}$. Here, $\varepsilon_\mathrm{K}$ is the cross-KITE capacitance between the two outer electrodes, which lowers the charging energy of the $\Delta$ mode. The simulated mutual inductance from the flux line varies from 1 to 2~pH.

To construct an inductor with the desired inductive energy $E_\mathrm{L}$, we fabricate an array of $N$ Josephson junctions, each with area $A$ and critical current density $J_c$, such that $E_\mathrm{L}=N\times E_\mathrm{JA}$, where $E_\mathrm{JA}=\phi_o J_c A$ is the Josephson energy of each junction. The charging energy of each junction corresponding to the self-capacitance $C_\mathrm{JA}$ can be estimated using the empirical formula $C_\square\approx45~\mathrm{fF/\mu m^2}$ for Al-AlO$_\mathrm{x}$-Al junctions. In addition, the effective capacitance to ground, $C_\mathrm{g}\approx 30-40~\mathrm{aF}$, is obtained from independent measurements~\cite{manucharyan2012evidence,pechenezhskiy2020superconducting}. The tunnel barrier of the junctions determines its critical current density $J_c$ and, in combination with the junction area $A$, its critical current $I_c$. Over the course of the experiment, the evaporation parameters can be tuned to change $J_c$, resulting in a range from 300 to 750~$\mathrm{nA/\mu m^2}$. By varying the junction area and number, we can achieve the desired inductors.

For example, with $J_c=300~\mathrm{nA/\mu m^2}$ and $A=0.5\mathrm{\mu m}\times 2\mathrm{\mu m}$, we obtain $E_\mathrm{CA}/h\approx 0.54~\mathrm{GHz}$, $E_\mathrm{JA}/h=149~\mathrm{GHz}$ (or $L_\mathrm{JA}\approx 1.1~\mathrm{nH}$)~\cite{junger2025implementation}. This allows us to fabricate an inductor with $E_\mathrm{L}/h=0.6~\mathrm{GHz}$ by chaining 250 junctions together. Meanwhile, with $J_c=750~\mathrm{nA/\mu m^2}$, we can miniaturize the junction to $A=0.25\mathrm{\mu m}\times 0.8\mathrm{\mu m}$, obtaining $E_\mathrm{CA}/h\approx 2.15~\mathrm{GHz}$ and $E_\mathrm{JA}/h\approx74.5~\mathrm{GHz}$ (or $L_\mathrm{JA}\approx 2.2~\mathrm{nH}$), thereby achieving the same inductance with half the number of junctions in the array.

\begin{figure}[t]
    \includegraphics[width=0.48\textwidth]{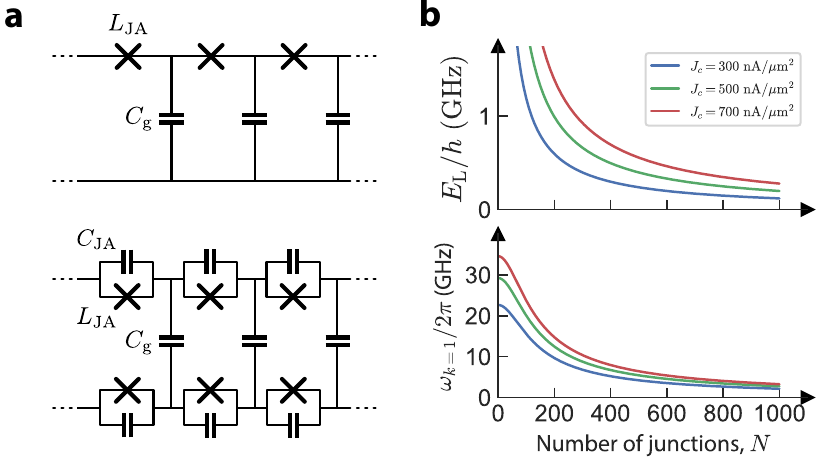}
    \caption{\label{figs_inductor} \textbf{Josephson junction array}. (\textbf{a}) Models of a distributed junction array. Each junction has a Josephson inductance $L_\mathrm{JA}$, self-capacitance $C_\mathrm{JA}$, and capacitance to ground $C_\mathrm{g}$. (\textbf{b}) Estimated frequency of the first array mode ($k=1$), below which the junction arrays behave as ideal lumped inductors. The junction area is taken as $A=0.5\mathrm{\mu m}\times 2\mathrm{\mu m}$.}
\end{figure}

At the same time, we require the array to behave as a lumped inductor in the frequency range of interest. Although the junction array may behave as a transmission line at high frequencies, it can be considered a lumped inductor below the first array mode frequency $\omega_{k=1}$. In other words, the lowest-order array mode of the chain of junctions must reside at a sufficiently high frequency. We estimate the array modes using~\cite{masluk2012microwave}
\begin{equation}\label{eqn:array_modes}
    \omega_{k}=\omega_0\sqrt{\frac{1-\cos(\pi k/N)}{C_g/(2C_\mathrm{JA})+(1-\cos(\pi k/N))}},
\end{equation}
where $\omega_0=(L_\mathrm{JA}C_\mathrm{JA})^{-1/2}$ is the plasma frequency of a single junction. The built superinductors should ideally reach the desired inductive energy while having a high-frequency array mode. Figure~\ref{figs_inductor} summarizes the effective array parameters and serves as a guideline to construct high impedance devices that operate as lumped-element circuits in the frequency range of interest.

Notably, coherent quantum phase slip can also occur along the superinductors, which may interfere with each other, causing unwanted decoherence via the Aharonov-Casher linewidth broadening effect~\cite{manucharyan2012evidence}. We estimate the pure dephasing rate due to this behavior stemming from the individual junction's phase slip energy $E_\mathrm{SA}$ as

\begin{equation}\label{eqn:array_ps}
\begin{split}
    \Gamma^\mathrm{PS}_\varphi&=\pi E_\mathrm{SA}\sqrt{N}|F_{\alpha\beta}(\varphi_\mathrm{ext})|, \\
    &\leq8\sqrt{\pi NE_\mathrm{JA}E_\mathrm{CA}}\sqrt[4]{\frac{8E_\mathrm{JA}}{E_\mathrm{CA}}}\exp\left(-\sqrt{\frac{8E_\mathrm{JA}}{E_\mathrm{CA}}}\right),
\end{split}
\end{equation}
where $F_{\alpha\beta}\leq 1$ characterizes the wavefunctions' overlap, which is dependent on the matrix elements of the device. We estimate the upper bound of the dephasing rate by considering $F_{\alpha\beta}=1$, noting that it is negligible for transitions with small phase matrix elements. Equation~\ref{eqn:array_ps} yields a dephasing time limit, $T_\varphi^\mathrm{PS}$, on the order of days for the first example, but below one microsecond for the second case considered above.

\subsection*{Supplementary Note 7 -- Fabrication procedure}

\noindent The gridium qubit is fabricated on a silicon substrate using standard photolithography and electron-beam lithography. Photolithography defines the capacitor pads, flux control line, and coplanar waveguide readout resonator on a sputtered niobium (Nb) base layer. Before Nb deposition, the silicon substrate undergoes a Piranha clean for 600~seconds followed by a 10:1 H$_2$O:HF dip for 60~seconds. The wafer is then rinsed, spun dry, and loaded into the Nb sputtering chamber. After pumping the loadlock for 15~hours, Nb is sputtered onto the silicon wafer with argon (Ar) plasma. A sputtering time of 720~seconds at 1.5~mTorr produces a Nb ground plane with compressive stress. 

To pattern the ground plane, AZ MIR 701 photoresist is first spin-coated to a thickness of 1~$\mathrm{\mu m}$. Optical lithography is performed using a Heidelberg MLA 150 with an exposure dose of 140~$\mathrm{mJ/cm^2}$. The resist is then developed in MF-26A solution. Before etching, a hard bake step is applied to prevent staining of the Nb ground plane.  Etching is carried out using BCL$_3$-Cl$_2$ plasma for 17~seconds. Finally, the photoresist is stripped in N-methylpyrrolidone (NMP) overnight at 80$^o\mathrm{C}$ which completes the ground plane fabrication. 

Next, electron-beam lithography is used to pattern the single Josephson junctions, junction arrays, and connecting leads. To remove the native silicon oxide, the wafer is dipped in a 5:1 buffered oxide etch for 30~seconds, then spin-rinsed with IPA and baked at 200$^\circ\mathrm{C}$ for 60~seconds. A bilayer resist stack of methyl methacrylate (MMA EL 13) and AR-P 6200 (CSAR 62) is then applied. The MMA EL 13 is spun at 2000~rpm for 90~seconds, baked at 150$^\circ\mathrm{C}$ for 90~seconds, and cooled with a nitrogen airgun for 60~seconds. Afterwards, CSAR 62 is spun at 1000~rpm for 60~seconds, baked at 150$^\circ\mathrm{C}$ for 60~seconds, and similarly cooled. The coated wafer is loaded into a Raith EBPG 5200 and the patterns are written using a 100~keV beam. We use a dose of 550~$\mathrm{\mu C/cm^2}$ to clear the entire bilayer stack and 150~$\mathrm{\mu C/cm^2}$ to selectively clear the MMA EL 13. The patterned wafer is then developed in two stages. First, CSAR 62 is developed in N-amyl acetate (NAA) for 60~seconds at 0$^\circ\mathrm{C}$. After an IPA rinse, MMA EL 13 is developed in a 1:1 solution of deionized water (DI H$_2$O) and IPA for 120~seconds. To prevent collapse of the Dolan bridge, sonication is avoided, and the wafer is spin-dried instead of blow-dried.

Following development, the wafer undergoes a gentle oxygen plasma descum (YES-G 500) before aluminum (Al) deposition in a Plassys MEB. After a 15-hour loadlock pump-down, Al is deposited using double-angle shadow evaporation, with an intermediate oxidation step at 20~mbar for 30~minutes to form the tunneling barrier. The resulting Al leads are 20~$\mathrm{\mu m}$ (bottom) and 100~$\mathrm{\mu m}$ (top) thick, with the large thickness difference suppressing quasiparticle tunneling across the junction barrier. Finally, all extraneous metal is lifted-off in acetone at 67$^\circ\mathrm{C}$ for over two hours.

To form robust electrical contact and eliminate spurious junctions, we deposit an Al bandaid layer on the overlapping region between the Nb capacitors and Al leads~\cite{grunhaupt2017argon}. All steps are equivalent to the junction fabrication, except for the development of the MMA EL 13, in which we sonicate the wafer for 60~seconds in a 3:1 IPA:H$_2$O solution. After deposition and liftoff, we gently ash the wafer in oxygen plasma and probe the resistances of all the samples across the wafer. While the resistance probing cannot determine the single junction and junction array resistances, it can rule out shorts, opens, and asymmetry in the gridium circuit. With scanning electron microscopy of test samples, the circuit parameters can be deduced from the measured resistances. 

Finally, we dice the wafer into individual chips, using soft-baked AZ MIR 701 as a protective layer. The individual chips are thoroughly cleaned with an overnight dip in NMP, followed by spray cleaning with DI water, acetone, and IPA. The samples undergo a final oxygen plasma ashing and are wirebonded onto a printed circuit board in an indium-sealed light-tight box.

\subsection*{Supplementary Note 8 -- Experimental setup}

\begin{figure}[t]
    \includegraphics[width=0.48\textwidth]{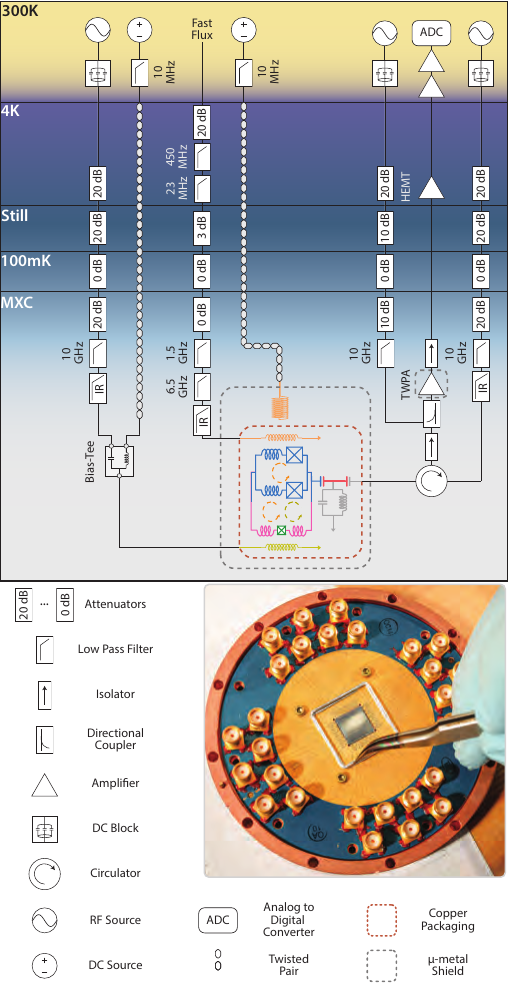}
    \caption{\label{figs_fridge} \textbf{Experimental setup}. Cryogenic wiring layout and device packaging utilized for the experiment. The lines connecting to the sample package are separated into four domains from left to right: RF+DC control, fast-flux-only, DC-only, and readout.}
\end{figure}

\noindent Prior to wirebonding, candidate devices are inspected using optical microscopy and resistance probing at room temperature. Peripheral and proxy dies are imaged with scanning-electron microscopy to further check for possible defects. Once a 1cm$\times$1cm chip is selected, it is wirebonded onto a printed circuit board that is integrated into a round copper box and packaged with indium seal, as shown in Fig.~\ref{figs_fridge}. In practice, the device contains other qubits such as transmons, fluxoniums, and inductively shunted KITE for testing and process qualification. Before being loaded into the fridge, the entire package is stored in a vacuum container. 

Once the cooldown time has been set, the package is taken out of vacuum and subsequently anchored to the designated cold finger in the mixing chamber plate. An external superconducting magnetic coil is then attached directly to the package along an axis perpendicular to the chip. The experiment is carried out either in an Oxford Triton 300 or a Bluefors LD 400 dilution refrigerator. The readout input line is attenuated by 20-dB at the 4K, still, and mixing chamber, with additional 10-GHz and Eccorsorb CR110 low pass filters located at the mixing chamber plate, as shown in Fig.~\ref{figs_fridge}. The RF driving line goes through similar attenuation, and is connected to the differential flux loop via a modified bias-Tee. The outgoing signal is amplified by a Josephson traveling wave amplifier (TWPA), which is pumped through a dedicated line. The DC control currents are applied through two dedicated twisted pairs made of NbTiN which superconduct and become lossless below 4K. The lines are further filtered at 10-MHz low-pass at room temperature. To perform fast flux measurement, we use another dedicated line that includes a 20-dB attenuator at 4K and a combination of low-pass filters as shown, with the lowest cutoff frequency at 25 MHz. 

We utilized the following instruments at room-temperature for the experiment. The readout input is generated using Holzworth HSX9000, while the qubit drives come from Rohde \& Schwarz SGMA SGS100A vector RF sources. The baseband pulses are generated by a Tektronix 5014C arbitrary waveform generator (AWG), which also triggers the AlazarTech ATS9373 analog-to-digital (digitizer) card. The TWPA is pumped using a Holzworth HSX9000. The DC currents/voltages are applied using Yokogawa GS200. Alternatively, we use a Zurich Instruments SHFQC+ to fulfill most of these roles. At room-temperature, the fast-flux line may include a bias-Tee to apply additional voltage/flux, or connected to a AWG channel with built-in DC digital-to-analog cards.
\subsection*{Supplementary Note 9 -- Proxy measurements}

\begin{figure}[t]
    \includegraphics[width=0.48\textwidth]{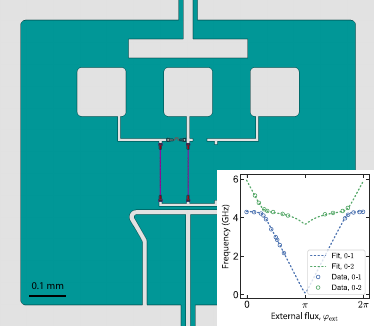}
    \caption{\label{figs_fluxonium} \textbf{Fluxonium experiment}. Design schematic of a fluxonium qubit implemented using the same Josephson junctions, capacitive electrodes, and external circuitry adapted from the gridium qubit architecture. Inset: fluxonium spectrum with theoretical fit used for parameter extraction.}
\end{figure}

\noindent Performing proxy experiments to extract design parameters offers a practical and efficient approach to characterizing complex quantum circuits. By isolating a simplified or well-understood subsystem that shares components with the target architecture, we can directly probe critical circuit parameters such as inductance, junction energy, or capacitance without the complications of the full device, thereby accelerating the experiment timeline. 

To this end, we incorporate standard qubits that share key components with the gridium architecture. Characterizing these auxiliary devices during each measurement cycle provides valuable design feedback and serves as an effective means of fabrication process control. As an example, Fig.~\ref{figs_fluxonium} shows the design schematic of a fluxonium qubit used to extract key parameters such as the charging energy, Josephson energy, inductive energy, coupling strengths, and flux tunability. The extracted values, $[E_\mathrm{J}, E_\mathrm{C}, E_\mathrm{L}]/h = [4, 0.76, 0.38]~\mathrm{GHz}$, show excellent agreement with target specifications.

\subsection*{Supplementary Note 10 -- Flux calibration}

\begin{figure}[t]
    \includegraphics[width=0.48\textwidth]{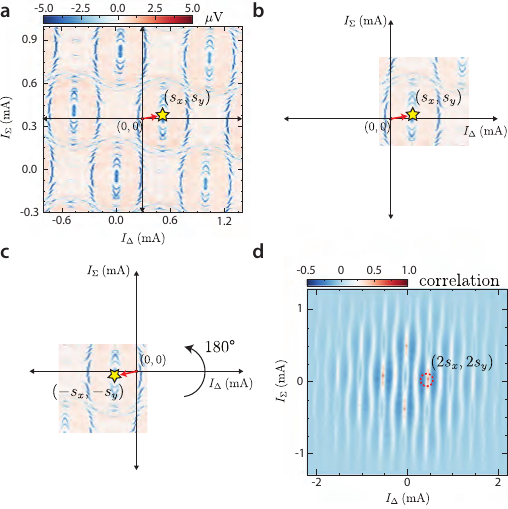}
    \caption{\label{figs6} \textbf{Calibration of DC flux control}. (\textbf{a}) The symmetry point of the reflected resonator signal presented in Fig.~\ref{fig3}\textbf{b} is indicated by the \textcolor{yellow}{yellow} star. This symmetry point corresponds to $(\vartheta_\mathrm{ext}=0, \varphi_\mathrm{ext}=0)$. The coordinate is given by the \textcolor{red}{red} vector $(s_x,s_y)$ with respect to the center $(0,0)$ of the image. (\textbf{b}) Cropped vicinity of the symmetry point. This reduced area is scanned for faster flux calibrations. (\textbf{c}) $180^\circ$ rotation of the data presented in panel \textbf{b}. The symmetry point is displaced to $(-s_x,-s_y)$. (\textbf{d}) Correlation of the data in panels \textbf{b} and \textbf{c}. The axes denote the relative translation of \textbf{c} with respect to \textbf{b}. Large correlation represents high overlap between the two images. Notably, the point of maximal correlation occurs when \textbf{c} is translated by  $(2s_x,2s_y)$ with respect to \textbf{b}. From here, the symmetry point can be extracted.}
\end{figure}
\noindent In order to calibrate the flux crosstalk and to determine the Cooper-quartet tunneling regime, we monitor the response of the readout resonator at 7.47~$\mathrm{GHz}$ while sweeping the currents of both the off-chip coil ($I_\Sigma$) and on-chip differential flux line ($I_\Delta$) (dual flux spectroscopy). Since the coil magnetic field threads the entire loop of the Cooper-quartet element while the differential flux line addresses the subloop containing the phase-slip element, we roughly expect the coil to tune both $\mathrm{\vartheta_{ext}}$ and $\mathrm{\varphi_{ext}}$, while the differential flux line solely tunes $\mathrm{\varphi_{ext}}$. Importantly, because the coil tunes $\mathrm{\varphi_{ext}}$ at nearly half the rate of $\mathrm{\vartheta_{ext}}$, we expect interleaved features in the dual flux spectroscopy. We calibrate the flux crosstalk by evaluating the autocorrelation of the two-dimensional flux tuning data. Each peak in the autocorrelation matrix represents a linear combination of the primitive translation vectors. From here, we are able to extract the following flux crosstalk matrix, 
\begin{align}
    \begin{bmatrix}
           \mathrm{\varphi_{ext}} \\
            \mathrm{\vartheta_{ext}} \\
     \end{bmatrix}  
 = 
    2\pi\begin{bmatrix}
           1.06 & 1.09 \\
           0.10 & 2.28 \\
     \end{bmatrix} 
     \begin{bmatrix}
           I_\Delta \\
           I_\Sigma\\
     \end{bmatrix},
\end{align}
where the currents are in milliamperes and each entry in the crosstalk matrix accounts for the mutual inductance between the circuit loop and magnetic field sources such that the currents translate to flux (i.e. units of $\mathrm{H/\Phi_0}$).  

Determining the Cooper-quartet tunneling regime is equivalent to finding the symmetry points in the 2D flux map, since these correspond to $\mathrm{\vartheta_{ext}\in \{0, \pi\}}$ and $\mathrm{\varphi_{ext}\in \{0, \pi\}}$ as shown in Fig.~\ref{figs6}\textbf{a}. For this, we evaluate the correlation of the flux mapping (Fig.~\ref{figs6}\textbf{b}) with its $180^\circ$-rotated self (Fig.~\ref{figs6}\textbf{c}). Let $\mathrm{(0,0)}$ denote the point on the image about which the rotation is performed, and let ${\vec{s} = (s_x, s_y)}$ denote the coordinate of the closest symmetry point relative to the rotation point. From Fig.~\ref{figs6}\textbf{d}, it can be seen that the peak of the correlation matrix lies at ${2\vec{s} = (2s_x, 2s_y)}$, from which ${\vec{s}}$ can be determined. The symmetry point corresponds to ${\vartheta_\mathrm{ext}=\pi}$ if it lies in the interleaved region of the repeating patterns of the 2D flux map, since here the features repeat at double the frequency along the $\mathrm{\varphi_{ext}}$ axis. This corresponds to Cooper-quartet tunneling. By determining the flux crosstalk and symmetry points, we are able to execute spectroscopy of the gridium qubit. 

To dynamically access different operating regimes, we employ fast-flux pulses (FFPs) shaped as square baseband envelopes. It is well established that these pulses experience significant distortion before reaching the qubit, due to the frequency-dependent transfer characteristics of the control line. At short timescales, the dominant distortion arises from low-pass filtering effects introduced by inductive elements along the line, which attenuate high-frequency components and round the pulse edges. At longer timescales, high-pass behavior—originating from capacitive coupling or combined inductive-resistive elements—introduces baseline drift and undershoot artifacts. To correct for these distortions, we apply finite impulse response (FIR) filters to compensate for short-time low-pass effects, and infinite impulse response (IIR) filters to correct long-timescale distortions.~\cite{hellings2025calibrating}. 

\begin{figure}[t]
    \includegraphics[width=0.48\textwidth]{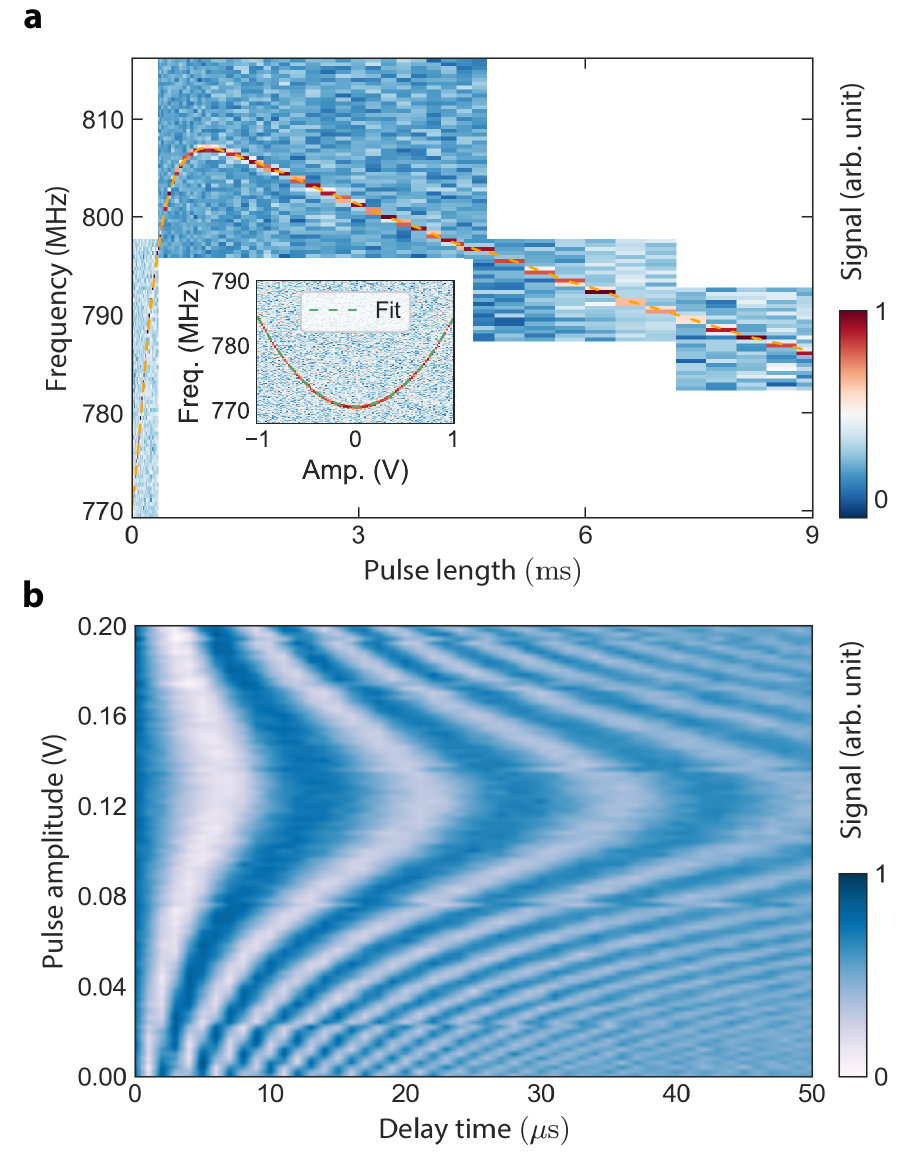}
    \caption{\label{figs_ff} \textbf{Fluxonium fast-flux experiment}. (\textbf{a}) Fast-flux spectroscopy, where the qubit is irradiated at the end of a flux pulse with varying length. (\textbf{b}) Fast-flux-assisted Ramsey spectroscopy, showing symmetric fringes as the qubit is driven across the flux inflection point.}
\end{figure}

To explore long-timescale distortions and validate our fast-flux pulse (FFP) corrections, we use a neighboring fluxonium qubit as a proxy probe. This qubit is coupled to the flux line via a finite mutual inductance, allowing it to sensitively track the local magnetic flux. The propagation of the baseband FFP through the cryogenic wiring to the probe qubit can be modeled as a linear time-invariant (LTI) system, with distortions described by a response function $\rho(t)$. The induced flux at the qubit translates into a time-dependent transition frequency, which can be approximated as $\omega_{01}(t)/2\pi=\lambda(t)$, where $\lambda$ is a nonlinear function of $\varphi_\mathrm{ext}(t)$. Focusing on long-timescale distortions, we extract the qubit frequency during the application of the FFP to infer $\rho(t)$, which can then be used to implement an IIR filter by inversion $\zeta(t)=\rho^{-1}(t)$.

To determine $\lambda(\varphi_\mathrm{ext})$, we apply a DC voltage to the same assembly using a bias-T at room-temperature, and perform regular two-tone spectroscopy at each flux point. Afterwards, we perform flux-assisted two-tone spectroscopy, where a 6~$\mathrm{\mu s}$-long dispersive readout procedure is performed at the end of the FFP. To ensure reset of the qubit, we lengthen the duty cycle of the sequence to 10~$\mathrm{ms}$. In addition, the starting length of the pulse is chosen to be 100-$\mathrm{ns}$ to exclude short-timescale distortions. Notably, the procedure can be applied to FFP with arbitrary length, which is useful to characterize distortion at millisecond timescale and correct long waveforms. We avoid performing this characterization near the sweet spot, where $\lambda(\varphi_\mathrm{ext})$ is ill-defined. 

In combination with $\lambda^{-1}(t)$, we sweep the length $\tau$ of the pulse and extract the step response of the flux line,
\begin{equation}
    \sigma(t)=\big( \alpha_0 + \sum_{i=1}^3\alpha_i e^{-\frac{t}{\tau_i}} \big)H(t),
\end{equation}
where $H(t)$ is the Heaviside time step function. Here, we include up to 3 exponential terms to avoid over-fitting to noisy data, and $\alpha_0$ is set to 0 (Fig.~\ref{figs_ff}\textbf{a}). We then implement a predistortion routine in the form of an impulse response function $\upsilon(t)$ using a Laplace transform, such that $\upsilon(t)\rho(t)\approx \mathrm{const}$~\cite{hellings2025calibrating}. After implementing the IIR filter, we again measure the FF spectrum, observing a residual frequency variation of less than 2~MHz. Treating the whole system as a nominal flux line with inherent distortion, we implement a second IIR filter, now setting $\alpha_0=1$, which allows us to decrease the residual error in term of qubit frequency to less than 1 MHz.

To benchmark the performance of the calibrated IIR filters, we characterize Ramsey fringes of a proxy qubit across the half-integer flux sweet spot. The qubit is biased at a flux-sensitive operating point, and Ramsey measurements are performed while applying the flux-fluctuation pulse (FFP). By gradually increasing the FFP amplitude to traverse the sweet spot, we observe a symmetric chevron pattern, which provides clear evidence of effective pulse predistortion (Fig.~\ref{figs_ff}\textbf{b}).

\begin{figure}[t]
    \includegraphics[width=0.48\textwidth]{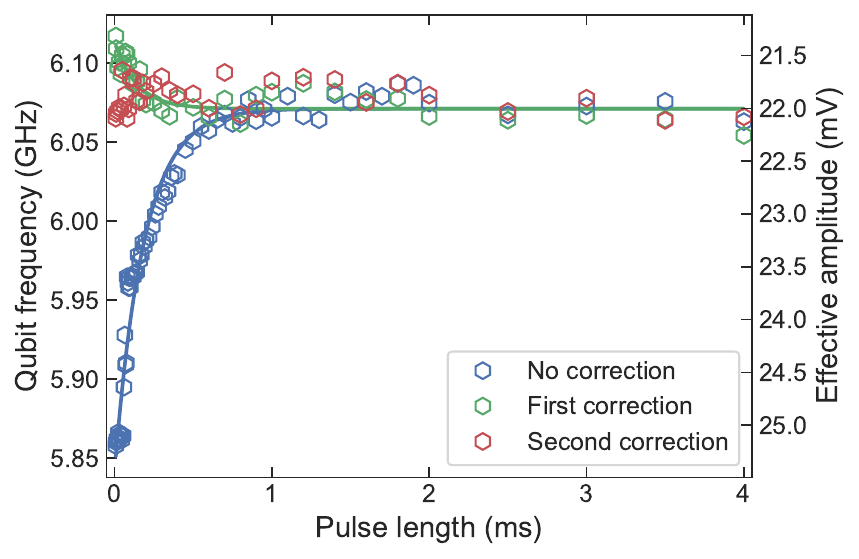}
    \caption{\label{figs_ff_gridium} \textbf{Gridium fast-flux correction}. Evolution of the qubit frequency and the corresponding effective pulse voltage as functions of pulse length, extracted from two-tone spectroscopy during the baseband pulse. The pulse amplitude is expected to shift the frequency from 7~GHz to 6.06~GHz. The data are fit to exponential functions (solid lines), whose coefficients are applied to predistort the flux pulse. A subsequent measurement with the predistorted pulse reveals an overshoot of the correction, which is mitigated by applying a second refinement to level the response across pulse length.}
\end{figure}

To perform fast-flux-assisted measurements of the gridium device, we adopt the procedure previously validated on fluxonium. Measurements are carried out in a spectral regime near the symmetric flux point, where the flux–frequency conversion is approximately linear. Fast-flux two-tone spectroscopy is then used to track the evolution of the flux threading the qubit loop, revealing a long stabilization time. The initial and final frequencies of this evolution are consistent with the applied dc flux bias, as verified independently. Exponential coefficients extracted from the data are subsequently used for pulse predistortion. The first correction, however, overshoots the target flux (Fig.~\ref{figs_ff_gridium}), necessitating a second refinement that restores the desired response.
\subsection*{Supplementary Note 11 -- Spectral fitting procedures}

\noindent We organize our spectral analysis into systematic measurement cycles. Each cycle begins by packaging and cooling a device containing gridium, fluxonium, and transmon qubits with distinct design parameters. Through spectral fitting, we extract the charging, inductive, and Josephson energies of these qubits. These measurements serve multiple purposes: they provide crucial feedback for circuit design optimization, help validate parameter targeting, and enable us to track coherence statistics. The statistical data allow us to evaluate fabrication consistency, confirm the reliability of the measurement setup, and optimize flux-pulse calibration.

The observed spectra are initially compared to extended Hamiltonian models representing ideal gridium and dualmon qubits. For systems with relatively small energy ratios and consequently less protection, we find good agreement between the low-frequency transitions and these ideal models. This aligns with our expectations, as the simplification from a 3-mode circuit to a 1-mode model assumes a low-frequency regime where the effects of other modes are negligible. The agreement progressively extends to higher-frequency regions as the energy ratios increase towards the ideal case (SI Note 5).  

Beyond the first few eigenstates, our analysis reveals good agreement between the measured spectra and the 3-mode Hamiltonian (given by Eq. \ref{eqn:gridium_3mode}) for frequencies below 5 GHz. Thus, to extract the parameters of the studied circuits, we fit the spectra below 5-GHz across the three regimes (along $\vartheta_\mathrm{ext}=0$, $\vartheta_\mathrm{ext}=\pi$, and $\varphi_\mathrm{ext}=0$). The flux mis-calibration in the gridium regime is accounted for by introducing a small linear term $\beta$ to the KITE flux bias, such that $\vartheta_\mathrm{ext}=\pi+\beta\varphi_\mathrm{ext}$. To avoid over-parameterizing the circuit model, we incorporate additional constraints in our fitting process, informed by feedback from the fluxonium and transmon qubits. For example, for the spectra shown in Fig.~3\textbf{c-e} corresponding to a less protected device, we impose $E_\mathrm{C}/h=0.76~\mathrm{GHz}$, $E_\mathrm{LK}\approx E_\mathrm{L}$, and $E_\mathrm{JS}\approx E_\mathrm{J}$. A simple least-square fit of all three spectra subsequently yields $[E_\mathrm{J}, E_\mathrm{L}, E_\mathrm{CS}, \varepsilon_\mathrm{K}]/h = [4.14, 0.77, 4.79, 2.48]~\mathrm{GHz}$, which all fall within the anticipated parameter range, validating our approach and circuit design. The visible flux mis-calibration parameter $\beta$ is also extracted and attributed to the resolution of the flux calibration.

Accurately modeling transitions above 5 GHz for this qubit necessitates the use of a 4-mode Hamiltonian that incorporates the parasitic capacitance $\varepsilon_\mathrm{P}$ (SI Note 4). Interestingly, the parameters extracted from this analysis fall outside our expected range. We attribute this discrepancy to additional high-frequency circuit modes not accounted for in our current model. For instance, we anticipate the array mode of this specific qubit to manifest in the 5.5-6.5 GHz range. We further note that fitting the spectrum below 5 GHz to the 4-mode model yields consistent result with the 3-mode model, with $\varepsilon_\mathrm{P}/h \geq 5.5~\mathrm{GHz}$. In cases where higher excited states are required to reproduce the observed spectrum—such as when sideband transitions involving array modes are present—a four-mode model must be employed. 

For instance, the spectrum of the protected device is captured by the parameter set $[E_\mathrm{J}, E_\mathrm{C},  E_\mathrm{L} , E_\mathrm{LK}, E_\mathrm{JS}, E_\mathrm{CS},\varepsilon_\mathrm{K}, \varepsilon_\mathrm{P}]/h = [6.67, 0.77, 0.3,0.53,1.25, 7.99, 2.07, 3.54]~\mathrm{GHz}$, with the array mode appearing at approximately 4.83 GHz. Looking ahead, future studies focusing on refining our understanding of circuit behaviors beyond the presented framework shall develop more comprehensive models that capture these high-frequency modes and their effects on qubit performance. In later devices, such as one shown in Extended Data Fig.~3, the array modes are lifted to above 8 GHz, resulting in simplification of the fitting. The 4-mode model fitting to the spectrum of this device yields the parameters $[E_\mathrm{J}, E_\mathrm{C},  E_\mathrm{L} , E_\mathrm{LK}, E_\mathrm{JS}, E_\mathrm{CS},\varepsilon_\mathrm{K}, \varepsilon_\mathrm{P}]/h = [7.21, 0.77, 0.73,0.73, 1.14, 8.13, 0.68, 1]~\mathrm{GHz}$. 

The per-cycle fitting times required for our procedures range from minutes with the extended GKP Hamiltonian to days with the 3-mode model and up to two weeks with the 4-mode model. This scaling underscores the importance of developing more efficient numerical methods, as optimized calculations will be essential for systematic exploration of multi-gridium devices and for guiding circuit design across increasingly complex parameter spaces.

\subsection*{Supplementary Note 12 -- Resistance probing}

\noindent In addition to characterizing device properties at cryogenic temperatures, our measurement cycle incorporates systematic probing of the room-temperature resistance between the qubit pads. This step is critical because the spectral fitting of the gridium does not reliably extract the asymmetry between the KITE components. Instead, we rely on room-temperature resistance measurements as part of our iterative design and fabrication refinement. By comparing the resistance between the two electrode pairs, we estimate the dispersion (or asymmetry) of the constituent circuit elements, defined as $\Delta R / R_\mathrm{avg}$.

In the first step, we measure inductively shunted KITE circuits having the same electrode design and finite asymmetry. Parameter extraction from cryogenic spectroscopy yields dispersions in the range of 1–10\%, in excellent agreement with the corresponding room-temperature resistance measurements (SI Note 3). Thus, we conclude that the probing at room temperature yields reliable signal on the circuit asymmetry. Subsequently, we perform resistance probing across all the gridium devices on the wafer, observing a dispersion ranging from under 1\% to over 15\%. Devices exhibiting the lowest asymmetry are prioritized for full cryogenic characterization.
\end{document}